\pdfoutput=1
\documentclass[12pt,english]{article}
\counterwithin{figure}{section}
\counterwithin{table}{section}
\counterwithin{equation}{section}
\usepackage[table]{xcolor}
\definecolor{mygray}{gray}{0.9}
\definecolor{beaublue}{rgb}{0.74, 0.83, 0.9}
\definecolor{junebud}{rgb}{0.74, 0.85, 0.34}
\definecolor{wisteria}{rgb}{0.79, 0.63, 0.86}
\usepackage[T1]{fontenc}
\usepackage[utf8]{inputenc}
\usepackage{booktabs}
\usepackage{threeparttable}
\definecolor{lavenderblue}{rgb}{0.8, 0.8, 1.0}
\usepackage{geometry}
\usepackage[english]{babel}
\usepackage{lastpage}
\usepackage{bbm}
\usepackage{upgreek}
\usepackage{iftex}
\usepackage{amsmath}
\usepackage{amsfonts}
\usepackage{longtable}
\usepackage{comment}
\usepackage{rotating}
\usepackage{multirow}
\usepackage{textcomp}
\usepackage{tikz}
\usetikzlibrary{shapes.geometric, arrows}

\tikzstyle{startstop} = [rectangle, rounded corners, 
minimum width=3cm, 
minimum height=1cm,
text centered, 
draw=black, 
fill=gray!20]

\tikzstyle{beginend} = [diamond, 
minimum width=1.5cm, 
minimum height=1cm,
text centered, 
draw=black, 
fill=gray!20]

\tikzstyle{io} = [rectangle, rounded corners, 
minimum width=3cm, 
minimum height=1cm, text centered, 
draw=black]

\tikzstyle{process} = [rectangle, 
minimum width=3cm, 
minimum height=1cm, 
text centered, 
text width=3cm, 
draw=black, 
fill=orange!30]

\tikzstyle{decision} = [diamond, 
minimum width=3cm, 
minimum height=1cm, 
text centered, 
draw=black, 
fill=green!30]
\tikzstyle{arrow} = [thick,->,>=stealth]
\usepackage{subcaption}
\usepackage{float}
\usepackage{setspace}
\usepackage{graphicx}
\usepackage{subcaption}
\usepackage{fancyhdr}
\fancypagestyle{plain}{%
	\fancyhf{}%
	\fancyfoot[C]{\thepage\ of \pageref*{LastPage}}}
\usepackage{array}
\usepackage{grffile}
\usepackage{pbox}
\usepackage{hhline}
\usepackage{etoolbox}
\definecolor{royalblue(traditional)}{rgb}{0.0, 0.14, 0.4}
\definecolor{steelblue}{rgb}{0.27, 0.51, 0.71}
\definecolor{mediumtealblue}{rgb}{0.0, 0.33, 0.71}
\usepackage{hyperref}
\hypersetup{
	colorlinks = true,
	allcolors = {mediumtealblue},
}

\usepackage{apacite}
\makeatletter
\usepackage{babel}
\usepackage{cleveref}
\title{Energy, economy, and emissions: A non-linear state space approach to projections}
\author{Mikkel Bennedsen\footnote{Department of Economics and Business Economics, Aarhus University, Fuglesangs Allé 4, 8210 Aarhus V, Denmark. }$\enspace$ Eric Hillebrand$^*$ and Jingying Zhou Lykke$^*$}
\begin{document}
	\maketitle
\begin{abstract}
	We propose a non-linear state-space model to examine the relationship between CO$_2$ emissions, energy sources, and macroeconomic activity, using data from 1971 to 2019. CO$_2$ emissions are modeled as a weighted sum of fossil fuel use, with emission conversion factors that evolve over time to reflect technological changes. GDP is expressed as the outcome of linearly increasing energy efficiency and total energy consumption. The model is estimated using CO$_2$ data from the Global Carbon Budget, GDP statistics from the World Bank, and energy data from the International Energy Agency (IEA). Projections for CO$_2$ emissions and GDP from 2020 to 2100 from the model are based on energy scenarios from the Shared Socioeconomic Pathways (SSP) and the IEA's Net Zero roadmap. Emissions projections from the model are consistent with these scenarios but predict lower GDP growth. An alternative model version, assuming exponential energy efficiency improvement, produces GDP growth rates more in line with the benchmark projections. Our results imply that if internationally agreed net-zero objectives are to be fulfilled and economic growth is to follow SSP or IEA scenarios, then drastic changes in energy efficiency, not consistent with historical trends, are needed.
\end{abstract}

\section{Introduction} 

Dramatically increasing amounts of energy have been produced and consumed worldwide since the pre-industrial era, which emit considerable amounts of  greenhouse gases (GHGs), especially CO$_2$, into the atmosphere. The massive energy use has driven rapid economic expansion both globally and regionally. Going forward, it is desirable to maintain economic prosperity while addressing the concerning global warming issue due to increasing atmospheric GHG concentration levels. However, the E3-nexus, i.e., the interaction between energy consumption, environmental preservation, and economic development poses challenges \shortcite{hamakawa2004background,mitic2023relationship}. As a response, the decoupling of economic growth from energy use and environmental impacts has come into focus in both academic research \shortcite<e.g.,>{stern2004rise,stern2004economic,csereklyei2015global,parrique2019decoupling,leitao2022new} and policy objectives such as the ``Sustainable Development Goals'' by the United Nations \cite{un2015}.

In this paper, we consider CO$_2$ emissions as a proxy for environmental impacts. We specify a non-linear state space model of the energy, economy, and emissions nexus, abbreviated as E3S2 model, which is a state space system of consumptions of the five primary energy sources (coal, oil, gas, nuclear, and renewables), GDP, and CO$_2$ emissions. 
 
The E3S2 model has several key features. First, it models time-varying emission factors. We represent CO$_2$ emissions as a weighted linear combination of fossil fuels. We allow the weights, emission conversion factors, to be time-varying. These factors are important because they affect the projection accuracy of CO$2$ emissions from fossil fuel use. Although it is recognized that they can change over time \cite<e.g.,>{hong1994carbon,roten2022co2}, many factor estimates are dated, and hence alignment with current conditions is no longer guaranteed \shortcite{omara2022methane,markovic2023tec}.  We derive and estimate the time variations of the emission conversion factors from historical data, which sheds light on how technological improvements or changes in the composites of fossil fuels impact these factors. 
 
Second, the E3S2 model captures non-stationarity.   The predominant nature of non-stationarity in the observations of variables poses challenges for statistical inference. Conventional approaches of handling non-stationarity in the context of the economy-emissions-energy nexus mainly include various cointegration methods such as panel cointegration \shortcite<e.g.,>{hamit2012greenhouse,lu2017greenhouse,zoundi2017co2}, panel vector error correction models  \shortcite<e.g.,>{pao2010co2,apergis2009co2}, and autoregressive distributed lags \shortcite<e.g.,>{yue2021role,adedoyin2021renewable}.  The state space methods employed in this paper allow for valid inference under non-stationarity \shortcite{caines1988}. As a point of departure, we model the energy productivity factor, i.e., the reciprocal of energy intensity,  and renewable energy consumption as a random walk with drift. Alternatively, a local linear trend form is employed if stochastic trends are found \cite{durbin2012time}. 
 
Third, the E3S2 model enables assessing parameter estimation uncertainty and incorporates both parameter and sampling uncertainty in the long-term projections and thus provides statistical confidence bands. A well-known class of models providing a comprehensive analysis of economy and environment is integrated assessment models (IAMs). IAMs adopt calibration to obtain parameter values and take averages of Monte Carlo runs of deterministic models with different parameter values drawn from a distribution to assess uncertainty \shortcite{crost2013optimal,nordhaus2018evolution}. However, as demonstrated by \shortciteA{crost2013optimal}, this approach can produce errors in both the magnitude and direction of the uncertainty.  In contrast, state space techniques obtain estimates of the unknown parameters from data using maximum likelihood estimation and allow for standard statistical inference.  

The remainder of the paper is structured as follows: Section \ref{model} introduces the three modules of the E3S2 model. Section \ref{simulation} describes the estimation of the E3S2 model using the extended Kalman filter and presents the results from a Monte Carlo simulation study as a verification of the estimation procedure. Section \ref{Data model selection} describes the data employed in this paper and introduces a data-driven model selection procedure. Section \ref{empirical} presents the results of the estimation of the E3S2 model to both regional and global data. Section \ref{scenario} presents long-term projections of CO$_2$ emissions, GDP, and the energy productivity factor conditional on the fuel mix pathways from the SSP 1.9 W$\mathrm{m^{-2}}$ scenario and the IEA Net Zero by 2050 roadmap. Section \ref{Conclusion} concludes and discusses future research directions.

\section{Specification of the E3S2  model}
\label{model}

This section presents the general specification of the E3S2 model, which nests many choices of structures. The model can be estimated on data at global, regional, or country level. The E3S2 model consists of three modules: energy, emissions, and economy. The energy and emissions modules consist of  linear equations, while the economy module contains a non-linear equation. 

\subsection{Energy}

The energy module contains specifications for each type of energy carrier in the fuel mix. We treat coal, oil, gas, and nuclear energy as exogenously given and set them equal to their observations. In the following, capital letters with asterisk $^*$ denote unobserved state variables and the same letters without asterisk their corresponding data observations. We model the various energy carriers as follows:
\begin{equation}
	\begin{aligned}
		C_{t}^{*} &=C_t,\enspace O_{t}^{*}=O_{t}, \enspace G_{t}^{*}=G_{t}, \enspace
		N_{t}^{*}=N_{t}, \\
			R_{t}^* & = d_{R,t-1}+ R_{t-1}^* + \eta_{R,t-1}, \\
			d_{R,t} & = d_{R,t-1} + \eta_{d_R,t-1},
	\end{aligned}
\label{energyState}
\end{equation}
where $(C_t, O_t, G_t, N_t, R_t)$ denote consumption quantities of the energy carriers coal, oil, gas, nuclear, and renewables, respectively. Renewable energy is specified as a random walk with time-varying drift. The drift process $d_{R,t}$ is a random walk itself, such that $R_t$ is modeled as a local linear trend \cite{durbin2012time}. The error processes $\eta_{R,t}$ and $ \eta_{d_R,t}$ are the state disturbances of  $R_t^*$ and $d_{R,t}$, and they follow the normal distributions $ \mathcal{N}\left(0,\sigma^2_{\eta_{R}}\right)$ and $ \mathcal{N}\left(0,\sigma^2_{\eta_{d_R}}\right)$, respectively. If the variance $\sigma_{\eta_{d_R}}^2$ of the error process of the drift is zero, the drift is a constant, and $R_t$ is a simple random walk with drift.

Treating the four fuel types $(C_t, O_t, G_t, N_t)$ as exogenous (i.e., setting them equal to their observations) allows the use of the E3S2 model for projections using future energy pathways of different common scenarios such as SSPs or the IEA Net Zero roadmap.  Modeling renewable energy instead of treating it as an exogenous variable not only allows for forecasting future renewable energy consumption conditional on trends inferred from historical data, but also facilitates conducting various scenario analyses regarding renewable energy, e.g., exploring the growth rate of renewable energy required to  maintain a certain level of GDP growth when assuming all fossil fuels are phased out.  See Section \ref{scenario} for more details.

The energy module is completed by a measurement equation that specifies the observations of renewable energy as the state $R^*$ plus a Gaussian error $\varepsilon_{R,t}$: 
\begin{equation}
	R_t=R_t^*+\varepsilon_{R,t}.
	\label{energyMeas}
\end{equation}

\subsection{Emissions}

We represent CO$_2$ emissions as a weighted average of consumptions of coal, oil, and gas \cite{marland1984carbon}. The weights are the so-called emission conversion factors, which are the amounts of CO$_2$ released into the atmosphere from combustion of one unit of a given fossil fuel type. The emission module is specified as follows:
\begin{equation}
	\begin{aligned}
		E^*_{t} &=\beta_{C,t} C_{t}^{*}+\beta_{O, t} O_{t}^{*}+\beta_{G, t} G_{t}^{*}+\eta_{E,t-1}, \\
	\beta_{C,t} & = \beta_{C,t-1}  +  \eta_{\beta_{C},t-1}, \\
	\beta_{O,t} & = \beta_{O,t-1}  +  \eta_{\beta_{O},t-1},\\
	\beta_{G,t}& =  \beta_{G,t-1}  +  \eta_{\beta_{G},t-1},
	\end{aligned}\\
		\label{emissionsState}
\end{equation}
where $\eta_{E,t} \sim \mathcal{N}\left(0,\sigma^2_{\eta_{E}}\right)$.
As specified in the first equation in system \eqref{energyState}, we equate states with data for $C^*_{t}=C_t$, $O^*_{t}=O_t$, and $G^*_{t}=G_t$, and the data act as regressors for $E_{t}^*$. The coefficients $\beta_{\cdot,t}$ denote the CO$_2$ emission conversion factors, which are time-varying and follow random walks, where $\eta_{\beta_\cdot,t} \sim \mathcal{N}\left(0,\sigma^2_{\eta_{\beta_\cdot}}\right)$. If the corresponding variances $\sigma_{\eta_{\beta_\cdot}}^2$ are set to or estimated as zero, the emission conversion factors are simply constants. Either constants or variances are estimated by maximum likelihood.

Allowing emission conversion factors $\beta_{\cdot,t}$ to be time-varying is a key feature of the E3S2 model. Although a baseline value can be derived from physics, the factor can vary temporally or across different countries or regions. This can be caused by differences in the composites of the fuel product \shortcite{hong1994carbon,roten2022co2} or by varying combustion efficiencies \cite{marland1984carbon}. Ignoring substantial time-varying behavior can result in misspecification and residual autocorrelation in the equation for $E^*_t$. A time-varying specification can result in overfitting, however, if time variation is minimal \shortcite{roten2022co2}. 

The emissions module is completed by a measurement equation:
\begin{equation}
	E_t=E_t^*+\varepsilon_{E,t},
	\label{emissionsMeas}
\end{equation}
where $E_{t}^*$ and $E_{t}$ represent the state variable and observations of CO$_2$, respectively, and $\varepsilon_{E,t} \sim \mathcal{N}\left(0,\sigma^2_{\varepsilon_E}\right)$.

\subsection{Economy}

In the economy module, the macroeconomic output state variable, $Y_t^*$, is specified as the product of the aggregation of the energy mix and the energy productivity factor $\beta_{Y,t}$: 
\begin{equation}
	\begin{aligned}
		Y_{t}^* &={ \left(C_{t}^{*}+ O_{t}^{*}+ G_{t}^{*}+ N_{t}^{*}+ R_{t}^{*}\right)\beta_{Y,t-1}}+\eta_{Y,t-1},  \\
		\beta_{Y,t} & = d_{\beta_{Y},t-1}+	\beta_{Y,t-1}  + \eta_{\beta_{Y},t-1},  \\
		d_{\beta_{Y}, t}&= d_{{\beta_{Y}}, t-1} +  \eta_{d_{\beta_{Y}}, t-1}.
	\end{aligned}
	\label{economyState}
\end{equation}
The process $\beta_{Y,t}$ is a measure of energy efficiency or energy productivity, i.e., it quantifies the amount of GDP produced by a unit of the energy mix. Summation of energy consumption is based on the assumption that economic growth does not distinguish the source of energy, i.e., all energy types share the same per-unit contribution to GDP. (See, however, \shortciteA{schurr1960} for a discussion of substitution of fossil fuels of different qualities and the effect on energy efficiency.) 

The residual processes $\eta_{Y,t}$, $\eta_{\beta_{Y},t}$, and $\eta_{d_{\beta_{Y}},t}$ are assumed to be Gaussian following $\mathcal{N}\left(0,\sigma^2_{\eta_Y}\right)$, $\mathcal{N}\left(0,\sigma^2_{\eta_{\beta_{Y}}}\right)$, and $\mathcal{N}\left(0,\sigma^2_{\eta_{d_{\beta_{Y}}}}\right)$, respectively.  Similar to $R_t^*$ in Equation \eqref{energyState}, $\beta_{Y,t}$ admits a flexible form of either a random walk ($\sigma_{\eta_{d_{\beta_Y}}}^2=0$) or a local linear trend ($\sigma_{\eta_{d_{\beta_Y}}}^2>0$). In the first equation in the system \eqref{economyState},  $R_{t}^{*}\beta_{Y,t-1}$ is a product of two state variables, rendering the E3S2 model a non-linear state space model. The product of two random walk states allows capturing a large class of non-linear functions, see \shortciteA[Supplement S1]{bennedsen2023multivariate}.

The economy module is completed by a measurement equation that specifies GDP observations as their unobserved state counterpart plus Gaussian noise,
\begin{equation}
	\begin{aligned}
		Y_{t} &=Y_{t}^{*}+\varepsilon_{Y,t}.
	\end{aligned}
	\label{economyMeas}
\end{equation}

In summary, the systems of equations \eqref{energyState}, \eqref{emissionsState}, and \eqref{economyState} comprise the non-linear state equations of the E3S2 model, and equations \eqref{energyMeas}, \eqref{emissionsMeas}, and \eqref{economyMeas} comprise  the measurement equations.

\section{Estimation and Monte Carlo simulation}
\label{simulation}

As the E3S2 model is a non-linear Gaussian state space model, we apply the extended Kalman filter, which is an approximate filter. It linearizes the state equations and measurement equations using Taylor series and applies the Kalman filter to the linearized model. See \citeA{durbin2012time} for a detailed description of the extended Kalman filter recursions. In Appendix \ref{EEEmatrixform}, we present the matrix form of the E3S2 model and the corresponding Jacobian matrix in the linearization of the non-linear state space model. 

We estimate the parameters using maximum likelihood evaluated by the Kalman filter. We maximize the log-likelihood using the Nelder-Mead simplex search method in the R function “optim” \cite{nelder1965simplex}. For non-stationary states, we adopt the “Big K” initialization technique to approximate the diffuse priors, where we set their initial variances to $K=10^6$ \cite{durbin2012time}. When an emission conversion factor is time-invariant, the corresponding state becomes a constant state, and we set the initial value of the mean to a constant and initial variance to zero. We conduct a Monte Carlo simulation study to assess the finite sample properties of the procedure. We consider the specification selected for OECD in Section \ref{Data model selection} and a sample size of 49, which equals the length of the historical sample size. The data-generating parameters are set equal to the parameter estimates obtained from the data, see Table \ref{estimates}.

In this specification, we assume that the emission conversion factor of gas is time-varying, and the  renewable energy state $R^*$ admits a stochastic trend. The parameters for this model include time-invariant emission conversion factors $\beta_{C}$ and $\beta_{O}$, the drift of $\beta_{Y_t}$, state disturbance variances $\sigma_{\eta_E}^2$, $\sigma^2_{\eta_{Y}}$, $\sigma^2_{\eta_{\beta_Y}}$, and $\sigma^2_{\eta_{R}}$, the variances of the error processes for the time-varying conversion factor of gas and stochastic trend of $R^*$, i.e.,  $\sigma_{\eta_{\beta_{ G}}}^2$  and  $\sigma_{\eta_{d_{{{R}}}}}^2$, and measurement error variances $\sigma_{\varepsilon_E}^2$, $\sigma^2_{\varepsilon_{Y}}$, and $\sigma^2_{{\varepsilon_R}}$. 
\begin{figure}[h]
	\centering
	\caption{\footnotesize Histogram of the differences of the parameter estimates from the data-generating parameter values in the Monte Carlo simulation. ``dX'' denotes the difference between the estimate of the parameter X and the corresponding data-generating value. The y-axis of each subfigure reports the relative frequencies. Solid yellow verticle lines and blue vertical lines represent the medians and means of dX, respectively. }
	\begin{subfigure}{\textwidth}
		\centering
		\includegraphics[width=\linewidth]{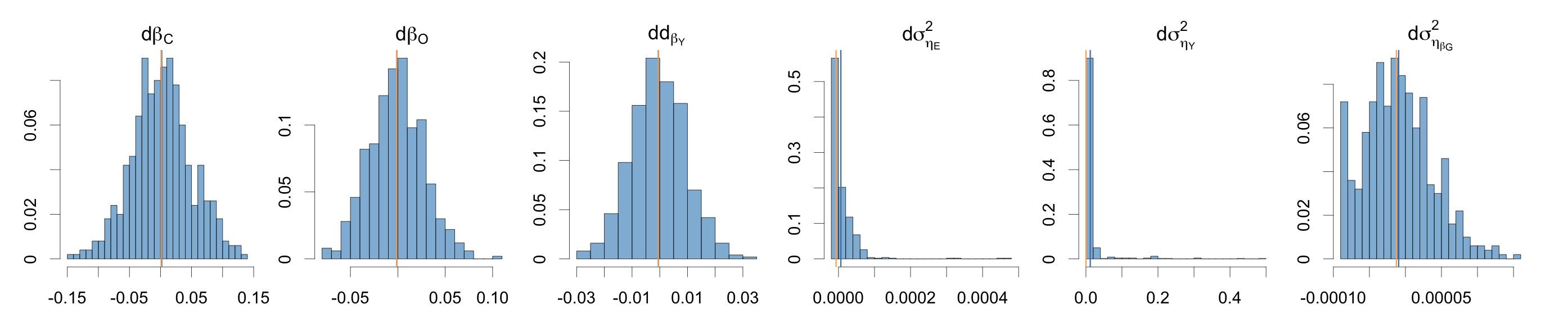}
	\end{subfigure}\\
	\begin{subfigure}{\textwidth}
		\centering
		\includegraphics[width=\linewidth]{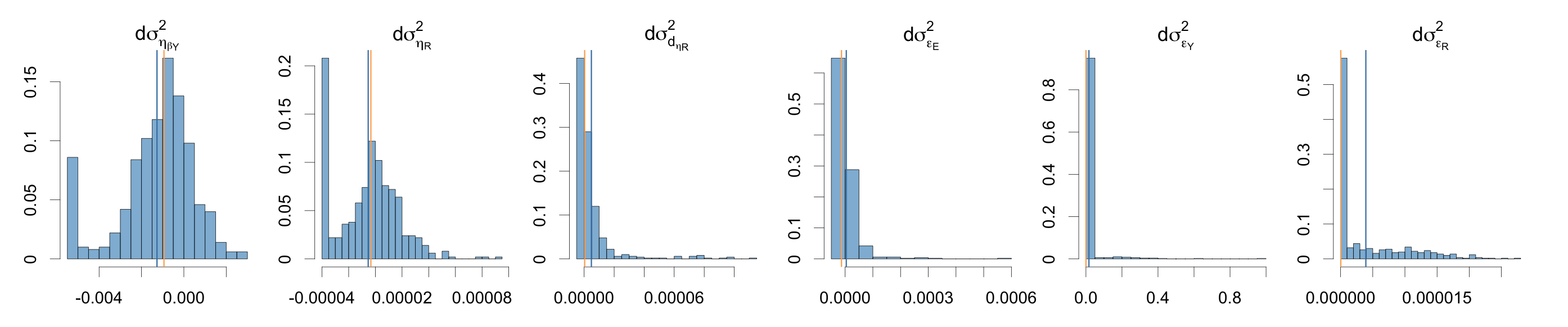}
	\end{subfigure}
	\label{Figuresimulation}
\end{figure}

Figure \ref{Figuresimulation} reports the distributions of parameter estimates in the Monte Carlo simulation across 500 runs. The differences d$\beta_{ C}$, d$\beta_{ O}$, and d$d_{\beta_{ Y}}$ of the parameter estimates from their data-generating values appear to be centered at zero and normally distributed. For the variances of state disturbances and measurement errors, the histograms cluster at zero, indicating that the model recovers the data-generating values despite their relatively small magnitudes.  Note that there are clusterings at $-0.004$ and $-0.00004$ for $d\sigma_{\eta_{\beta_{ Y}}}^2$ and $d\sigma_{\eta_{R}}^2$, as they are truncated to be non-negative in the algorithm and the data-generating values are 0.0053 and 0.000036, respectively. This is the so-called "pile-up" issue \cite<e.g.,>{shephard1993distribution}. In summary, the simulation exercise demonstrates that our estimation procedure has good finite-sample of properties.

\section{Data and model selection}
\label{Data model selection}

\subsection{Regions and data}

We estimate the E3S2 model at both global and regional levels. For comparability with the regional analysis in SSPs, we adopt the region definition from the SSPs and consider the five regions OECD, REF, ASIA, MAF, and LAM, see Table \ref{SSPregions} \shortcite{Riahi2017}.\footnote{A detailed list of countries in each region can be found at \url{https://tntcat.iiasa.ac.at/SspDb/dsd?Action=htmlpage\string&page=10\#regiondefs}.}
\begin{table}[h]
	\setlength{\tabcolsep}{3pt} 
	\renewcommand{\arraystretch}{1.3}
	\centering
	\caption{\footnotesize Five regions defined by SSPs.}
	\begin{tabular}{l|l}
		\hline \hline
\textbf{Acronym}	& \textbf{Description } \\
\hline
\textbf{OECD} & OECD and European Union (EU) member states and candidates \\
\textbf{REF}&  Eastern Europe and former Soviet Union \\
\textbf{ASIA} & most Asian countries (excl. the Middle East, Japan and former Soviet Union) \\
\textbf{MAF}& Middle East and Africa\\
\textbf{LAM}& Latin America and the Caribbean \\
		\hline\hline
	\end{tabular}
\label{SSPregions}
\end{table}

We collect country-level and global-level data of CO$_2$ emissions from the Global Carbon Budget\footnote{\url{https://www.globalcarbonproject.org/carbonbudget/22/data.htm}, last accessed on October 2, 2023.} \cite{friedlingstein2022global} and of total energy supply\footnote{Energy consumption data are not directly available.  We use the total energy supply, which equals production + imports - exports - international marine bunkers - international aviation bunkers + stock changes.} from the International Energy Agency (IEA).\footnote{\url{https://www.iea.org/data-and-statistics/data-sets}, last accessed on September 29, 2023.}  Country-level GDP data are from the World Bank.\footnote{\url{https://data.worldbank.org/indicator/NY.GDP.MKTP.CD}, last accessed on September 29, 2023.}  We aggregate country-level data into regions. The five regions defined by SSP contain 190 countries in total, while due to the smaller data coverage of the energy dataset, we only include 146 countries in our analysis. See Appendix \ref{listofcountries} for a list of countries that are excluded in this paper. The countries we do consider contribute approximately 96\%  of global emissions from fossil fuels. 

The CO$_2$ emissions data are production-based emissions estimates from fossil fuels. They include emissions from fossil fuel combustion,  oxidation, and cement production, and they exclude emissions from bunker fuels that are used for international aviation and maritime transport. Total energy supply data also exclude international marine and aviation bunkers. We use data series GDP (current local currency (LCU))(code: NY.GDP.MKTP.CN), Purchasing Power Parity (PPP) conversion factor (code: PA.NUS.PPP), and GDP deflator (code: NY.GDP.DEFL.ZS) to obtain GDP data measured in billions of 2005 USD (PPP adjusted). This enables the comparison of GDP data across different years and different countries. We obtain GDP data at the global level by summing up the country-level series. 
\begin{figure}[h]
	\centering
	\caption{\footnotesize Data of the five main sources of primary energy (Panels (a) -- (e)), CO$_2$ emissions (Panel (f)), and GDP (Panel (g)) during 1971 -- 2019. The units of energy data and emissions data are million tonnes of oil equivalent and gigatonne ($10^9$ tonne), respectively. The GDP data are in billion 2005 USD (PPP adjusted).}
	\begin{subfigure}{\textwidth}
		\centering
		\includegraphics[width=\linewidth]{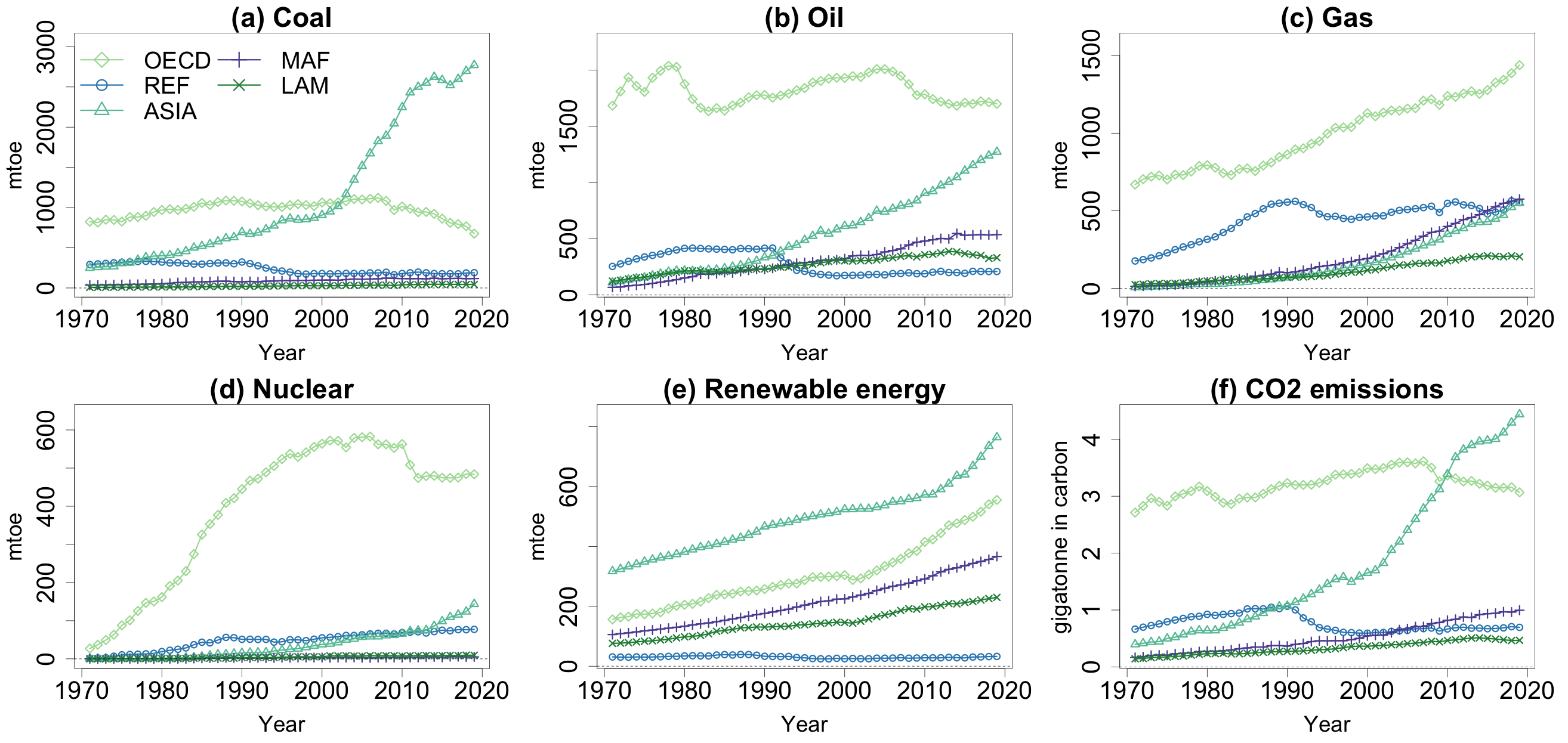}
	\end{subfigure}\\
	\begin{subfigure}{0.33\textwidth}
		\centering
		\includegraphics[width=\linewidth]{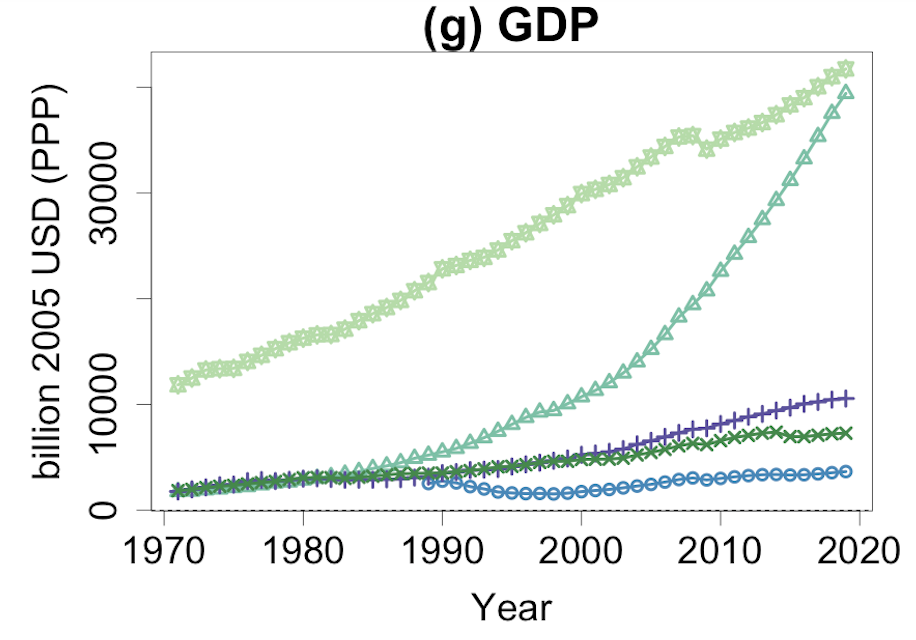}
	\end{subfigure}
	\label{data}
\end{figure}
\begin{table}[h]
	\setlength{\tabcolsep}{6pt} 
\renewcommand{\arraystretch}{1.3}
\centering
\caption{Share of each energy type in percentages for the five regions and the world in 2019. The values of each row sum up to 100. }
\begin{tabular}{l|cccccc}
	\hline\hline
\textbf{Region}  & \textbf{Coal} &  \textbf{Oil} &  \textbf{Gas} &  \textbf{Nuclear} &  \textbf{Renewable} \\ 
  \hline
\textbf{OECD} & 13.92 & 35.03 & 29.64 & 9.97 & 11.44 \\ 
  \textbf{REF} & 17.87 & 19.65 & 52.14 & 7.27 & 3.08 \\ 
  \textbf{ASIA} & 50.36 & 23.13 & 10.00 & 2.61 & 13.90 \\ 
  \textbf{MAF} & 7.32 & 33.53 & 35.94 & 0.31 & 22.90 \\ 
  \textbf{LAM} & 5.36 & 40.45 & 24.97 & 1.14 & 28.07 \\ 
    \textbf{WORLD} &26.67& 31.42 & 23.17 & 5.02& 13.72 \\ 
  \hline \hline
  \end{tabular}
\label{share}
\end{table}

Figure \ref{data} shows the data from 1971 to 2019, comprising 7 variables across 5 regions and the world. The data records for REF during 1971 -- 1989 come from the region ``Former Soviet Union'' in the IEA database. This region includes 15 post-Soviet countries, while REF includes only 12 post-Soviet countries.\footnote{Estonia, Latvia, and Lithuania are European Union (EU) member countries and are included in region OECD.}  In Panels (a) -- (e) in Figure \ref{data}, the five main sources of primary energy have been transformed to the same unit of million tonnes of oil equivalent (mtoe).

In terms of regional magnitudes of energy use, as shown in Panels (a), (b), and (d), the OECD region has been the largest consumer of oil, gas, and nuclear energy, and it used to be the leading coal consumer until it was overtaken by ASIA in 2003, which continuously experiences high growth rates in coal usage afterward. The total supplies of coal and oil in the OECD region have shown a decreasing trend since 2008 and 2006, respectively. This is in line with the observation that CO$_2$ emissions in Panel (f) start to decline in the same period. Meanwhile, as shown in Panel (g), GDP in the OECD region continuously grows with the exception of two years due to the financial crisis. This is a sign of decoupling of economic growth and emissions, i.e., CO$_2$ emissions decrease, while economic growth continues.\footnote{This finding, however, does not hold true for consumption-based emissions, see \shortciteA{bennedsen2023neural}.}  Renewable energy use has increased for all regions, with ASIA and OECD experiencing accelerating growth after 2000, MAF and LAM visually linear growth, and REF remaining relatively constant. At the global level, all of the seven variables are trending upward over the sample period, with renewable energy and GDP rising at increasing rates over the past two decades. 

Table \ref{share} presents the contributions in percentage from each of the primary energy sources for the five regions and the world in 2019.  Fossil fuels are still the major source of energy for the world (taking up more than 80\%) and all regions up to 2019. Coal is the largest contributor for ASIA, oil for OECD and LAM, and gas for REF and MAF. Notably, more than a quarter of the primary energy in LAM comes from renewables. Renewable energy also takes up a high share in MAF, as many African countries are abundant in renewables such as sunlight \shortcite{ieaafrica}.

\subsection{Model selection}
\label{model selection}

As shown in Figure \ref{data}, the primary energy supplies, CO$_2$ emissions, and GDP exhibit different characteristics across regions. Therefore, it is not feasible to incorporates all these features in a single statistical model simultaneously. In this section, we introduce a data-driven model selection procedure. It starts from a base model without time-varying parameter processes and hones in on a specification appropriate for the regional data.  We estimate the base model and test the residuals for serial correlation to decide which parameters to specify as time-varying. 

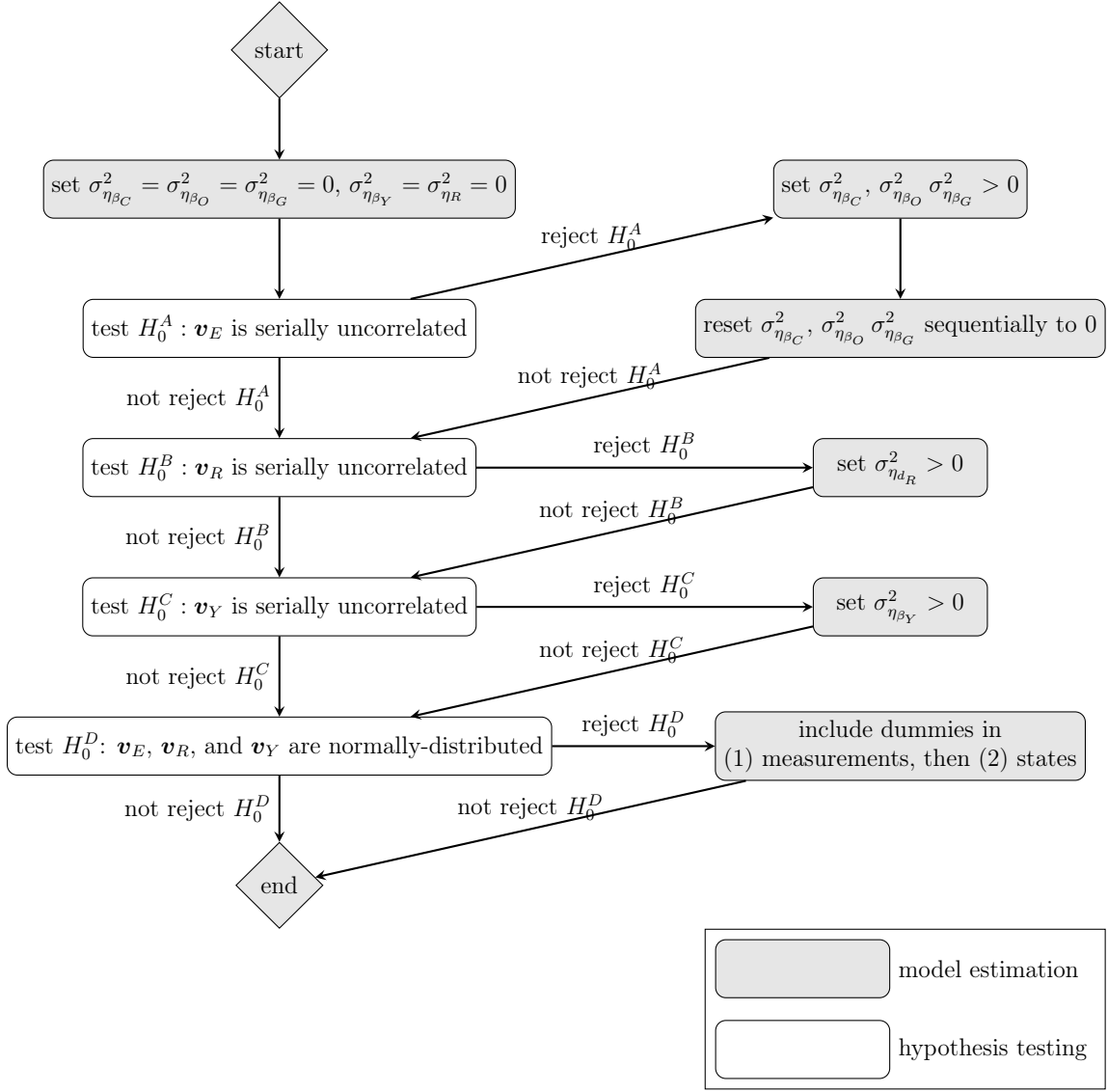
\begin{figure}[H]
	\centering
	\caption{\footnotesize Flow chart of the data-driven model selection procedure}
	\begin{tikzpicture}[node distance=2.4cm,scale=0.8,transform shape]
		\centering
		\node (begin) [beginend] {start};
		\node (start) [startstop, below of=begin] {set $\sigma_{\eta_{\beta_C}}^2=\sigma_{\eta_{\beta_O}}^2=\sigma_{\eta_{\beta_G}}^2=0$, $\sigma_{\eta_{\beta_Y}}^2=\sigma_{\eta_{R}}^2=0$};
		\node (in1) [io, below of=start] {test $H_0^A: \boldsymbol{v}_{E}$ is serially uncorrelated};
		\node (allTV) [startstop, right of=start, xshift=8.3cm] [align=center]{set $\sigma_{\eta_{\beta_C}}^2,\,\sigma_{\eta_{\beta_O}}^2\,\sigma_{\eta_{\beta_G}}^2>0$};
		\node (trial) [startstop, below of=allTV] [align=center]{reset $\sigma_{\eta_{\beta_C}}^2,\,\sigma_{\eta_{\beta_O}}^2\,\sigma_{\eta_{\beta_G}}^2$ sequentially to 0};
		\node (pro1) [io, below of=in1] {test $H_0^B: \boldsymbol{v}_{R}$ is serially uncorrelated};
		\node (pro2) [io, below of=pro1] {test $H_0^C: \boldsymbol{v}_{Y}$ is serially uncorrelated};
		\node (RLL) [startstop, right of=pro1, xshift=8.3cm] {set $\sigma_{\eta_{d_R}}^2>0$};
		\node (YLL) [startstop, right of=pro2, xshift=8.3cm] {set $\sigma_{\eta_{\beta_Y}}^2>0$};
		\node (pro3) [io, below of=pro2] {test $H_0^D$: $\boldsymbol{v}_{E}$, $\boldsymbol{v}_{R}$, and $\boldsymbol{v}_{Y}$ are normally-distributed};
		\node (end) [beginend, below of=pro3] {end};
		\node (dummy) [align=center][startstop, right of=pro3, xshift=8.3cm] {include dummies in \\ (1) measurements, then (2) states};
		\draw [arrow] (begin) -- (start);
		\draw [arrow] (start) -- (in1);
		\draw [arrow] (allTV) -- (trial);
		\draw [arrow] (in1) -- node[anchor=east]{not reject $H_0^A$}(pro1);
		\draw [arrow] (trial) -- node[anchor=south]{not reject $H_0^A$}(pro1);
		\draw [arrow] (in1) -- node[anchor=south]{reject $H_0^A$}(allTV);
		\draw [arrow] (pro1) -- node[anchor=south]{ reject $H_0^B$}(RLL);
		\draw [arrow] (pro1) -- node[anchor=east]{not reject $H_0^B$}(pro2);
		\draw [arrow] (pro2) -- node[anchor=south]{reject $H_0^C$}(YLL);
		\draw [arrow] (pro2) -- node[anchor=east]{not reject $H_0^C$}(pro3);
		\draw [arrow] (RLL) -- node[anchor=south]{not reject $H_0^B$}(pro2);
		\draw [arrow] (YLL) -- node[anchor=south]{not reject $H_0^C$}(pro3);
		\draw [arrow] (pro3) -- node[anchor=south]{reject $H_0^D$}(dummy);
		\draw [arrow] (pro3) -- node[anchor=east]{ not reject $H_0^D$}(end);
		\draw [arrow] (dummy) -- node[anchor=south]{ not reject $H_0^D$}(end);
		\matrix [draw,column sep=1ex,below left,row sep=3mm,nodes={scale=0.8}] at (current bounding box.south east) {
			\node [startstop,label=right:model estimation] {}; \\
			\node [io,label=right:hypothesis testing] {}; \\
		};
	\end{tikzpicture}
	\label{flowchart}
\end{figure}

Figure \ref{flowchart} presents a diagram of the model selection procedure. The point of departure is the base specification estimated on the data, and four hypothesis tests A, B, C, and D are conducted sequentially on the residuals, which are the standardized one-step ahead prediction errors series $\boldsymbol{v}_X$, where $X\in \left\{E,Y,R\right\}$. The first three hypothesis tests A, B, and C are Ljung-Box tests for autocorrelation \cite{box1970distribution,ljung1978measure}. We consider lag orders 1 and 5 and significance levels $1\%$ and $5\%$.  The null hypothesis $H_0$ is the absence of serial correlation. If the null of test A, $H_0^A$, is rejected, constant emission conversion factors cannot capture changes in the relationship between emissions and fossil fuel consumption. Then, all conversion factors are set to be time-varying, and the model is re-estimated. The smoothed conversion factors that remain almost constants are reset to constants. The procedure is repeated until we find a specification where $H_0^A$ cannot be rejected. 

In tests B and C, we test whether autocorrelation remains in the residuals of renewables and GDP, respectively. Upon rejection of the null we incorporate a stochastic trend for $R^*$ in Equation \eqref{energyState} or/and $\beta_Y$ in Equation \eqref{economyState}. We test $v_R$ first because $R^*$ is an element in the equation for GDP. Modeling $R^*$ using a local linear trend model can help remove autocorrelation in the residuals of GDP, if $R^*$ is the only source of a stochastic trend in GDP. 

Test D employs the Jarque-Bera \cite{jarque1980efficient} test for non-normality of the three residual processes $v_E$, $v_Y$, and $v_R$. In case of rejection, we introduce dummy variables for large spikes first in the corresponding measurement equations, then in the state equations, until the test fails to reject. We ignore the issue of multiple testing since the nominal size of the tests in this model building cycle is not material.

We apply the model selection procedure to the five regions and the world and obtain a specification for each. Table \ref{specification} summarizes the specifications. 
\begin{table}[h]
	\setlength{\tabcolsep}{7pt} 
	\renewcommand{\arraystretch}{1.2}
	\centering
	\caption{\footnotesize Summary of specifications nested in the general form of E3S2 model. ``$\circ$'' indicates a particular structure is needed. ``LL'' denotes the local linear trend model ($\sigma_{\eta_{\beta_Y}}^2,\,\sigma_{\eta_{R}}^2>0$).}
	\begin{tabular}{l|cccccccccccccc}
		\hline \hline
		\multirow{2}{*}{	\textbf{Region}}		& 	\textbf{time-varying } & 	\textbf{time-varying }  & 	\textbf{time-varying }  & \textbf{LL in}  & \textbf{LL in } & \textbf{No. of}\\
		& $\boldsymbol{\beta}_{C}$ & $\boldsymbol{\beta}_O$ & $\boldsymbol{\beta}_G$ &$\boldsymbol{\beta}_Y$ & $\boldsymbol{R^*}$& \textbf{dummies}\\
		\hline
		\textbf{World}&$\circ$ & &&$\circ$ &$\circ$ & 3 \\
	\textbf{OECD} &  &  & $\circ$  & & $\circ$& 1\\ 
	\textbf{REF}&$\circ$ &  & &$\circ$  &&  0\\
\textbf{ASIA}&$\circ$ & &&$\circ$ &$\circ$ & 2 \\
	\textbf{MAF} & $\circ$ &  &  & $\circ$& $\circ$ &  6 \\
	\textbf{LAM}&$\circ$ & & & &  &  0  \\
		\hline\hline
	\end{tabular}
\label{specification}
\end{table}
Note that four types of specifications are selected: (i) $\beta_{C}$ is time-varying, (ii) $\beta_{C}$ is time-varying, $\beta_{Y}$ is local linear trend, (iii) $\beta_{G}$ is time-varying, $R^*$ is local linear trend, and dummy variables are added, (iv) $\beta_{C}$ is time-varying, both $\beta_{Y}$ and $R^*$ are local linear trends, and dummy variables are added.

\section{Estimation on global and regional data}
\label{empirical}

Employing the specifications selected in Section \ref{model selection}, we fit the models to the data series for each region and estimate the parameters using maximum likelihood via the output from the extended Kalman filter. Table \ref{estimates} presents the results, where we estimate the regionally averaged and globally averaged conversion factors empirically.\footnote{We refrain from reporting the estimated variances. They are available upon request.}  Emission conversion factors represent the amount of CO$_2$ emissions per unit of combusted energy carrier; smaller values indicate that the energy carrier is less polluting.  

Table \ref{estimates} shows that the majority of time-invariant emission conversion factor estimates are in line with numbers reported in the literature and presented in Panel E. Emission conversion factor estimates vary across regions, consistent with the statements in \citeA{hong1994carbon} and \citeA{roten2022co2}. An exception is ASIA with a higher $\widehat{{\beta}}_O$ and lower $\widehat{{\beta}}_G$, but these two estimates also have larger standard errors. Considering the uncertainty and confidence intervals, they are in good agreement with the benchmarks. 

\begin{table}[H]
	\centering
	\small
	\setlength{\tabcolsep}{5pt} 
	\renewcommand{\arraystretch}{1.1} 
	\caption{\footnotesize Point estimates and standard errors (in parentheses) of CO$_2$ emission conversion factors $\beta_{ C}$, $\beta_{O}$, and $\beta_{G}$, drifts $d_{\beta_{Y}}$ and $d_R$, and coefficients of dummies $D$. The superscript of $D$ indicates whether the dummy is included in the state equation ($S$) or the measurement equation ($M$), and the subscript refers to the variable name and time index of the dummy. Emission conversion factors are expressed in units of tonne CO$_2$ equivalent per tonne of oil equivalent (tonne CO$_2$e/ toe). Estimates of the time-varying emission conversion factors (labeled ``TV'') are reported in the figure below the table. Panel E reports the emission conversion factor estimates from \protect \citeA{marland1984carbon} and IPCC emission factor database \protect \shortcite{manginoestablishment} for comparison. }
	\begin{threeparttable}
		\small
		\begin{tabular}{l|cccccccccc}
			\hline \hline
			\multicolumn{9}{c}{\textbf{A. $\beta_{ C}$ is time-varying}} &
			\\
			\hline
			& $\widehat{\boldsymbol{\beta}}_{C}$ & $\widehat{\boldsymbol{\beta}}_{O}$ & $\widehat{\boldsymbol{\beta}}_{G}$ & $\widehat{\boldsymbol{d}}_{\beta_Y}$ & $\widehat{\boldsymbol{d}}_R$\\
			\hline
			\multirow{2}{*}{\textbf{LAM}}	& 	\multirow{2}{*}{{TV}}&3.3775 & 2.3005 & 0.0196 & 0.0032  \\ 
			&  &(0.2291) & (0.3002) & (0.0199) & (0.0004) \\ 
			\hline
			\multicolumn{9}{c}{\textbf{B. $\beta_{ C}$ is time-varying and $\beta_{Y}$ has a local linear trend form}} &
			\\
			\hline
			& $\widehat{\boldsymbol{\beta}}_{C}$ & $\widehat{\boldsymbol{\beta}}_{O}$ & $\widehat{\boldsymbol{\beta}}_{G}$ &  $\widehat{\boldsymbol{d}}_R$\\
			\hline
			\multirow{2}{*}{\textbf{REF}}	& 	\multirow{2}{*}{{TV}} &3.7799 & 1.9663 & 0.00004 &&\\ 
			& &(1.6623) & (0.3399)  & (0.0002)\\ 
			\hline
			\multicolumn{9}{c}{\textbf{C. $\beta_{G}$ is time-varying, $R^*$ has a local linear trend form, and dummies needed}} 
			\\
			\hline
			&  $\widehat{\boldsymbol{\beta}}_{C}$ & $\widehat{\boldsymbol{\beta}}_{O}$ & $\widehat{\boldsymbol{\beta}}_{G}$ & $\widehat{\boldsymbol{d}}_{\beta_Y}$ & $\widehat{\boldsymbol{D}}^S_{E,1990}$  & \\
			\hline
			\multirow{2}{*}{\textbf{OECD}}	&3.9852 & 2.6785& 	\multirow{2}{*}{{TV}} & 0.1075  & 0.036  \\ 
			& (0.5086) & (0.2098) &  & (0.0106) & (0.009) && \\ 
			\hline 
			\multicolumn{10}{c}{\textbf{D. $\beta_{G}$ is time-varying, $\beta_{Y}$ and $R^*$ have a local linear trend form, and dummies needed}} 
			\\
			\hline
			& $\widehat{\boldsymbol{\beta}}_{C}$ & $\widehat{\boldsymbol{\beta}}_{O}$ & $\widehat{\boldsymbol{\beta}}_{G}$&  $\widehat{\boldsymbol{D}}^S_{E,2004}$ & $\widehat{\boldsymbol{D}}^M_{E,1990}$\\
			\hline
			\multirow{2}{*}{\textbf{ASIA}} & 	\multirow{2}{*}{{TV}} &  4.2428 & 1.3744 & -0.0778 & -0.0809\\ 
			& & (1.9129) & (2.473)  & (0.0227) & (0.0119)  \\ 
			\hline 
			&   $\widehat{\boldsymbol{\beta}}_{C}$ & $\widehat{\boldsymbol{\beta}}_{O}$ & $\widehat{\boldsymbol{\beta}}_{G}$	& $\widehat{\boldsymbol{D}}^S_{Y,1982}$ &$\widehat{\boldsymbol{D}}^M_{R,2000}$&$\widehat{\boldsymbol{D}}^M_{R,2004}$&$\widehat{\boldsymbol{D}}^M_{R,2011}$&$\widehat{\boldsymbol{D}}^M_{R,2012}$& $\widehat{\boldsymbol{D}}^M_{R,2013}$\\
			\hline
			\multirow{2}{*}{\textbf{MAF}}&	\multirow{2}{*}{{TV}} & 1.918 & 2.992& -0.1103 &-0.0026& 0.0014& 0.0017 & 0.0053 & 0.0040\\ 
			& & (1.2612) & (0.7784) & (0.0834) & (0.0007) & (0.0007) & (0.0009) & (0.0010) & (0.0009) \\ 
			\hline 
			&  $\widehat{\boldsymbol{\beta}}_{C}$ & $\widehat{\boldsymbol{\beta}}_{O}$ & $\widehat{\boldsymbol{\beta}}_{G}$	& $\widehat{\boldsymbol{D}}^S_{E,1991}$ & $\widehat{\boldsymbol{D}}^S_{Y,1990}$ &$\widehat{\boldsymbol{D}}^M_{Y,1989}$\\
			\hline
			\multirow{2}{*}{\textbf{WORLD}}&	\multirow{2}{*}{{TV}} &2.8984 & 2.8848  & 0.149 &1.0447 & 1.1089\\ 
			& & (0.4132) & (0.4328) & (0.020) & (0.2531) & (0.2612) \\ 
			\hline
		\end{tabular}
		\centering
	\begin{subtable}{\textwidth}
			\begin{tabular}{l|cccccccccc}
			\hline 
			\multicolumn{5}{c}{\textbf{E. Emission conversion factor estimates from other literature}} 
			\\
			\hline
			& $\widehat{\boldsymbol{\beta}}_{C}$ & $\widehat{\boldsymbol{\beta}}_{O}$ & $\widehat{\boldsymbol{\beta}}_{G}$&  \\
			\hline
			\protect \citeA{marland1984carbon}	& 2.6841 & 2.8590 & 2.0560 \\
			\hline \shortciteA{manginoestablishment}$^\text{a}$
			& 4.1112 & 3.0681 & 2.3470 \\
			\hline\hline
		\end{tabular}
	\end{subtable}
		\begin{tablenotes}
			\item[a] IPCC emission factor database (EFDB).
		\end{tablenotes}
		\centering
		\begin{subfigure}{\textwidth}
			\subcaption*{Smoothed states of the time-varying emission conversion factors. Confidence bands are pointwise at the 90\% level.}
			\includegraphics[width=\linewidth]{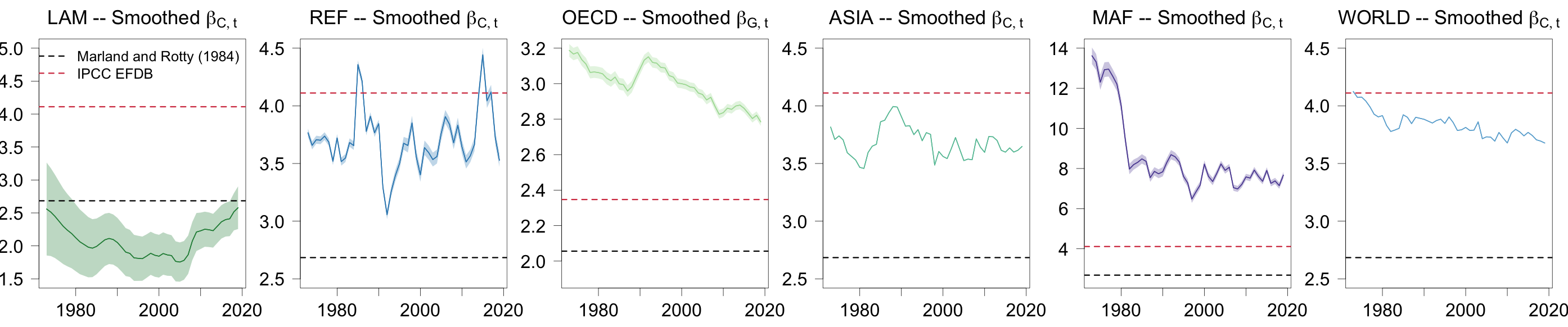}
		\end{subfigure}
	\end{threeparttable}
	\label{estimates}
\end{table}
\begin{table}[H]
	\centering
	\caption{Diagnostic statistics of the standardized one-step ahead prediction errors of emissions, GDP, and renewables. It reports the first four moments (mean, standard deviation (Std), skewness (Skew), and excess kurtosis (Kurt) as well the test statistics of the Jarque-Bera test for non-normality \protect \cite{jarque1980efficient} (JB) and of the Ljung-Box test for autocorrelation \protect \cite{box1970distribution,ljung1978measure} (Q(1) corresponds to a lag order of 1 and Q(5) corresponds to a lag order of 5.) We consider significance levels of 1\% (**) and 5\% (*). The null hypothesis of the Jarque-Bera is skewness equal 0 and kurtosis equal 3. The null hypothesis of the Ljung Box test is that the residuals are serially uncorrelated when a fixed number of lags are included. }
	\setlength{\tabcolsep}{5pt} 
	\renewcommand{\arraystretch}{1.1} 
	\begin{tabular}{l|lcccccccccccccc}
		\hline \hline
		\multicolumn{11}{c}{\textbf{A. $\beta_{ C}$ is time-varying}} \\
		\hline 
		&	& Mean & Std & Skew & Kurt & JB  & Q(1)   & Q(5) \\ 
		\hline
		\multirow{3}{*}{\textbf{LAM}}	&	Emissions & -0.113 & 0.993 & 0.435 & -0.294 & 1.673 & 2.331 & 3.889 \\ 
		&	GDP & 0.000 & 1.000 & -0.229 & 0.225 & 0.712 & 0.780 & 3.255 \\ 
		&	Renewables & 0.001 & 1.011 & 0.158 & -0.633 & 0.779 & 1.779 & 7.673 \\ 
		\hline
		\multicolumn{11}{c}{\textbf{B. $\beta_{ C}$ is time-varying, and $\beta_{ Y}$ has a local linear trend form}} \\
		\hline 
		&	& Mean & Std & Skew & Kurt & JB  & Q(1)   & Q(5)  \\ 
		\hline
		\multirow{3}{*}{\textbf{REF}}	&	Emissions &-0.027 & 1.000 & 0.220 & 0.905 & 2.714 & 0.177 & 5.108  \\ 
		&	GDP &-0.035 & 0.977 & -0.648 & -0.202 & 2.391 & 0.723 & 5.653 \\ 
		&	Renewables & 0.001 & 1.011 & -0.461 & 0.502 & 2.659 & 0.702 & 3.993  \\ 
		\hline
		\multicolumn{11}{c}{\textbf{C. $\beta_{G}$ is time-varying, $R^*$ has a local linear trend form, and dummies needed}} \\
		\hline
		&	& Mean & Std & Skew & Kurt & JB  & Q(1)   & Q(5)  \\ 
		\hline
		\multirow{3}{*}{\textbf{OECD}}	&	Emissions	 &-0.266 & 0.963 & 0.200 & -0.891 & 1.624 & 0.278 & 3.814\\
		&	GDP & 0.016 & 0.957 & 0.283 & -0.075 & 0.686 & 2.498 & 4.326 \\ 
		&	Renewables& 0.123 & 0.992 & -0.095 & 0.561 & 1.097 & 0.064 & 1.109 \\ 
		\hline
		\multicolumn{11}{c}{\textbf{D. $\beta_{C}$ is time-varying; $\beta^Y$ and $R^*$ have local linear trend form, and dummies needed}} 
		\\
		\hline
		&	& Mean & Std & Skew & Kurt & JB  & Q(1)   & Q(5)  \\ 
		\hline
		\multirow{3}{*}{\textbf{ASIA}}	&	Emissions & -0.043 & 0.999 & -0.040 & 0.973 & 2.629 & 0.317 & 4.164&  \\ 
		&	GDP& 0.027 & 0.983 & 0.256 & 0.553 & 1.554 & 1.127 & 6.273 \\ 
		&	Renewables & 0.169 & 0.983 & 0.106 & 0.609 & 1.264 & 0.007 & 13.553$^*$ \\
		\hline 
		\multirow{3}{*}{\textbf{MAF}}&	Emissions	& -0.183 & 0.983 & -0.733 & 0.303 & 4.976 & 0.997 & 2.634  \\ 
 	&	GDP &-0.016 & 0.988 & -0.088 & 0.902 & 2.355 & 0.260 & 1.481  \\ 
		&	Renewables & 0.169 & 0.987 & -0.334 & 0.951 & 3.464 & 0.053 & 18.945$^{**}$ \\ 
		\hline 
		\multirow{3}{*}{\textbf{WORLD}}&Emissions & -0.203 & 0.979 & 0.212 & 0.949 & 2.888 & 1.550 & 9.694  \\ 
		& GDP & 0.053 & 0.974 & 0.415 & 0.095 & 1.569 & 0.254 & 9.433   \\ 
		& Renewables  & 0.181 & 0.984 & 0.210 & -0.291 & 0.437 & 0.019 & 2.148   \\ 
		\hline\hline
	\end{tabular}
	\label{Diagnostics}
\end{table}

Five out of six time-varying emission factors are moving within the ranges comparable to the benchmarks of either \citeA{marland1984carbon} or the IPCC emission factor database (EFDB) \shortcite{manginoestablishment}. MAF has a high estimated $\beta_{C,t}$ starting from over 12 tonnes CO$_2$e/ toe but rapidly falling to eight. LAM, MAF, and WORLD exhibit downward trends in $\beta_{C,t}$, indicating ``cleaner'' coal usage in terms of CO$_2$ emissions over the years. The decreasing smoothed $\beta_{G,t}$ suggests gas becomes less polluting over time in the OECD region. 

Table \ref{Diagnostics} reports diagnostics on the residuals to evaluate the performance of the model fits. Except for ASIA and MAF where autocorrelation cannot be rejected at a lag order of 5, all the other residual processes appear to be normally distributed and serially uncorrelated. This indicates that the specifications provide a good fit to the data, and no missing dynamics need to be accounted for. The smoothed states and residuals are shown in Appendix \ref{KFoutput}. 
	\begin{figure}[H]
	\caption{\footnotesize CO$_2$ emissions (y- axis, gigatonne in carbon) against GDP (x-axis, billion 2005 USD (PPP)). The solid lines represent smoothed states, while circles represent historical data. }
	\includegraphics[width=\linewidth]{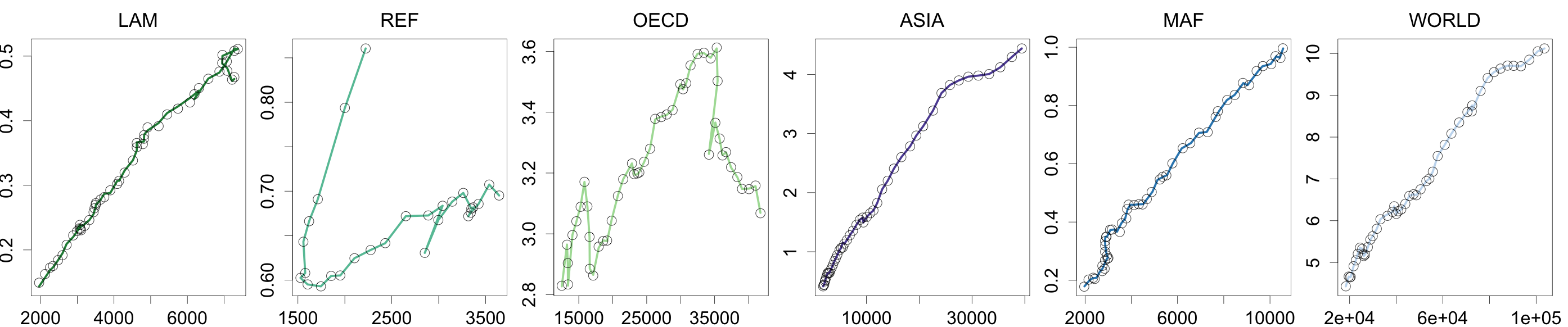}
	\label{EKC}
\end{figure}
Figure \ref{EKC} shows the smoothed states and historical data of CO$_2$ emissions on the y-axis and of GDP on the x-axis. For the OECD region, the graph shows an inverse U-shape. This is the environmental Kuznets curve (EKC), reflecting a relationship between economic growth and environmental impact  \cite{grossman1991environmental,castle2020climate}, here estimated from the output of the Kalman smoothing recursions. The hypothesis is that in the early stages of economic growth, environmental externalities increase, in this case increases in CO$_2$ emissions. Once a turning point is reached, economic expansion improves environmental quality. Note that CO$_2$ emissions data employed in this paper are production-based. None of the other regions display an EKC shape. The picture for the OECD changes when using consumption-based data, as discussed in  \shortciteA{bennedsen2023neural}.

\section{Scenario analysis}
\label{scenario}

In this section, we conduct scenario analysis with the E3S2 model conditional on energy pathways from the SSPs and the IEA Net Zero Roadmap.

\subsection{Long-term projections}
\label{approach}

\begin{figure}[h!]
	\centering
	\caption{\footnotesize SSP 1.9  W$\mathrm{m^{-2}}$ projections of primary energy use (unit: million tonnes of oil equivalent (mtoe)) at 10-year frequency over the period 2020 -- 2100 \protect \cite{rogelj2018scenarios}.}
	\begin{subfigure}{\textwidth}
		\centering
		\subcaption*{SSP1-1.9}
		\includegraphics[width=0.9\linewidth]{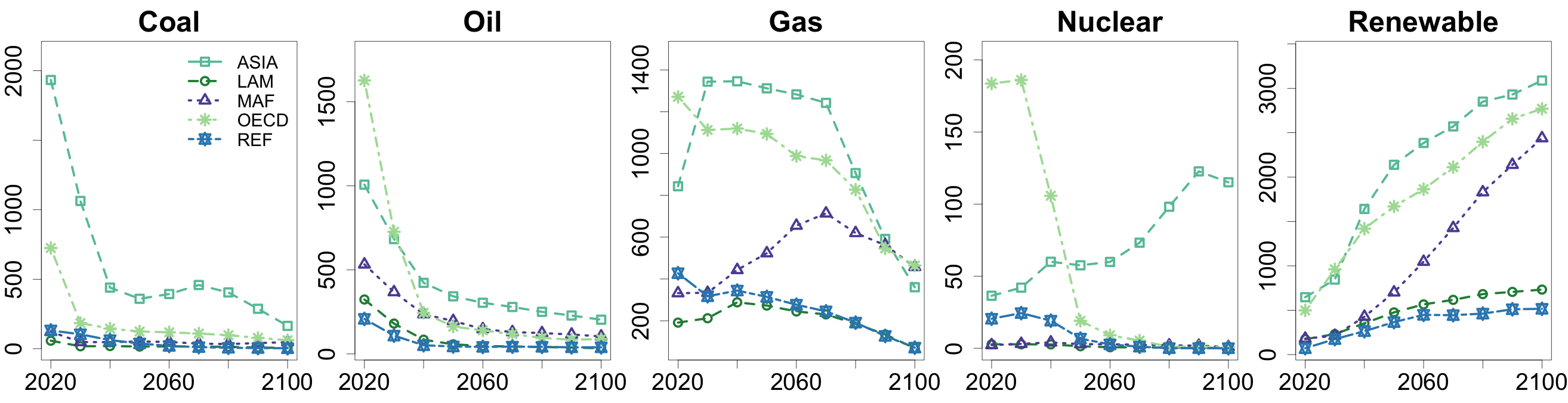}
	\end{subfigure} 
	\begin{subfigure}{\textwidth}
		\centering
		\subcaption*{SSP2-1.9}
		\includegraphics[width=0.9\linewidth]{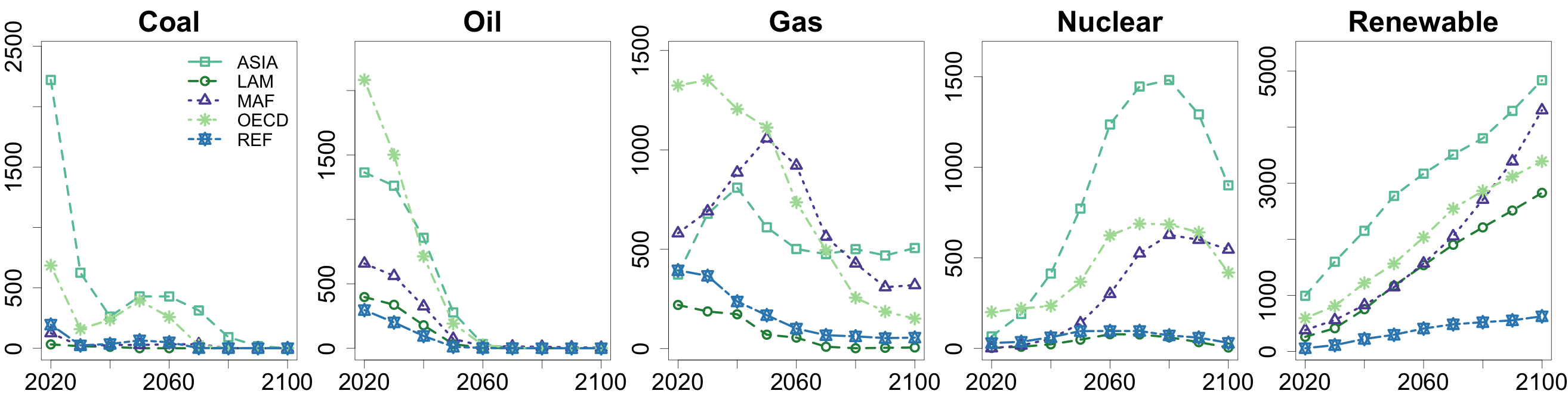}
	\end{subfigure}
	\begin{subfigure}{\textwidth}
		\centering
		\subcaption*{SSP5-1.9}
		\includegraphics[width=0.9\linewidth]{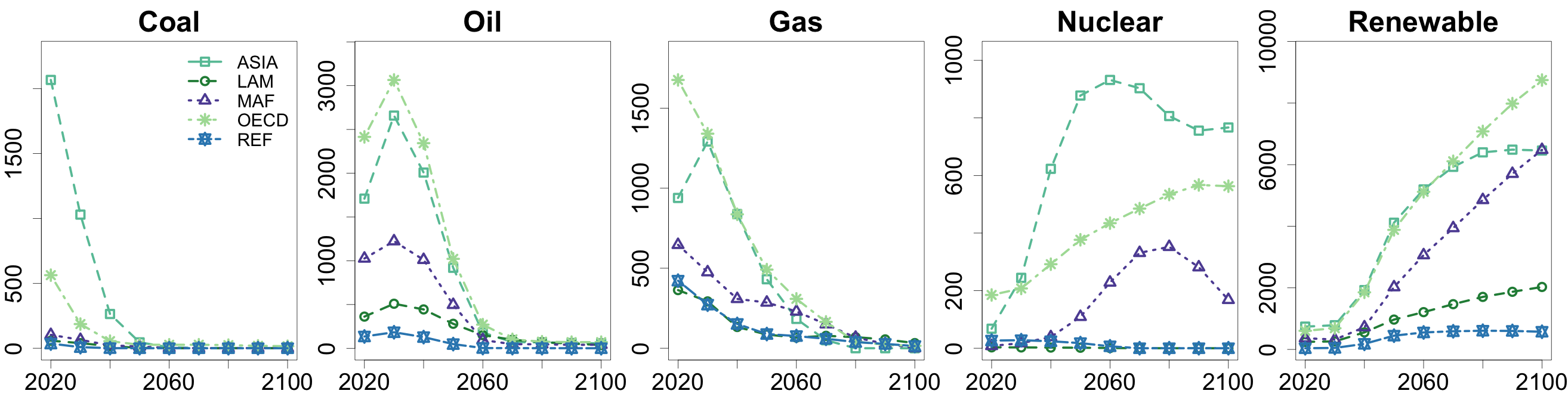}
	\end{subfigure}
\label{SSP1.9energy}
\end{figure}

We perform statistical projections of CO$_2$ emissions and GDP with the E3S2 model, using the estimated parameters from the empirical exercise in Section \ref{empirical} for each region. Fossil and nuclear energy sources are exogenous in the E3S2 model. In the projection, they are given by the fuel mix series from SSP 1.9 W$\mathrm{m}^{-2}$ (shown in Figure \ref{SSP1.9energy}).  

In contrast to IAMs, we consider parameter uncertainty and construct confidence bands accordingly. We start with the vector of five parameters $\boldsymbol{\uptheta} =(\beta_{ C},\beta_{ O},\beta_{ G},d_{\beta_{ Y}}, d_R)'$. We randomly draw a vector of parameters from the distribution $\mathcal{N}\left(\widehat{\boldsymbol{\uptheta}}, \widehat{\boldsymbol{\Omega}}\right)$, where $\widehat{\boldsymbol{\uptheta}}$ denotes the estimated mean of $\boldsymbol{\uptheta}$ and $\widehat{\boldsymbol{\Omega}}$ denotes the estimated variance-covariance matrix of $\boldsymbol{\uptheta}$, respectively. We insert the drawn parameter vector into the model and keep other parameters such as variances of measurement errors constant at their point estimates. For time-varying emission conversion factors, we set the mean and the variance to their smoothed values in 2019. For $\beta_Y$ and $R_t$, we set the mean and variance equal to the median from the period 2015-2019.  As the SSP database does not contain confidence bands of the energy pathways, we do not consider uncertainty of the energy trajectories. We extract the filtered states of $E^*$ and $Y^*$ conditional on the SSP 1.9 W$\mathrm{m}^{-2}$ energy trajectories as projected emissions and GDP over the period 2020 -- 2100 at a 10-year frequency. We iterate this process and generate 10,000 trajectories.  From these, we obtain the 5th percentile, the median, and the 95th percentile, from which we draw the 90\% confidence bands. These confidence bands capture parameter estimation uncertainty from the draw of the parameter vectors and sampling uncertainty from the 10,000 Monte Carlo simulations. In Sections \ref{regional} and \ref{global}, we present the scenario analysis at both the regional and global levels.

\subsection{Regional scenario analysis conditional on SSP 1.9 $W\mathrm{m}^{-2}$}
\label{regional}

\subsubsection{SSP 1.9 $\mathrm{Wm^{-2}}$}

The SSP scenarios provide a framework of five narratives with distinct socio-economic assumptions for the future until 2100 to complement the GHG concentration trajectories from the Representative Concentration Pathways. The five storylines are labeled: SSP1 -- sustainability, SSP2  -- middle of the road, SSP3 -- regional rivalry, SSP4 -- inequality, and SSP5 -- fossil-fueled development. 

Six IAMs participate in quantifying the SSP narratives: the Asia–pacific Integrated Model (AIM) \cite{fujimori2017ssp3}; the Global Change Assessment Model (GCAM4) \cite{calvin2017ssp4}; the Integrated Model to Assess the Global Environment (IMAGE) \cite{van2017energy}; the Model for Energy Supply Strategy Alternatives and their General Environmental Impact combined with the Global Biosphere Management Model (MESSAGE-GLOBIOM) \cite{fricko2017marker}; the Regionalized Model of Investments and Development combined with the Model of Agricultural Production and its Impact on the Environment (REMIND-MAgPIE) \cite{kriegler2017fossil}; and the World Induced Technical Change Hybrid model combined with GLOBIOM (WITCH-GLOBIOM) \cite{emmerling2016witch}. Each SSP storyline is implemented in multiple IAMs, and one of the IAMs is selected as a representative ``marker'' of the narrative. Within each of the main SSP storylines 1--5, different scenarios for radiative forcing levels by 2100 are considered, some aligning with the goals of the Paris Agreement, others not.

In this paper, we focus on the 1.9 $W\text{m}^{-2}$ scenario, which restricts radiative forcing level to 1.9 $W\text{m}^{-2}$ by the end of the twenty-first century. Under this scenario, the goal set by the Paris Agreement to limit the global mean temperature rise relative to the pre-industrial level  below 1.5 ${ }^{\circ} \text{C}$ is achievable \cite{rogelj2018scenarios}. Across all 1.9 $W\text{m}^{-2}$ scenarios, the probability of limiting peak warming below 1.5 ${ }^{\circ} \text{C}$ ranges from 20\% to 48\%, with the probability inversely related to yearly GHG emissions in 2030 \cite{rogelj2018scenarios}. 

Following \citeA{rogelj2018scenarios}, we use the name SSPx-1.9, where x$\in\left\{1, 2, 3, 4, 5\right\}$ refers to the specific SSP storyline. Note that due to high regional heterogeneity, all models under SSP3 fail to generate a solution that achieves limiting the end-of-century forcing to 1.9 $W\text{m}^{-2}$. The goal is also challenging for SSP4 and SSP5, although it is reached by the marker model in SSP5 \cite{rogelj2018scenarios}. For this reason, we restrict our focus to SSP1-1.9, SSP2-1.9, and SSP5-1.9. 
\begin{table}[h]
	\setlength{\tabcolsep}{3pt} 
	\renewcommand{\arraystretch}{1.2}
	\centering
	\footnotesize
	\caption{\footnotesize Key energy characteristics for the marker in SSP 1.9 $W\text{m}^{-2}$ scenario \protect \cite{rogelj2018scenarios}. Non-biomass renewables include solar, wind, hydro and geothermal energy.}
	\begin{tabular}{l|cccccc}
		\hline \hline
		\textbf{Scenario}	& 	\textbf{SSP1-1.9 } & 		\textbf{SSP2-1.9 } & 	\textbf{SSP5-1.9 }\\
			\textbf{Narrative}	& 	{Sustainability} & 		{Middle of the road} & 	{Fossil-fueled development}\\
		\textbf{Marker Scenario} & {IMAGE} &	{MESSAGE-GLOBIOM} & {REMIND-MAgPIE} 
		\\
			\hline 
			Fossil fuels & {phase-out} &{phase-out} &{phase-out} 
			\\
		\multirow{2}{6cm}{{Final energy reduction over 2020 -- 2100 relative to  the baseline }} &	\multirow{2}{*}{37\%} 
		&\multirow{2}{*}{30\%}  & \multirow{2}{*}{34\%} &\\
		& & & & & \\
			\multirow{2}{6.5cm}{{Global annual final energy intensity rate of change over 2020 -- 2050}} &	\multirow{2}{*}{-0.9\%} 
		&\multirow{2}{*}{-1.2\%}  & \multirow{2}{*}{-1.4\%} &\\
		& & & & & \\
			{Share of non-biomass renewables in 2050} & 23\% & 32\% &24\%\\
			{Share of non-biomass renewables in 2100} & 30\% & 55\% &54\%\\
		\hline\hline
	\end{tabular}
	\label{SSP1.9}
\end{table}

Table \ref{SSP1.9} presents an overview of some key features of the energy sector for SSP1-1.9, SSP2-1.9, and SSP5-1.9, where the marker scenario achieves the goal of 1.9 $W\text{m}^{-2}$. Despite SSP5-1.9 having a fossil-fueled storyline, it assumes a phase-out of all fossil fuels and the largest annual reduction in energy intensity. SSP1-1.9 features the highest final energy reduction, aligning with its core narrative of adopting a sustainable path. All scenarios exhibit an increase in the share of non-biomass energy \cite<e.g.,>{creutzig2015bioenergy}. 

We collect SSP 1.9 projections of the energy mix at a regional level from the SSP Database \cite{rogelj2018scenarios}. The regional realizations of each energy source are generally in line with the key features in Table \ref{SSP1.9}. All regions shift away from fossil fuels and exhibit rapid scale-up in renewable energy use. However, SSP1-1.9 and SSP2-1.9 do not completely phase out fossil fuels for some regions. For instance, ASIA still uses a small amount of coal and oil by the end of the century, mostly with abatement (i.e., with carbon capture and storage, CCS). SSP1-1.9 deviates considerably from other two scenarios in nuclear energy use. Four out of the five regions in SSP1-1.9 phase out nuclear energy in the period 2050 -- 2070, including OECD, whereas in SSP2-1.9 and SSP5-1.9, the regions OECD, ASIA, and MAF initially increase nuclear use and reduce only later towards the end of the century. 

\subsubsection{Projections with linear growth in energy productivity}
\label{EEElinear regional}

We follow the approach described in Section \ref{approach} and generate projections from the E3S2 model. For the time-varying emission conversion factors, we use the smoothed values from 2019 and assume them as constant conversion factors in the projection. If $\beta_{ Y}$ is found to have a stochastic trend $d_{\beta_{ Y},t}$, we calculate the median of the smoothed $d_{\beta_{ Y},t}$ over the five-year period 2015 -- 2019 and assume it as the linear trend slope for $\beta_{ Y}$ in the future. The same approach is used if $R^*$ is found to have a stochastic trend. 

\begin{figure}[h!]
	\centering
	\caption{\footnotesize Comparison of CO$_2$ emissions projected by the E3S2 model plus carbon capture and storage (CCS) from SSP scenarios and CO$_2$ projections under SSP 1.9 W$\mathrm{m^{-2}}$ scenarios. Confidence bands are pointwise at the 90\% level.}
	\begin{subfigure}{0.5\textwidth}
		\centering
		\includegraphics[width=\linewidth]{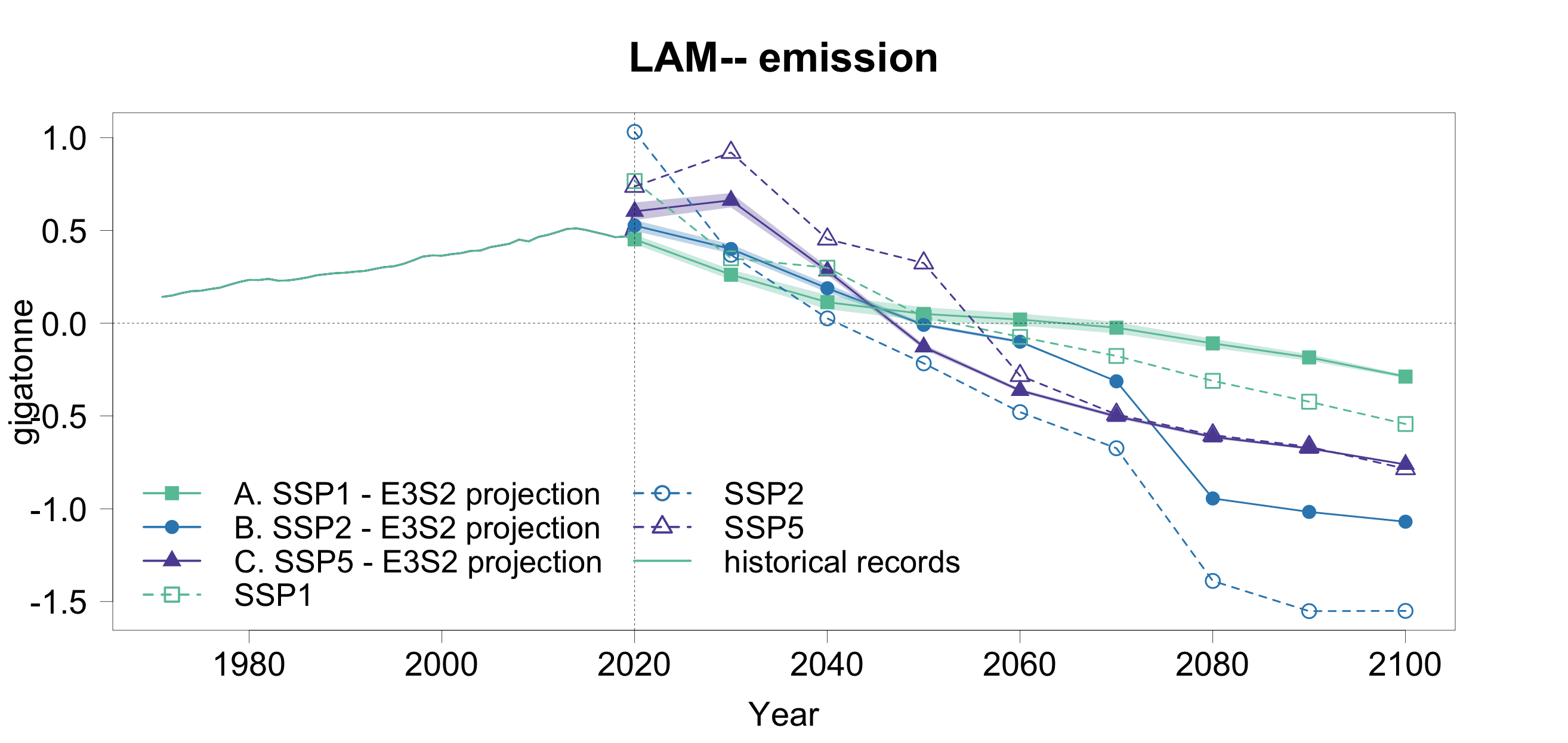}
	\end{subfigure}\hfill
	\begin{subfigure}{0.5\textwidth}
		\centering
		\includegraphics[width=\linewidth]{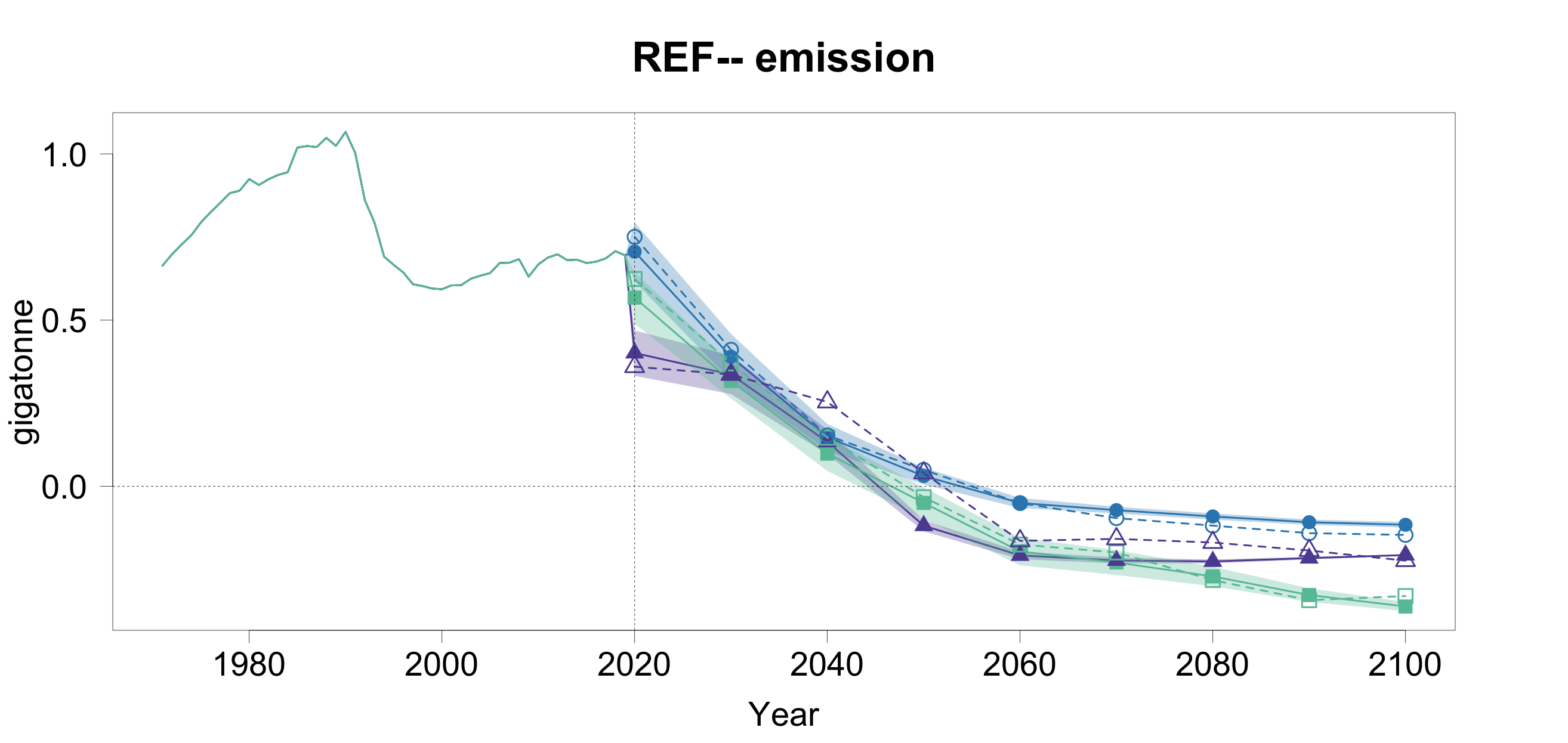}
	\end{subfigure}\\
	\begin{subfigure}{0.5\textwidth}
	\centering
	\includegraphics[width=\linewidth]{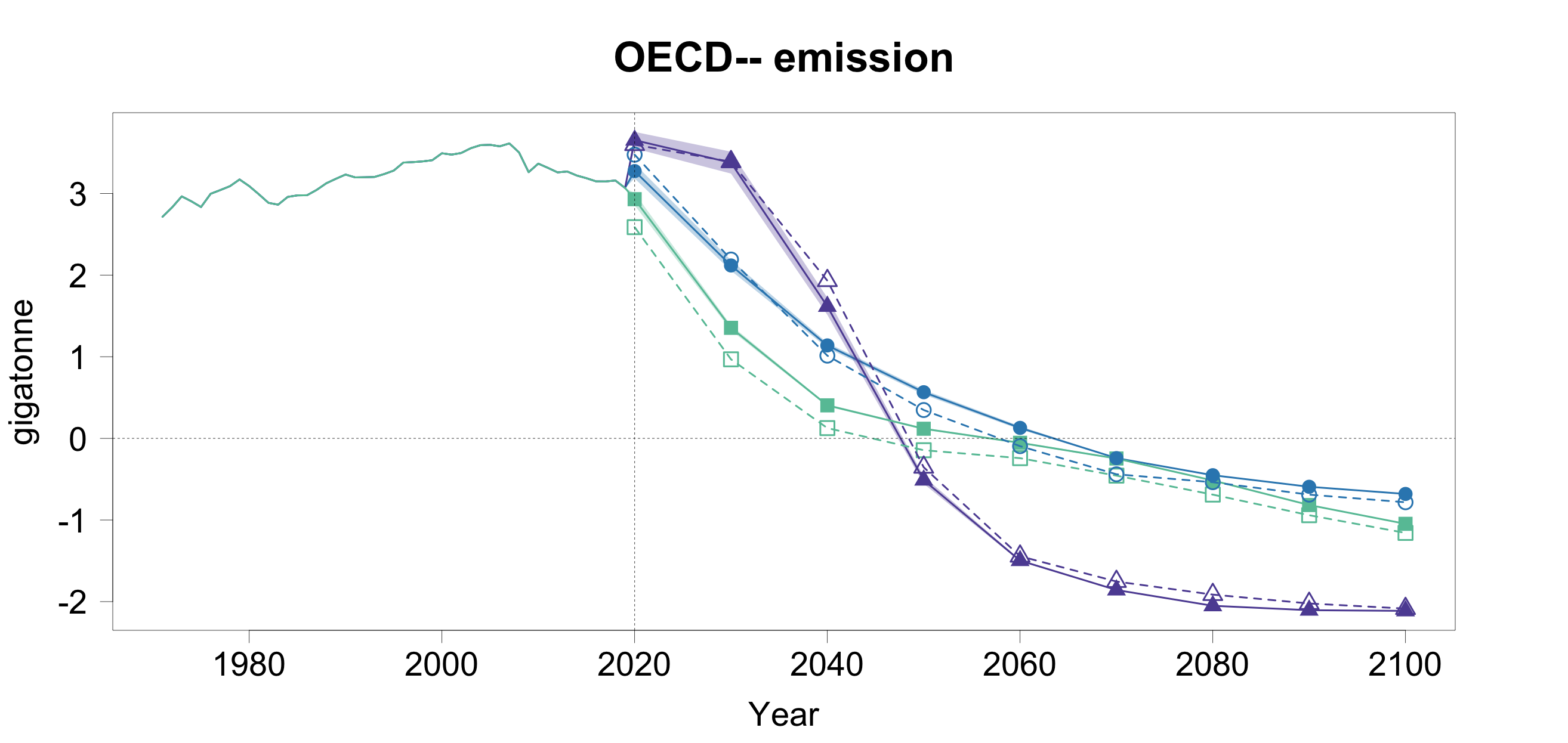}
\end{subfigure}\\
\begin{subfigure}{0.5\textwidth}
	\centering
	\includegraphics[width=\linewidth]{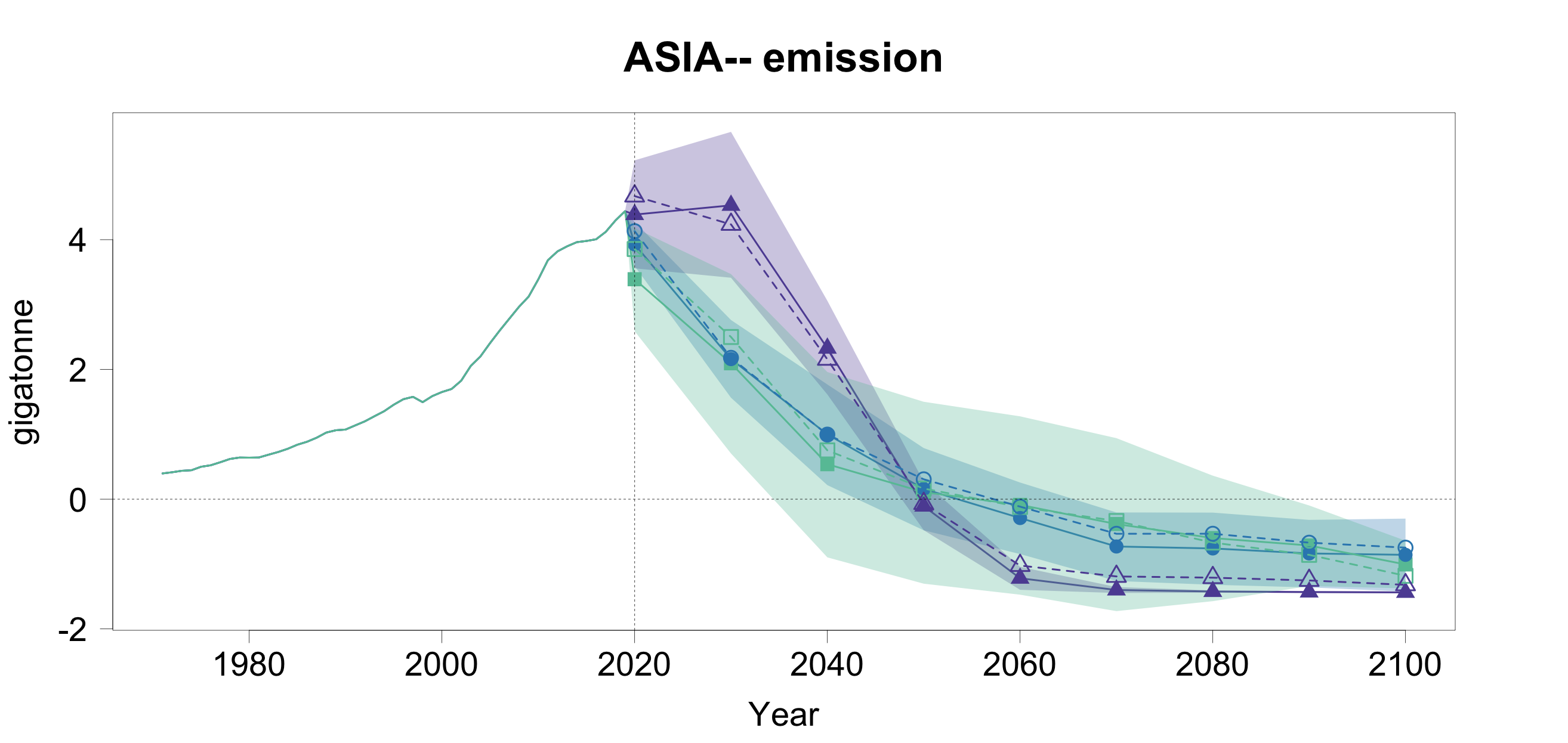}
\end{subfigure}\hfill
\begin{subfigure}{0.5\textwidth}
	\centering
	\includegraphics[width=\linewidth]{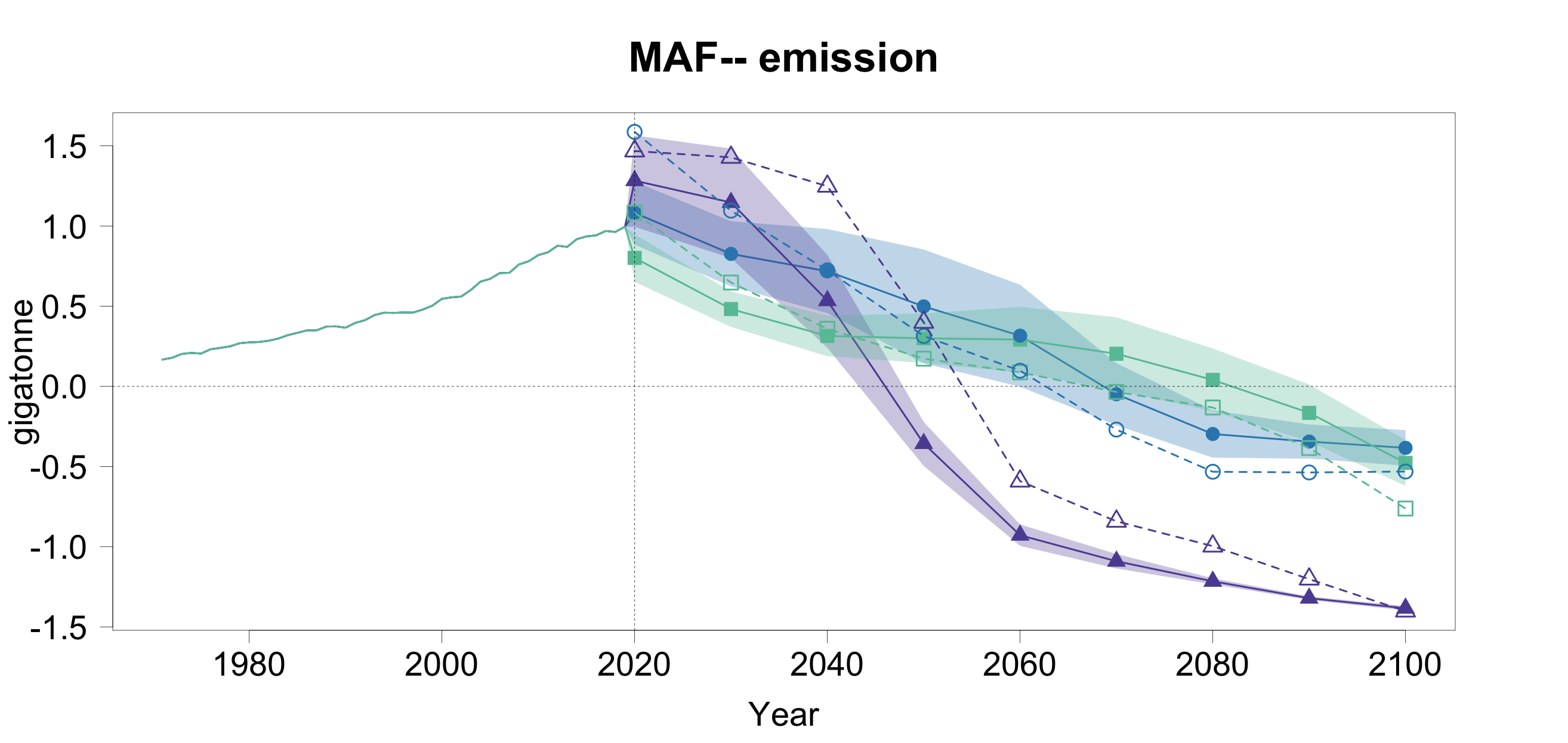}
\end{subfigure}
\label{emission}
\end{figure}

Figure \ref{emission} presents model-projected CO$_2$ emissions compared to the emissions trajectories from SSP1-1.9, SSP2-1.9, and SSP5-1.9 marker scenarios. As the SSP trajectories are adjusted by CCS, and this is not featured in our model, we collect the SSP CCS pathways and subtract them from our model output to obtain projections comparable to the SSP trajectories. As shown in Figure \ref{emission}, the CO$_2$ emissions projections generated from the E3S2 model have good agreement with the SSP 1.9 trajectories. LAM and MAF exhibit some discrepancies, primarily because our datasets cover fewer countries than the SSP scenarios for these regions. See Appendix \ref{listofcountries} for a list of countries that are not included in our definitions of LAM and MAF. ASIA shows a wider 90\% confidence band compared to other regions, reflecting larger uncertainty in the emission conversion factors (Table \ref{estimates}). 
\begin{figure}[h!]
	\centering
	\caption{\footnotesize Comparison of GDP projected by the E3S2 model and GDP under SSP 1.9 W$\mathrm{m^{-2}}$ scenarios. Confidence bands are pointwise at the 90\% level.}
	\begin{subfigure}{0.5\textwidth}
		\centering
		\includegraphics[width=\linewidth]{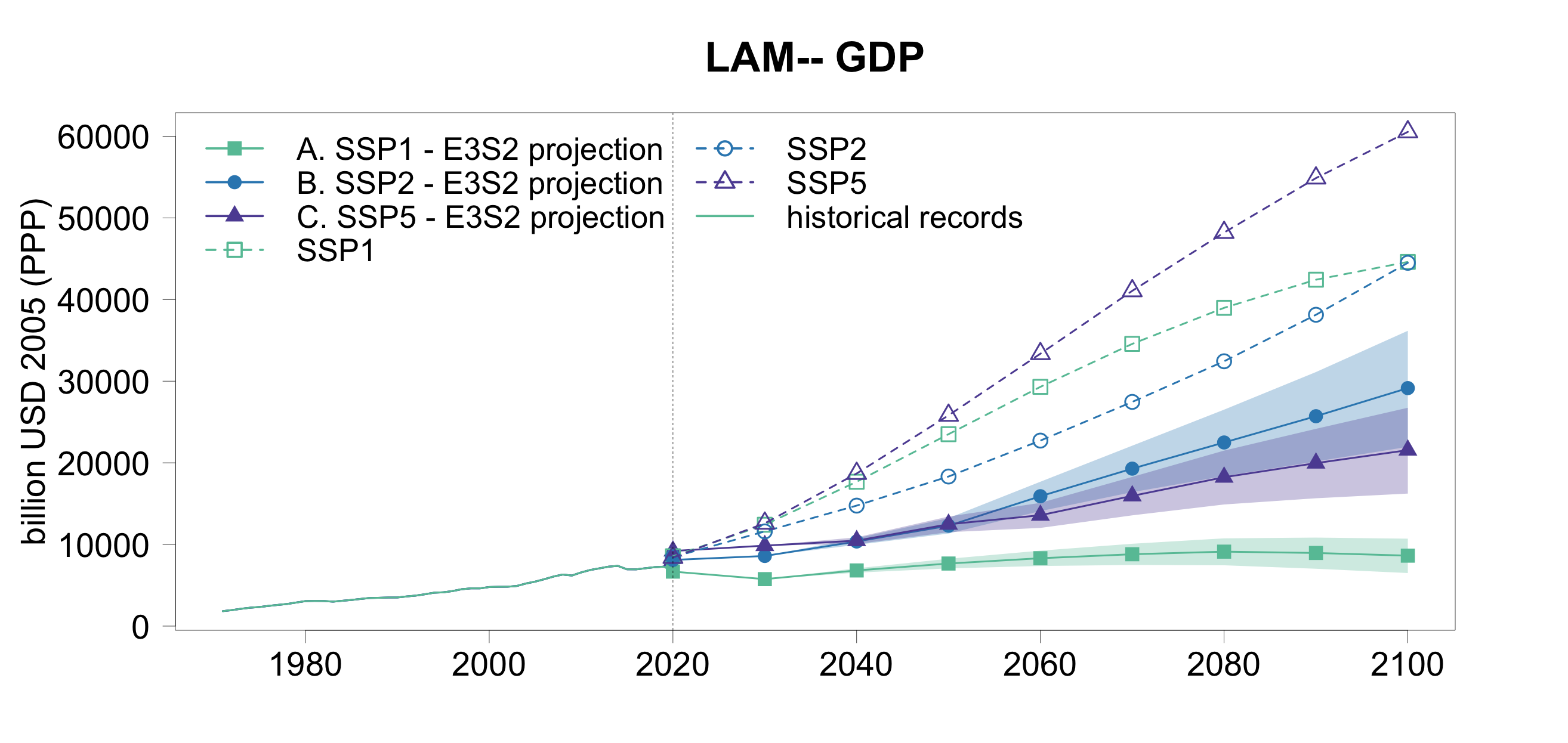}
	\end{subfigure}\hfill
	\begin{subfigure}{0.5\textwidth}
		\centering
		\includegraphics[width=\linewidth]{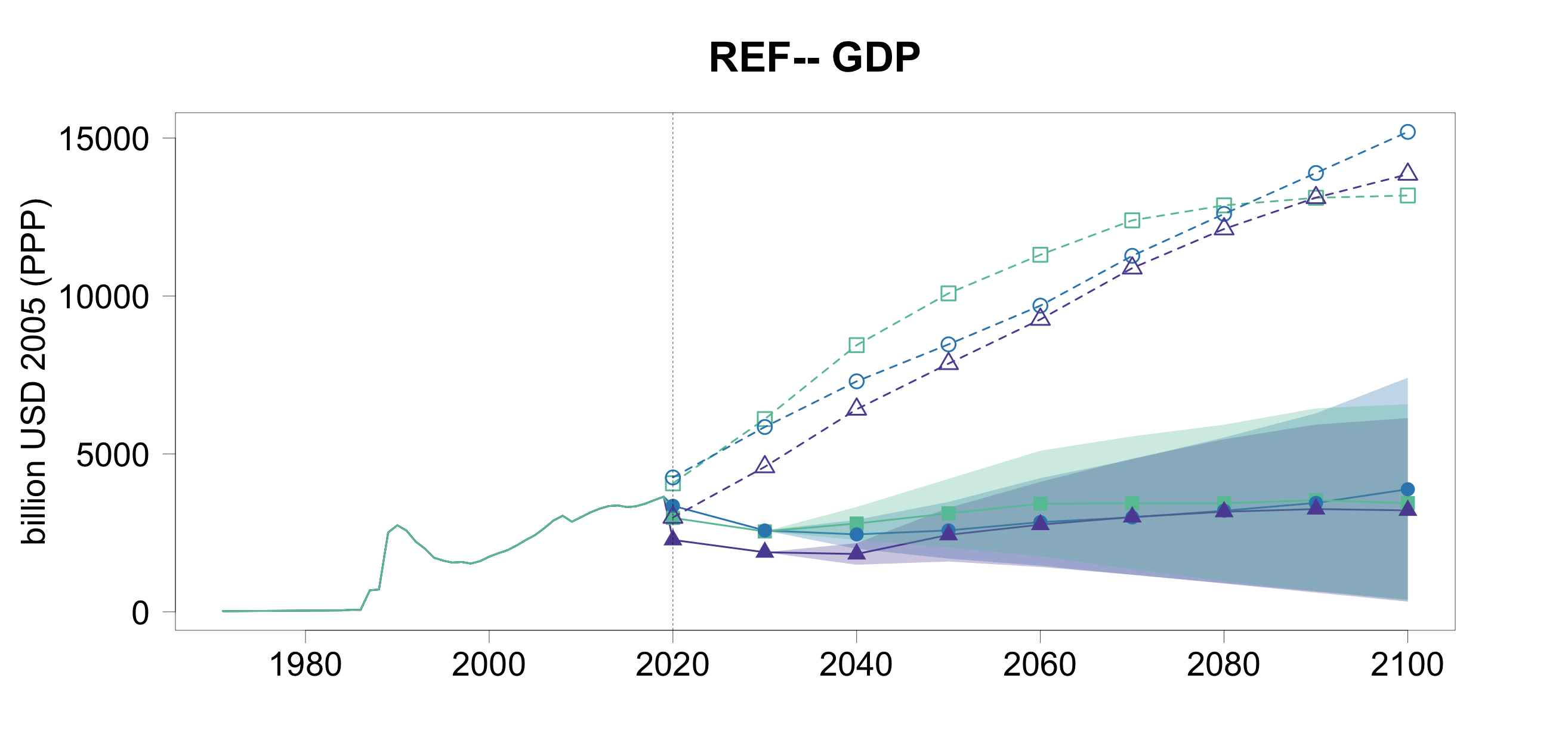}
	\end{subfigure}\\
	\begin{subfigure}{0.5\textwidth}
		\centering
		\includegraphics[width=\linewidth]{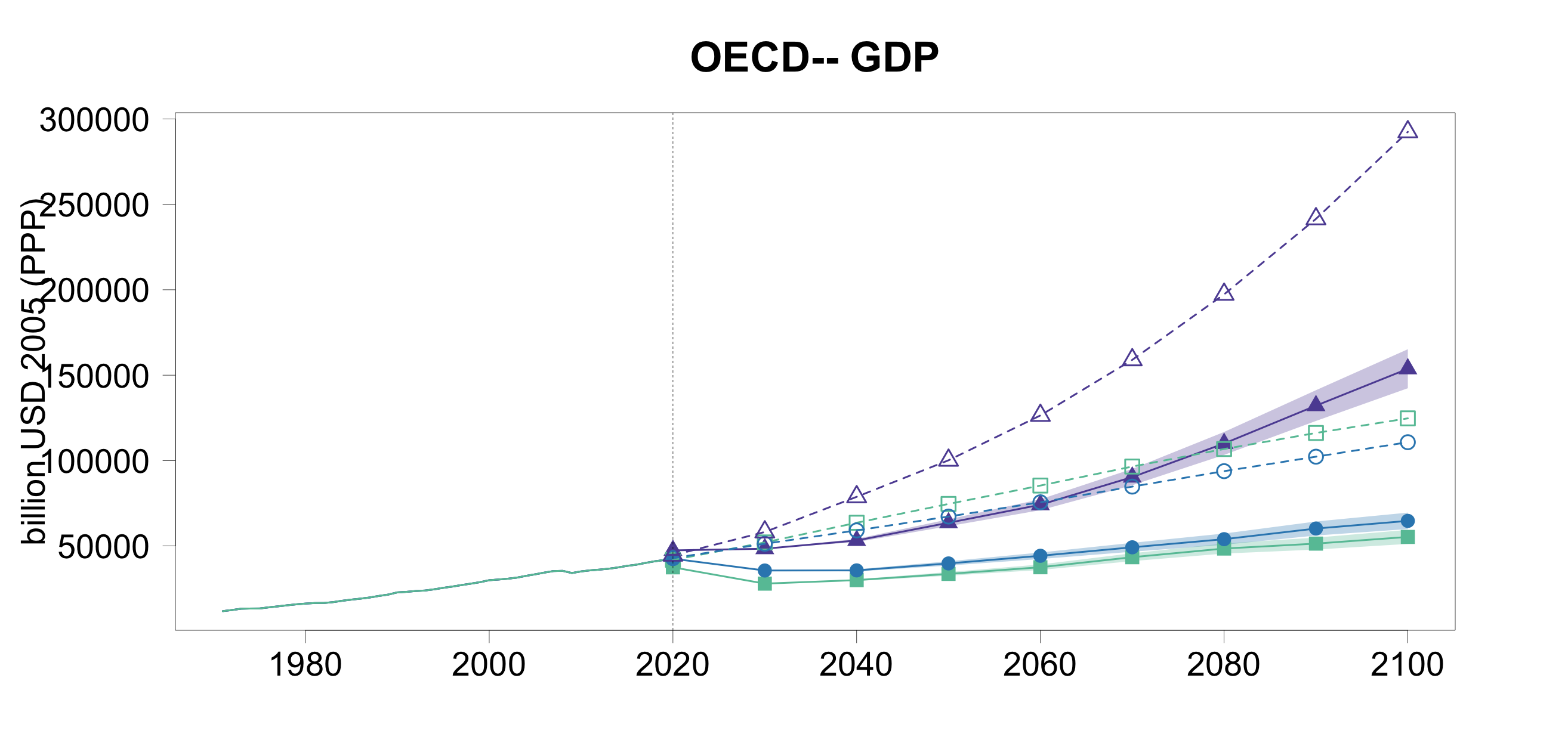}
	\end{subfigure}\\
	\begin{subfigure}{0.5\textwidth}
		\centering
		\includegraphics[width=\linewidth]{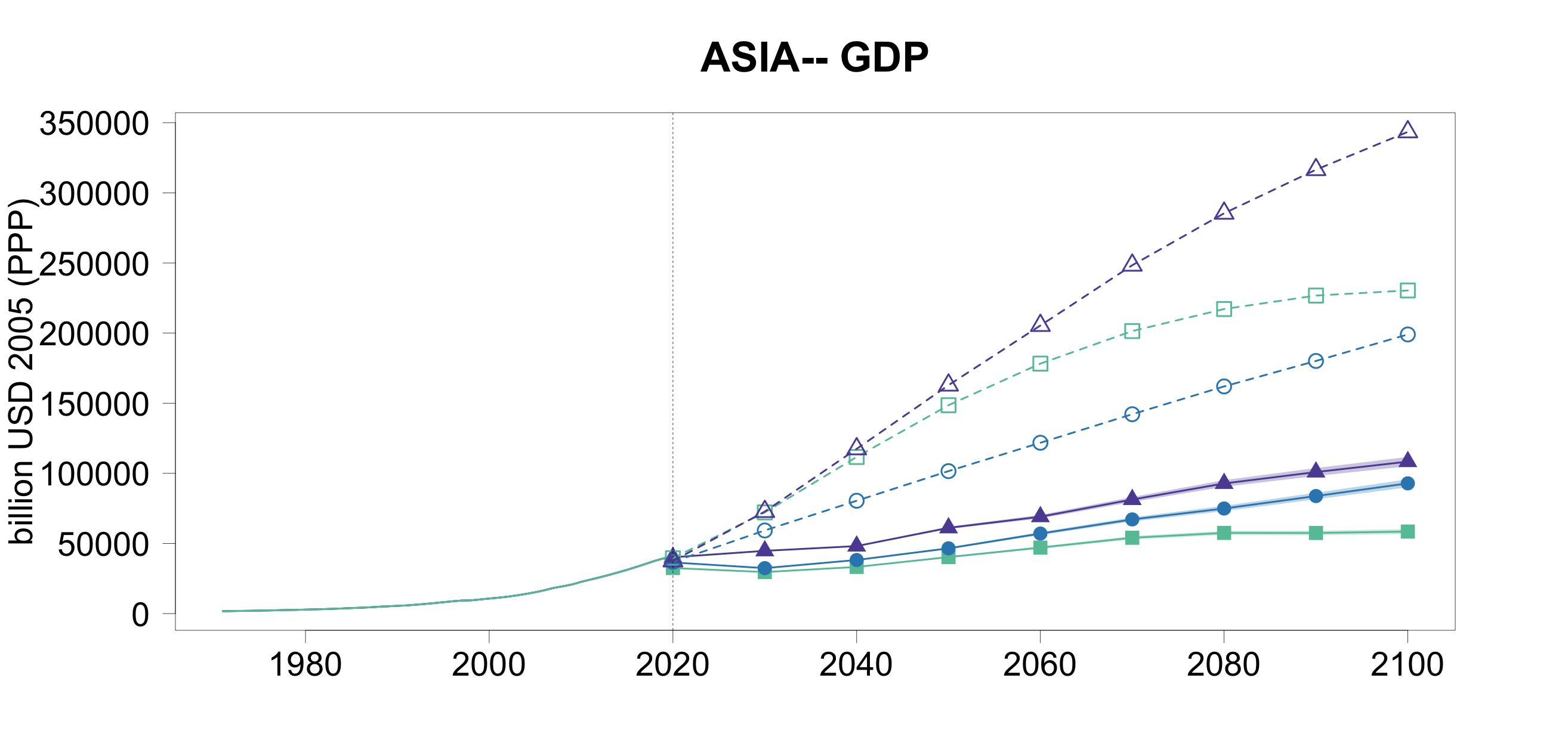}
	\end{subfigure}\hfill
	\begin{subfigure}{0.5\textwidth}
		\centering
		\includegraphics[width=\linewidth]{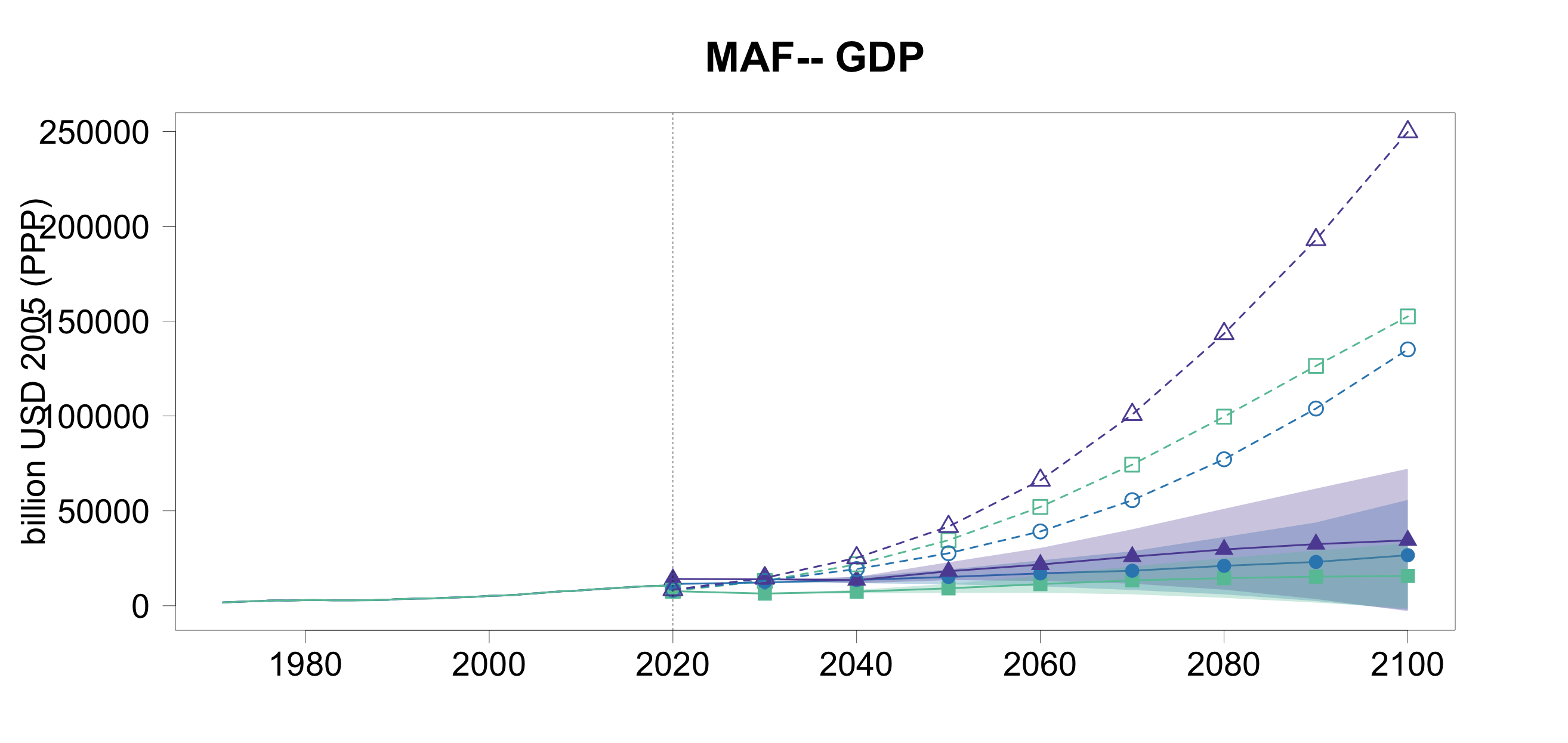}
	\end{subfigure}
	\label{GDP}
\end{figure} 

While our emissions projections largely agree with those from the SSP scenarios, our GDP projections appear considerably lower than the SSP projections, as illustrated in Figure \ref{GDP}. The significant discrepancies arise from differences in the fundamental assumptions and methodologies between our approach and the SSP studies. 

Three models perform the economic projection for the SSPs.  These models are developed by research teams from the International Institute for Applied Systems Analysis (IIASA) \shortcite{cuaresma2017income}, the Organization for Economic Co-operation and Development (OECD) \shortcite{dellink2017long}, and the Potsdam Institute for Climate Impact Research (PIK) \shortcite{leimbach2017future}, respectively. All three models share the harmonized assumptions regarding the main drivers of economic growth but they differ in methodology. 
 
For example, the ENV-Growth model adopts the ``conditional convergence growth'' methodology \shortcite{dellink2017long}, wherein energy functions both as a production input and as an income generator, particularly for countries rich in oil and gas. The model assumes that the energy efficiency of each country (equivalent to the energy productivity in our paper) drives energy demand and converges to that of the leading economies in the projection. There are six factors contributing to long-term economic growth: physical capital, employment, human capital, energy demand, extraction and processing of the fossil fuel resources oil and gas, and total factor productivity. The model is then calibrated using historical data. The ENV-Growth model incorporates numerous determinants of the economy, along with various assumptions on endogenous convergence mechanisms and exogenous factors across regions. Consequently, the projections are subject to a high level of uncertainty \shortcite{dellink2017long}. 
 \begin{figure}[h!]
 	\centering
 	\caption{\footnotesize Comparison of $\beta_Y$ projected by the E3S2 model with $\beta_Y$ implied by SSP 1.9 projections. The title of each panel reports the region and the estimated growth rate of $\beta_{ Y}$. For REF, ASIA, and MAF, where $\beta_{Y}$ follows a local linear trend model, we use median($\hat{d}_{{\beta_Y,2015}}:\hat{d}_{{\beta_Y,2019}}$), which represents the median of the smoothed $d_{\beta_{ Y},t}$ over 2015 -- 2019, as the linear trend for $\beta_{ Y}$ in the projection. Confidence bands are pointwise at the 90\% level.}
 	\begin{subfigure}{0.5\textwidth}
 		\subcaption{LAM: $\hat{d}_{\beta_Y}=0.0120 (0.020)$}
 		\centering
 		\includegraphics[width=\linewidth]{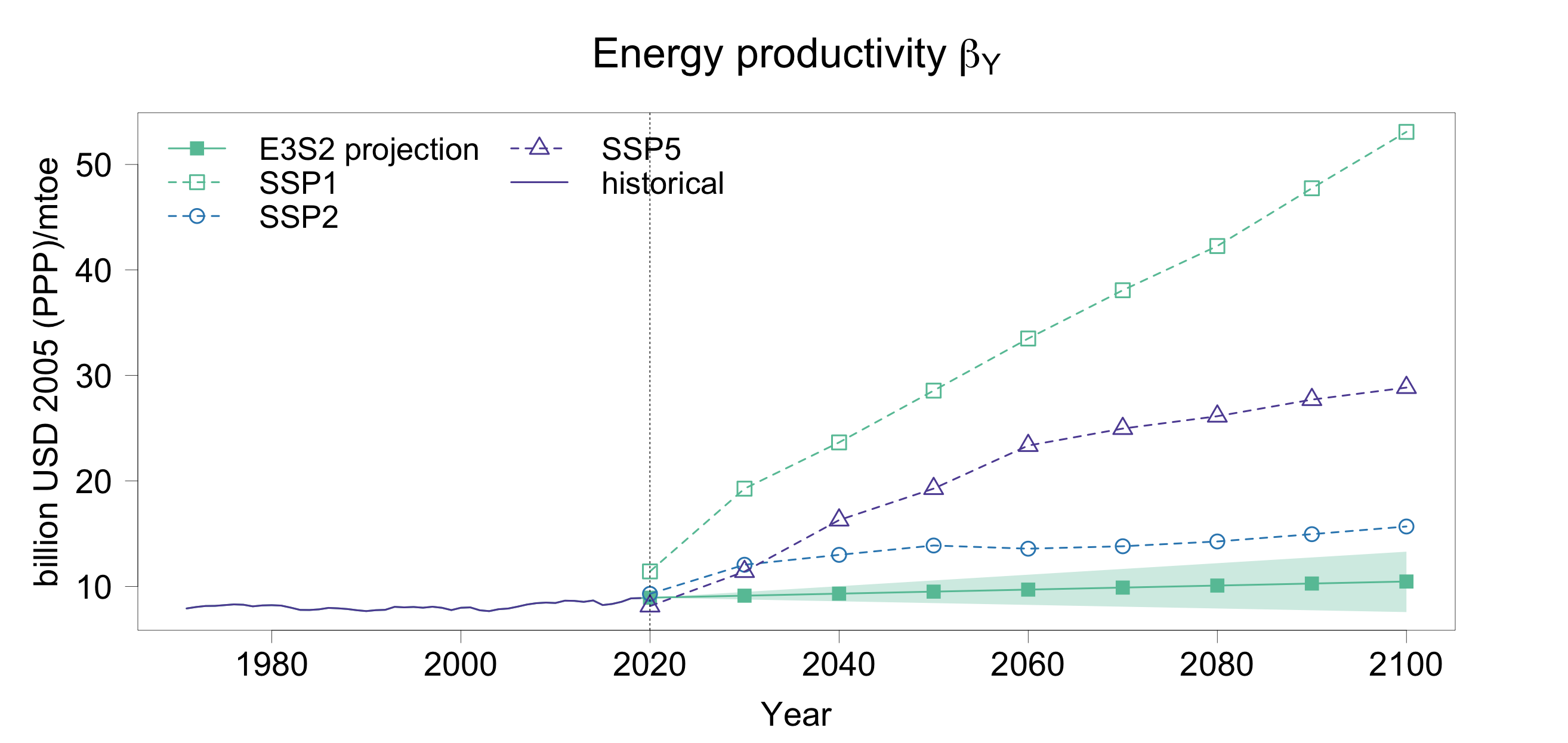}
 	\end{subfigure}\hfill
 	\begin{subfigure}{0.5\textwidth}
 		\subcaption{REF: median($\hat{d}_{{\beta_Y,2015}}:\hat{d}_{{\beta_Y,2019}}$)=0.029}
 		\centering
 		\includegraphics[width=\linewidth]{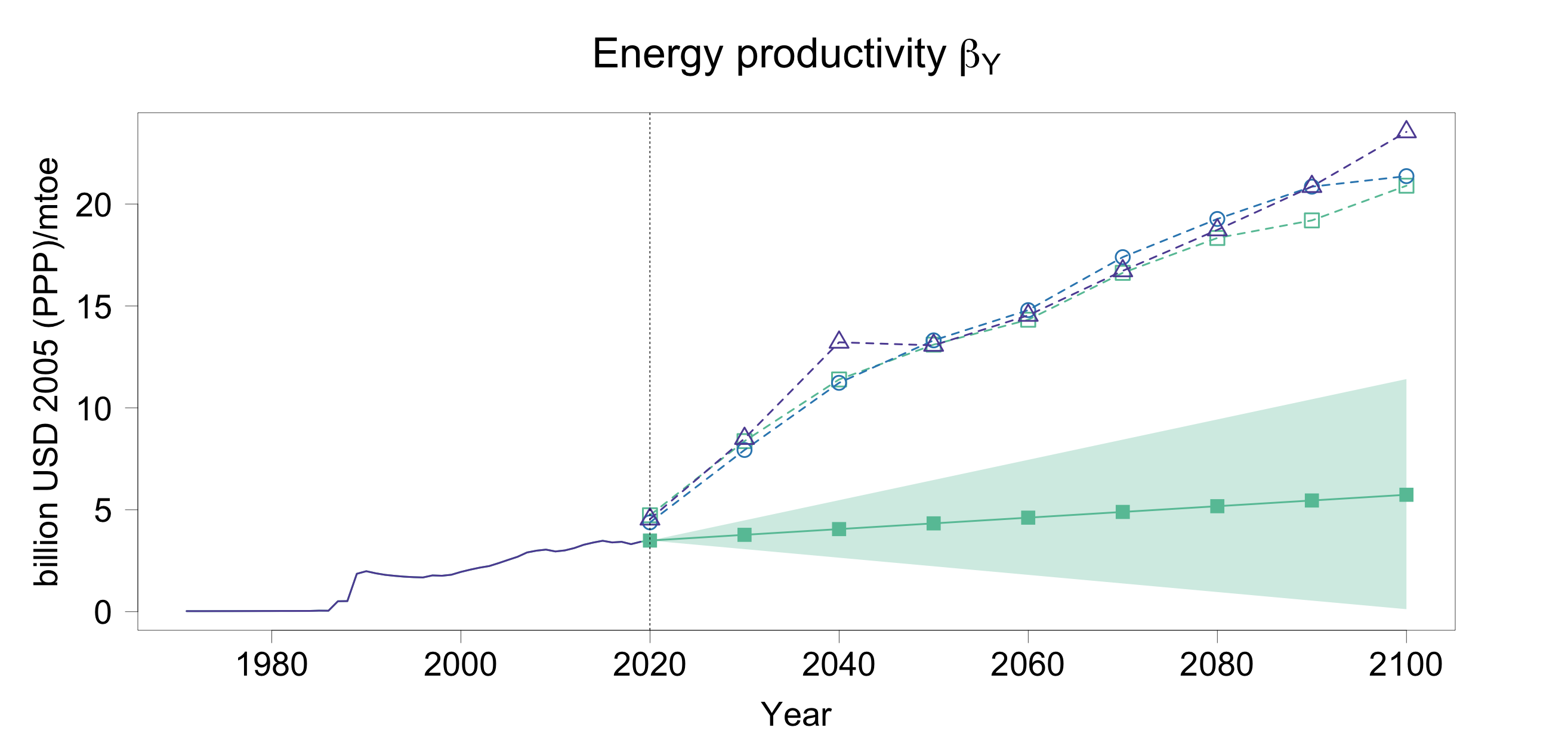}
 	\end{subfigure}\\
 	\begin{subfigure}{0.5\textwidth}
 		\centering
 		\subcaption{OECD: $\hat{d}_{\beta^Y}=0.1080 (0.011)$}
 		\includegraphics[width=\linewidth]{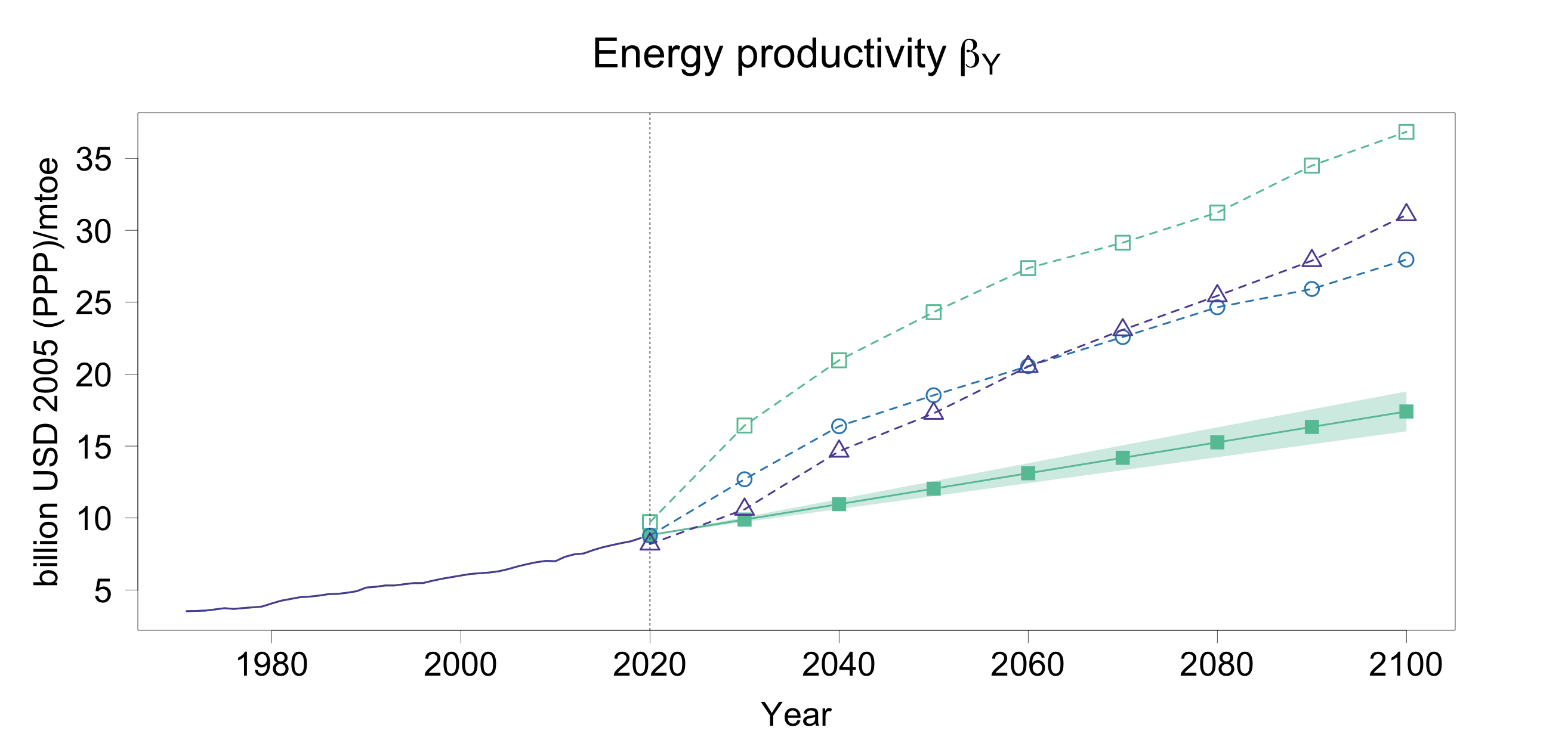}
 	\end{subfigure}\\
 	\begin{subfigure}{0.5\textwidth}
 		\centering
 		\subcaption{ASIA: median($\hat{d}_{{\beta_Y,2015}}:\hat{d}_{{\beta_Y,2019}}$)=0.106}
 		\includegraphics[width=\linewidth]{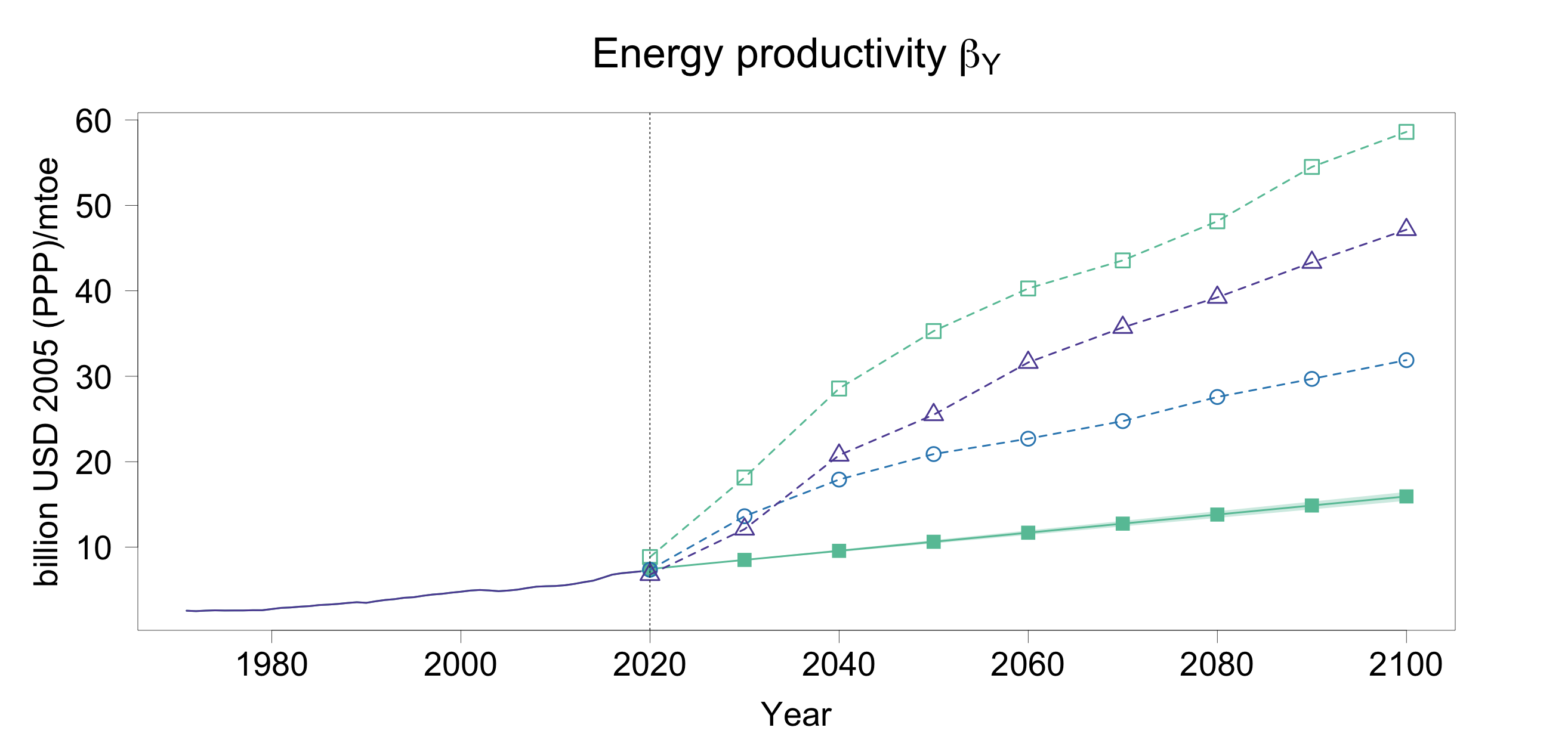}
 	\end{subfigure}\hfill
 	\begin{subfigure}{0.5\textwidth}
 		\centering
 		\subcaption{MAF: median($\hat{d}_{{\beta_Y,2015}}:\hat{d}_{{\beta_Y,2019}}$)=-0.061}
 		\includegraphics[width=\linewidth]{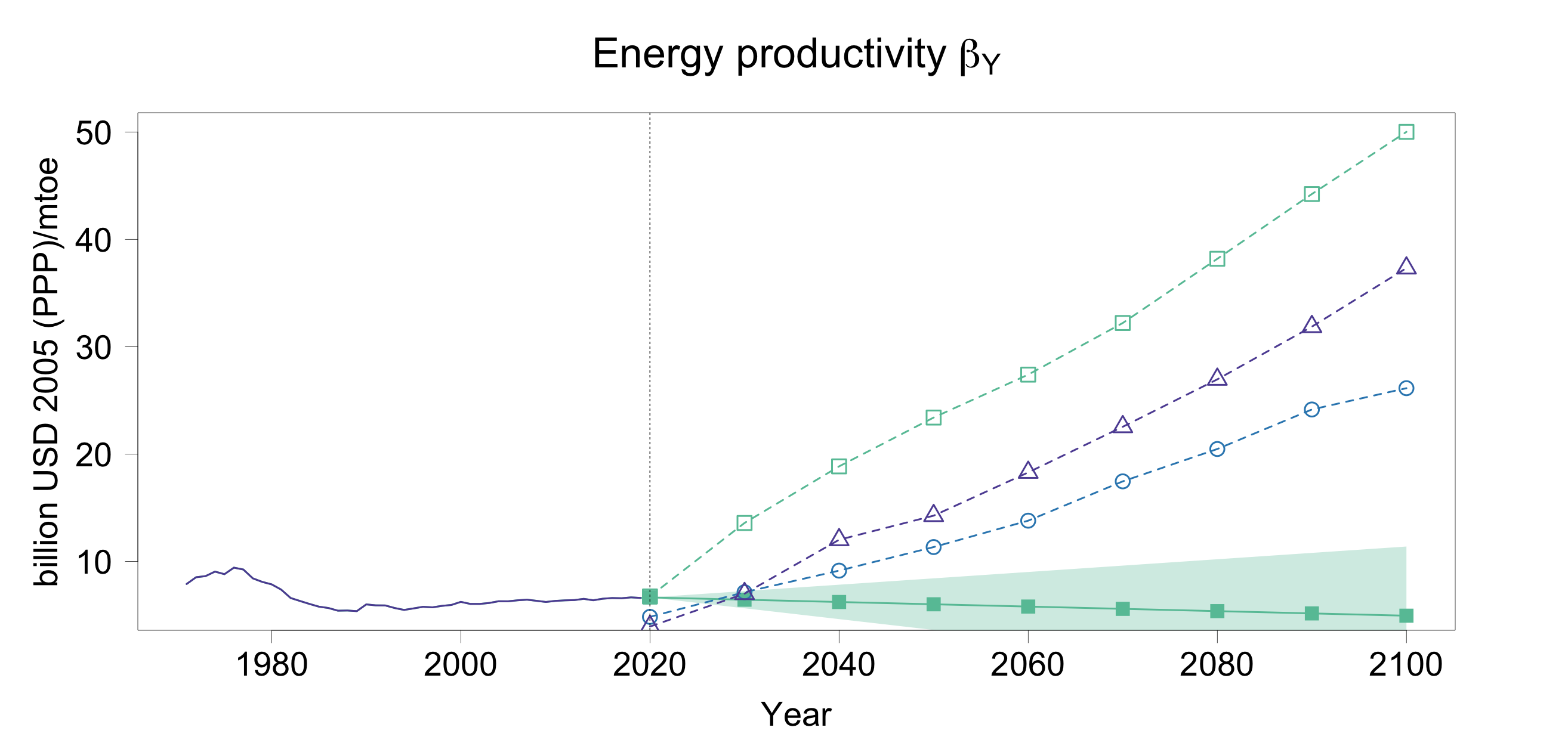}
 	\end{subfigure}
 	\label{beta}
 \end{figure} 

In our model, time-varying energy productivity captures changes in technology. We model energy productivity ($\beta_{Y}$ in Equation \eqref{economyMeas}) either as a deterministic trend ($\sigma^2_{\beta_{ Y}}=0$) or a stochastic trend ($\sigma^2_{\beta_{ Y}}>0$). In the projections, $\beta_{Y}$ is the only endogenous factor determining economic growth when all energy source series are  given from SSP scenarios. The dynamics of $\beta_{Y}$ are inferred statistically from historical data as shown in Table \ref{Diagnostics}.

Figure \ref{beta} shows that the energy productivity implied by the SSP projections, calculated by dividing SSP-projected GDP by total primary energy use, can be more than four times as large as the energy productivity $\beta_Y$ implied by the E3S2 model. This shows that in order to generate GDP trajectories comparable to the SSPs, non-linear dynamics of energy productivity $\beta_{ Y}$ need to be assumed that lie outside of the support of historical data.

\subsubsection{Trends of renewable energy and energy productivity implied in SSP GDP projections}

\begin{table}[ht]
	\setlength{\tabcolsep}{5pt} 
	\renewcommand{\arraystretch}{1.2}
\centering
\caption{Comparison of $d_R$ (Panel A) and $d_{\beta_{ Y}}$ (Panel B) implied by SSP 1.9 GDP projections with estimates from historical data. Standard errors are reported in parentheses. If $R^*_t$ or $\beta_{Y}$ admits a stochastic trend $d_{R,t}$, the values reported for the historical period are the medians of the smoothed $d_{R,t}$ from 2015 to 2019.}
\begingroup
\begin{tabular}{lcccccc}
	  \hline   \hline 
	  \multicolumn{7}{c}{\textbf{A. estimates of $d_R$ (unit: mtoe yr$^{-1}$)}}\\
	  \hline
  & \textbf{LAM }& \textbf{REF}& \textbf{OECD} &\textbf{ASIA} &\textbf{MAF} & \textbf{WORLD} \\ 
  \hline
 \multirow{2}{*}{\textbf{SSP1-1.9}} &   47.701 & 17.275 & 41.855 & 130.303 & 382.815 & 58.666 \\ 
   & (5.679) & (9.959) & (6.634) & (83.694) & (61.499) & (14.49) \\ 
   \hline
 \multirow{2}{*}{\textbf{SSP2-1.9}}    & 51.34 & 24.574 & 47.629 & 85.363 & 340.03 & 61.079 \\ 
   & (0.598) & (5.333) & (5.679) & (11.175) & (74.721) & (3.49) \\ 
     \multirow{2}{*}{\textbf{SSP5-1.9}}   & 76.689 & 25.894 & 221.723 & 276.818 & 660.879 & 133.494 \\ 
   & (6.199) & (8.196) & (8.169) & (92.381) & (129.291) & (62.583) \\ 
    \hline 
      \multirow{2}{*}{\textbf{historical period}}  & 3.164 & 0.035 & 16.35 & 32.289& 7.485 & 60.45 \\ 
    & (0.385) & (0.197) & (2.741) & (1.640) & (0.382) & (4.528) \\ 
    \hline 
      \hline 
      	  \multicolumn{7}{c}{\textbf{B. estimates of $d_{\beta_{ Y}}$ (unit: billion USD 2005 (PPP) /mtoe yr$^{-1}$)}} \\
      \hline
      & \textbf{LAM }& \textbf{REF}& \textbf{OECD} &\textbf{ASIA} &\textbf{MAF} & \textbf{WORLD} \\ 
      \hline
 \multirow{2}{*}{\textbf{SSP1-1.9}}     & 0.483 & 0.161 & 0.256 & 0.485 & 0.519 & 0.399 \\ 
   & (0.012) & (0.007) & (0.011) & (0.027) & (0.054) & (0.013) \\ 
   \hline
\multirow{2}{*}{\textbf{SSP2-1.9}}  & 0.043 & 0.169 & 0.191 & 0.23 & 0.299 & 0.192 \\ 
   & (0.012) & (0.031) & (0.006) & (0.006) & (0.013) & (0.003) \\ 
   \hline
  \multirow{2}{*}{\textbf{SSP5-1.9}}  & 0.189 & 0.175 & 0.269 & 0.433 & 0.439 & 0.331 \\ 
   & (0.024) & (0.053) & (0.007) & (0.024) & (0.057) & (0.008) \\ 
      \hline 
      \multirow{2}{*}{\textbf{historical period}}  & 0.020 & 0.029 & 0.108 &0.106  & -0.061 & 0.103 \\ 
      & (0.020) & (0.0001) & (0.011) & (0.0032) & (0.036) & (0.0015) \\ 
      \hline 
      \hline 
  \end{tabular}
\endgroup
\label{implied}
\end{table}

In this section, we examine the necessary growth rates in renewable energy $R_t$ and energy productivity $\beta_{Y}$, respectively, that would be required to achieve the GDP growth implied in the SSP scenarios from 2020 to 2100. This analysis is based on the assumption that all other factors will remain the same. We set the emission conversion factors equal to their estimated values. In case of time-varying emission conversion factors, we set them constantly equal to their smoothed value in 2019. 

In the first analysis, we only vary $R_t$, meaning we assume constant energy productivity, and all reductions in fossil sources have to be compensated by growth in renewables. We insert the estimated $d_{\beta_{ Y}}$ reported in Table \ref{estimates} as the linear trend for $\beta_{ Y}$ in the future. If $\beta_{Y}$ is specified as a local linear trend model, we use the median of the smoothed $d_{\beta_Y,t}$ over the period 2015 -- 2019. We assume that all other energy sources follow the trajectories in SSP 1.9. We assume a linear trend $d_R$ for $R^*_t$ and estimate it on the SSP 1.9 data.

In the second analysis, we assume that all energy sources, including renewables, follow the pathways in SSP 1.9. Economic growth is obtained by linear energy productivity improvements brought by technological advancements. Assuming $\sigma^2_{\beta_{ Y}}=0$ in Equation \ref{economyState}, we estimate the linear trend $d_{\beta_{ Y}}$ of $\beta_{ Y}$ on the SSP data. 

Table \ref{implied} shows the estimates from the two analyses and compares with the estimates from the historical period. The table shows that the growth implied by the SSPs in either $R_t$ or $\beta_{Y,t}$ largely exceeds the data estimates, the only exception being the ASIA region. 

\begin{figure}[ht]
	\centering
	\caption{\footnotesize Comparison of $\beta_Y$ projected by the E3S2-g model and $\beta_Y$ implied by SSP 1.9 projections. The title of each panel reports the region and growth rate used for projection, which is estimated using historical data. The standard errors are reported in parentheses. $\hat{g}_2$ in MAF represents the growth rate estimated for the second stage starting from 1990. Confidence bands are pointwise at the 90\% level.}
	\begin{subfigure}{0.5\textwidth}
		\centering
			\subcaption{LAM: $\hat{g}=0.002(0.003)$}
		\includegraphics[width=\linewidth]{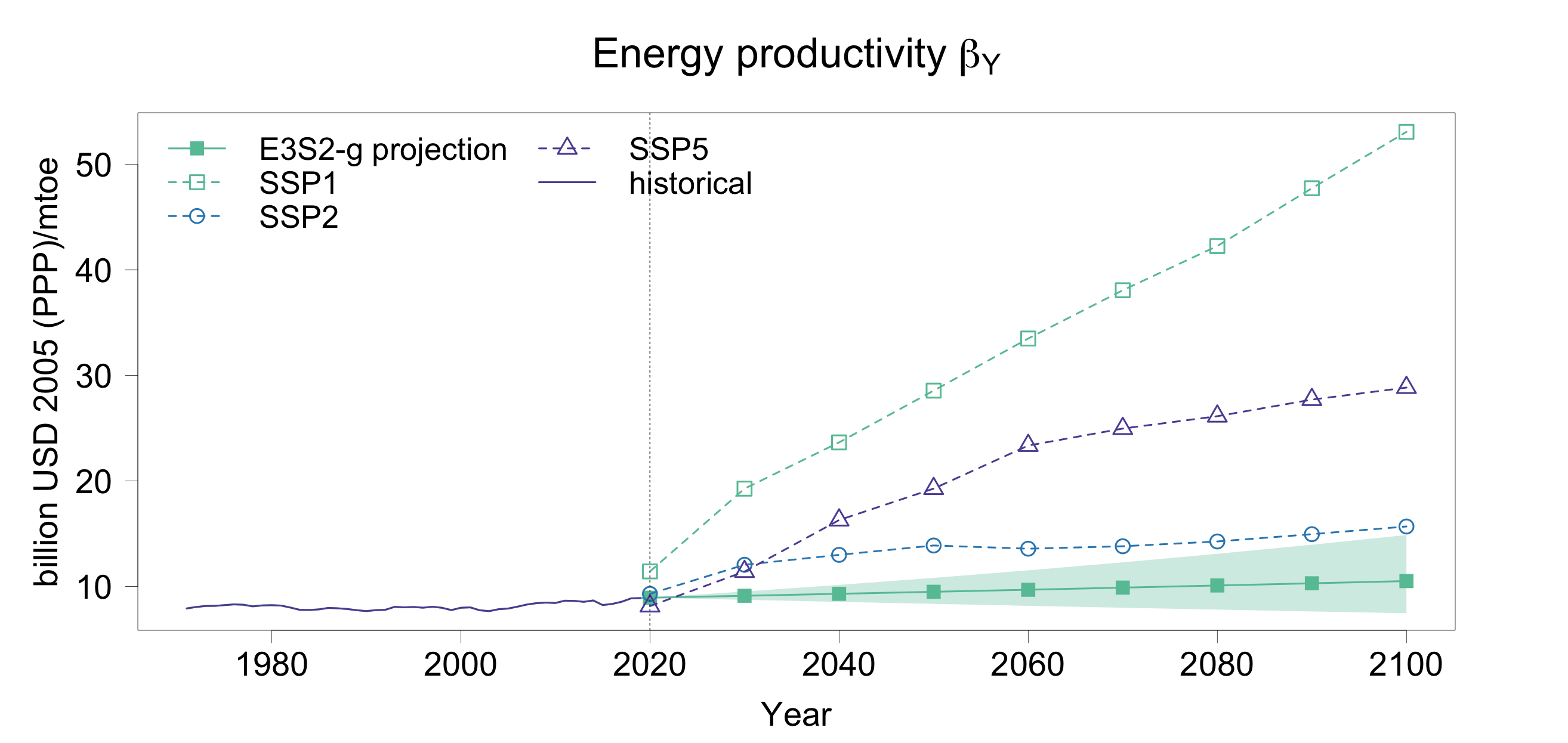}
	\end{subfigure}\hfill
		\begin{subfigure}{0.5\textwidth}
		\subcaption{REF: $\hat{g}=0.023 (0.006)$}
	\centering
		\includegraphics[width=\linewidth]{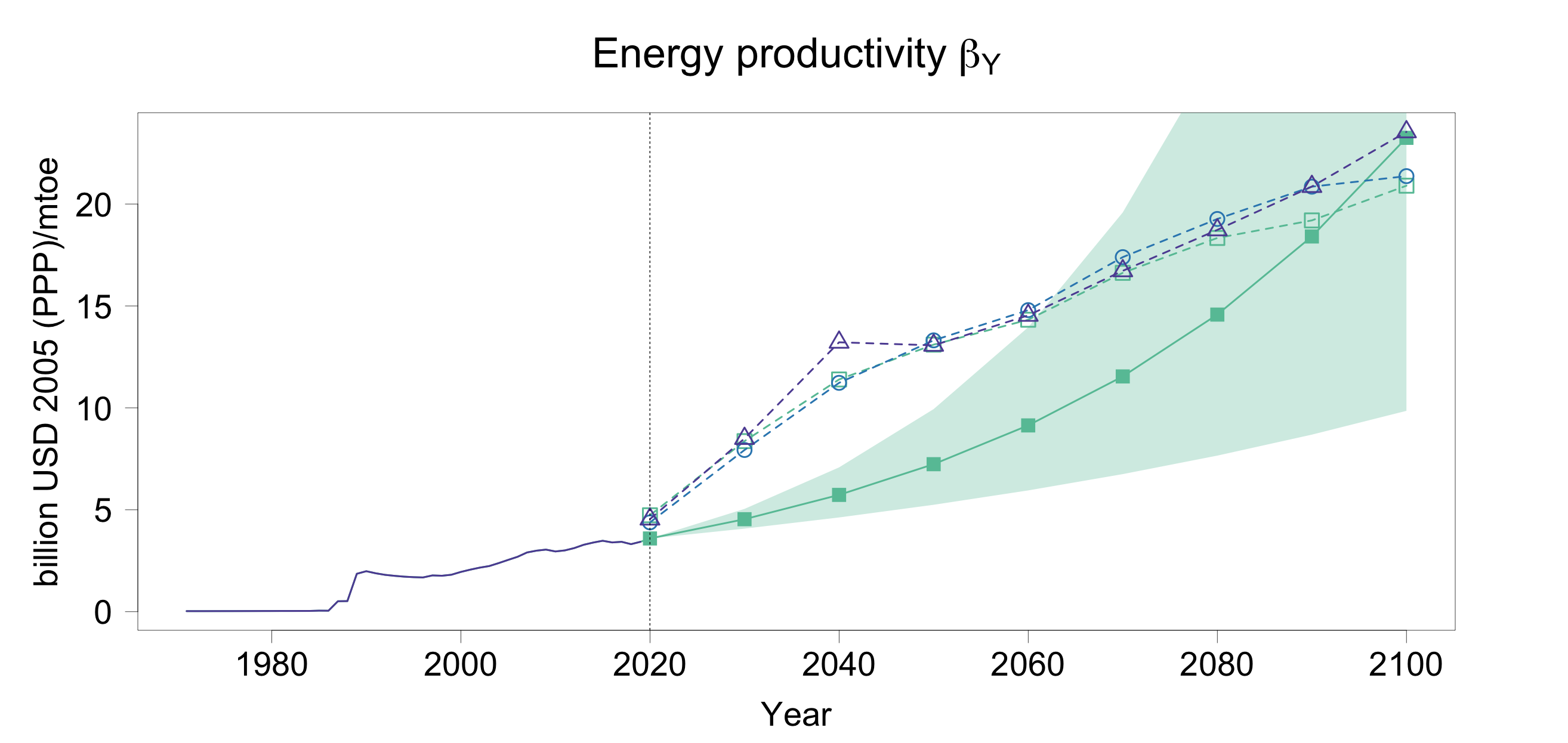}
		\end{subfigure}\\
	\begin{subfigure}{0.5\textwidth}
		\centering
			\subcaption{OECD: $\hat{g}=0.019 (0.002)$}
		\includegraphics[width=\linewidth]{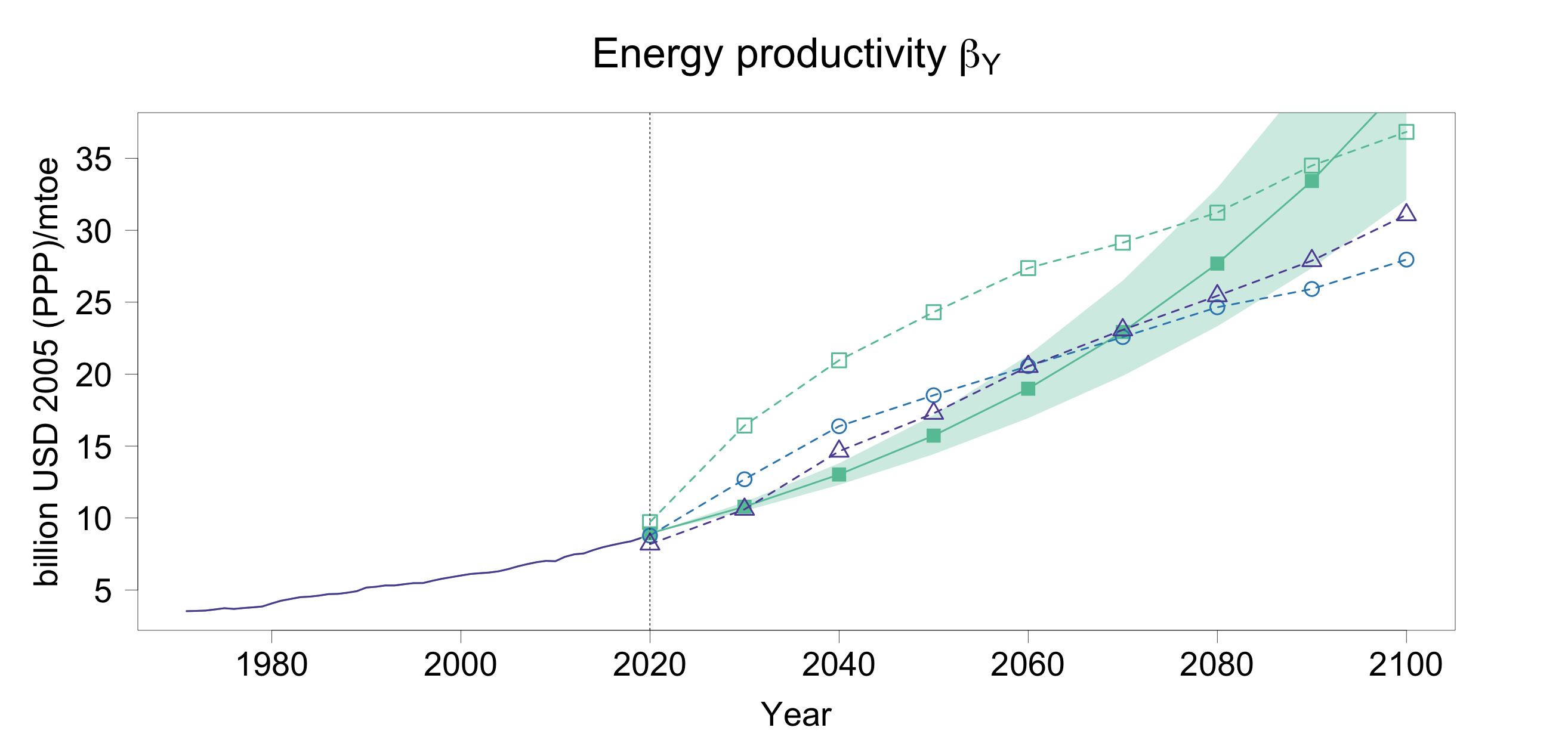}
	\end{subfigure}\\
	\begin{subfigure}{0.5\textwidth}
		\centering
			\subcaption{ASIA: $\hat{g}=0.022 (0.003)$}
		\includegraphics[width=\linewidth]{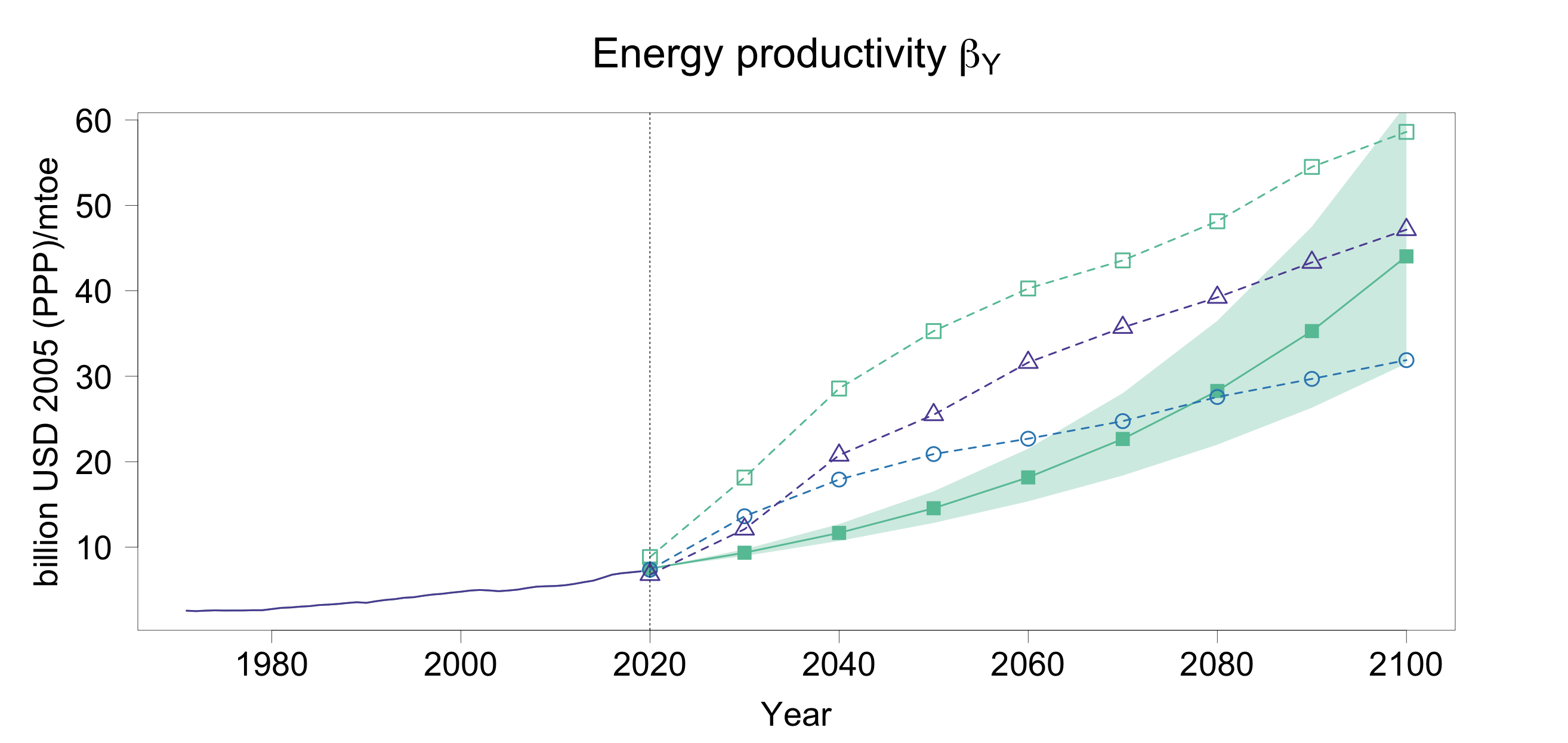}
	\end{subfigure}\hfill
	\begin{subfigure}{0.5\textwidth}
		\centering
			\subcaption{MAF: $\hat{g}_2=0.004 (0.006)$}
		\includegraphics[width=\linewidth]{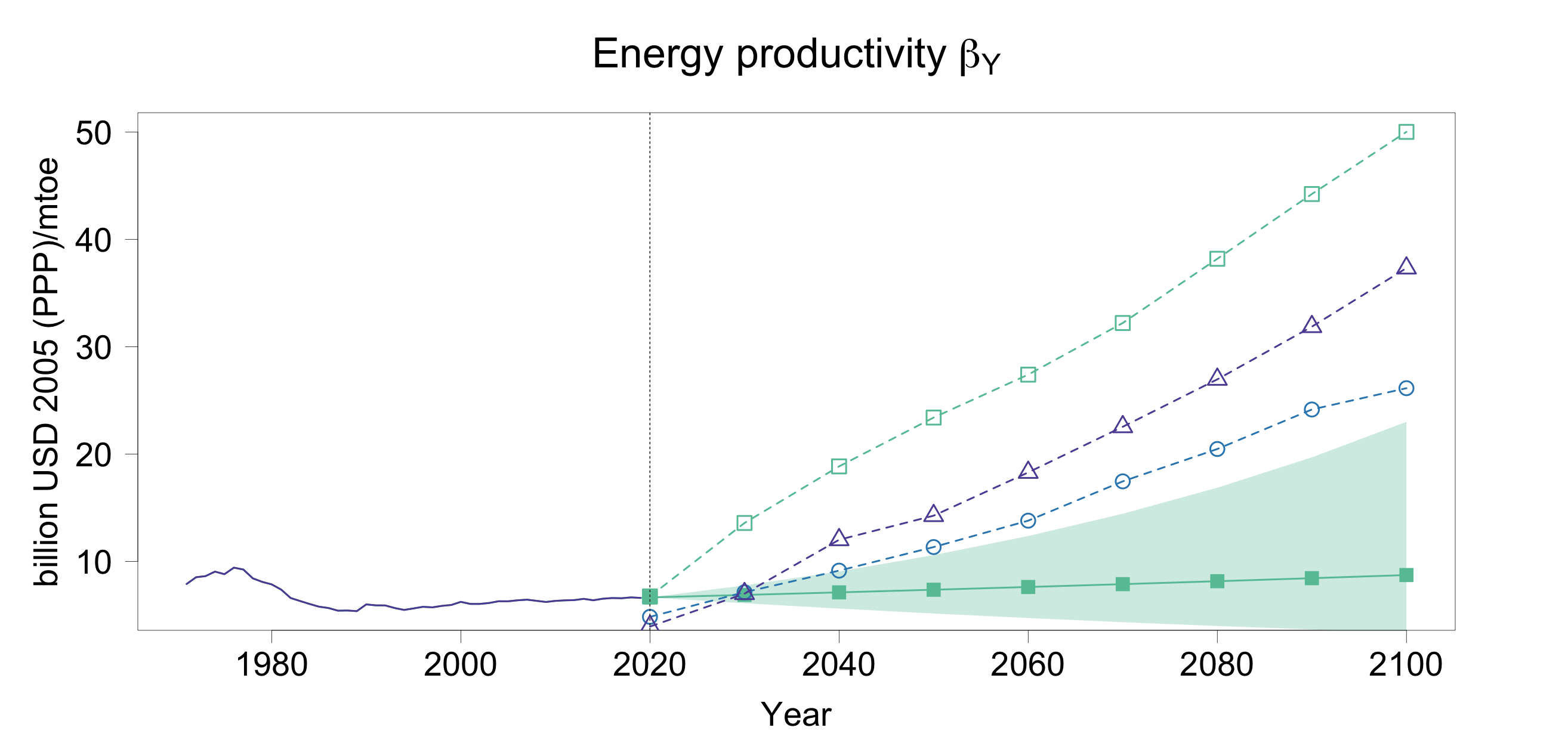}
	\end{subfigure}
\label{beta-g}
\end{figure}

\subsubsection{Projections with geometric growth in energy productivity}

Motivated by the large discrepancies of GDP trajectories from SSPs when assuming linear growth in energy productivity, in this section we consider a variant of the E3S2 model that admits geometric growth in $\beta_{Y}$ (E3S2-g model). We model the logarithm of $\beta_{Y}$ as the state variable and specify $\log \beta_{Y}$ as:
\begin{equation}
	\log \beta_{Y,t}=g+\log \beta_{Y,t-1}+\eta_{\beta_Y,t-1}.
	\label{geometric}
\end{equation}

Equation \eqref{geometric} specifies $\log \beta_{Y,t}-\log \beta_{Y,t-1}$ as the growth rate $g$ plus an error term $\eta_{\beta,t}$. To simplify the numerical optimization, we set all emission conversion factors as time-invariant and instead introduce an AR(1) structure in the measurement error process to account for autocorrelation. We also set $R^*_t=R_t$ and thus treat renewables as exogenous. The matrix form of the E3S2-g model is presented in Appendix \ref{EEEg}. The energy productivity $\beta_{ Y}$ of MAF shows a two-stage pattern, whereby $\beta_{ Y}$ declines in the first two decades, followed by a steady rise since the 1990s. In order to avoid distortions from using a single geometric growth rate to capture this pattern, we estimate the growth rates of these two stages separately. We use the estimate of the second growth rate in the projection. We use the gradient-based algorithm ``Limited-memory Broyden-Fletcher-Goldfarb-Shanno with Bound constraints'' (L-BFGS-B) \shortcite{byrd1995limited} embedded in the R function ``optim'' to maximize the log-likelihood , which performs more stably in the geometric growth specification when compared to the ``Nelder-Mead'' method. 

Figure \ref{beta-g} shows that when the energy productivity $\beta_{Y}$ is assumed to grow geometrically, three out of the five regions (REF, OECD, and ASIA) have comparable $\beta_{Y}$ projections to the SSPs. For the regions LAM and MAF, the E3S2-g model still undershoots by a large margin due to low estimated growth rates $g$. Detailed results from the E3S2-g model are presented in Appendix \ref{estimates_EEE-g}, \ref{Diagnostis EEE-g}, and \ref{projection EEE-g}, which report parameter estimates, diagnostic statistics, and projections comparison, respectively.

\subsection{Global scenario analysis}
\label{global}

In this section, we study projections at a global level. Apart from the SSP 1.9 $W\mathrm{m}^{-2}$ scenario we describe in the regional analysis, we also consider an alternative scenario, the IEA Net Zero road map (IEA NZE) \cite{bouckaert2021net}.

	\begin{figure}[h]
	\centering
	\caption{\footnotesize Comparison between global total energy supply pathways of the primary energy sources in the NZE scenario over the period 2020 -- 2050 and global SSP 1.9 W$\mathrm{m^{-2}}$ energy trajectories over the period 2020 -- 2100. NZE projections are available in 2020, 2030, 2035, 2040, and 2050, while the SSP trajectories provide data at a 10-year frequency from 2020 to 2100. }
	\includegraphics[width=\linewidth]{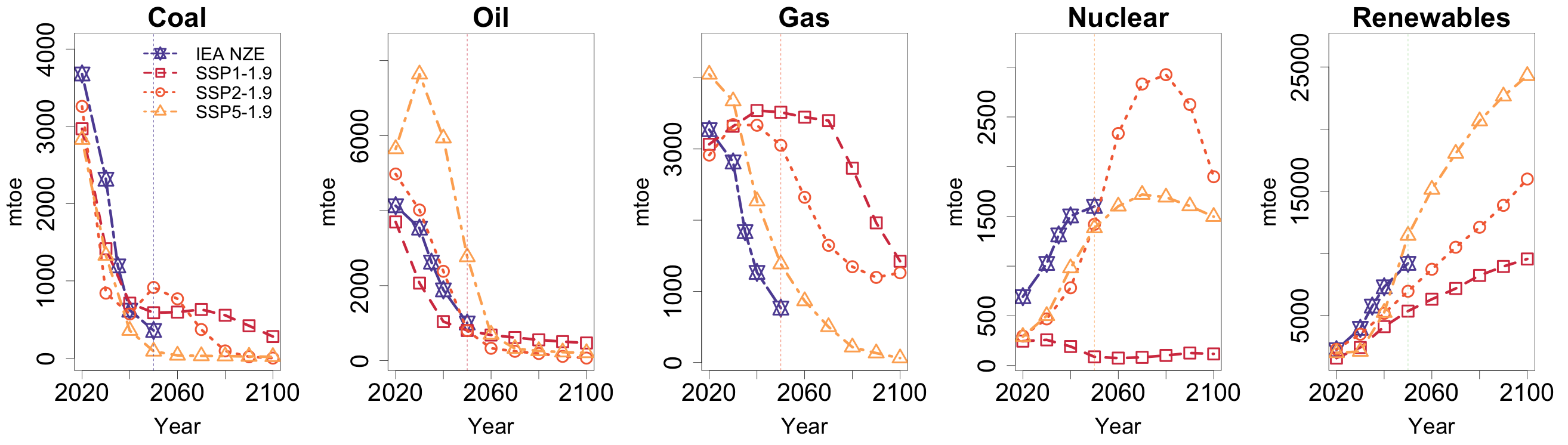}
	\label{IEANZandSSP1.9}
\end{figure}

IEA NZE explores the transition at the global level to a net zero emission  energy system by 2050 while adopting a cost-effective pathway and maintaining robust economic growth. It provides an alternative to SSP 1.9 $W\text{m}^{-2}$ in limiting global mean temperature rise to $1.5^{\circ} \mathrm{C}$. Figure \ref{IEANZandSSP1.9} compares the pathways of the five primary energy sources from SSP 1.9 $W\mathrm{m}^{-2}$ and IEA NZE. Figure \ref{IEANZandSSP1.9} shows that IEA NZE trajectories are most comparable to SSP5-1.9 in magnitude and variation pattern for coal, gas, nuclear, and renewables, while for oil, it is closer to SSP2-1.9. 

\begin{figure}[h!]
	\centering
	\caption{\footnotesize Comparison of global projections of CO$_2$ emission, GDP, and $\beta_Y$ generated from the E3S2 model and the trajectories from SSP 1.9 W$\mathrm{m^{-2}}$ and IEA NZE scenarios. Confidence bands are pointwise at the 90\% level.}
	\begin{subfigure}{0.5\textwidth}
		\subcaption{Emissions}
		\centering
		\includegraphics[width=\linewidth]{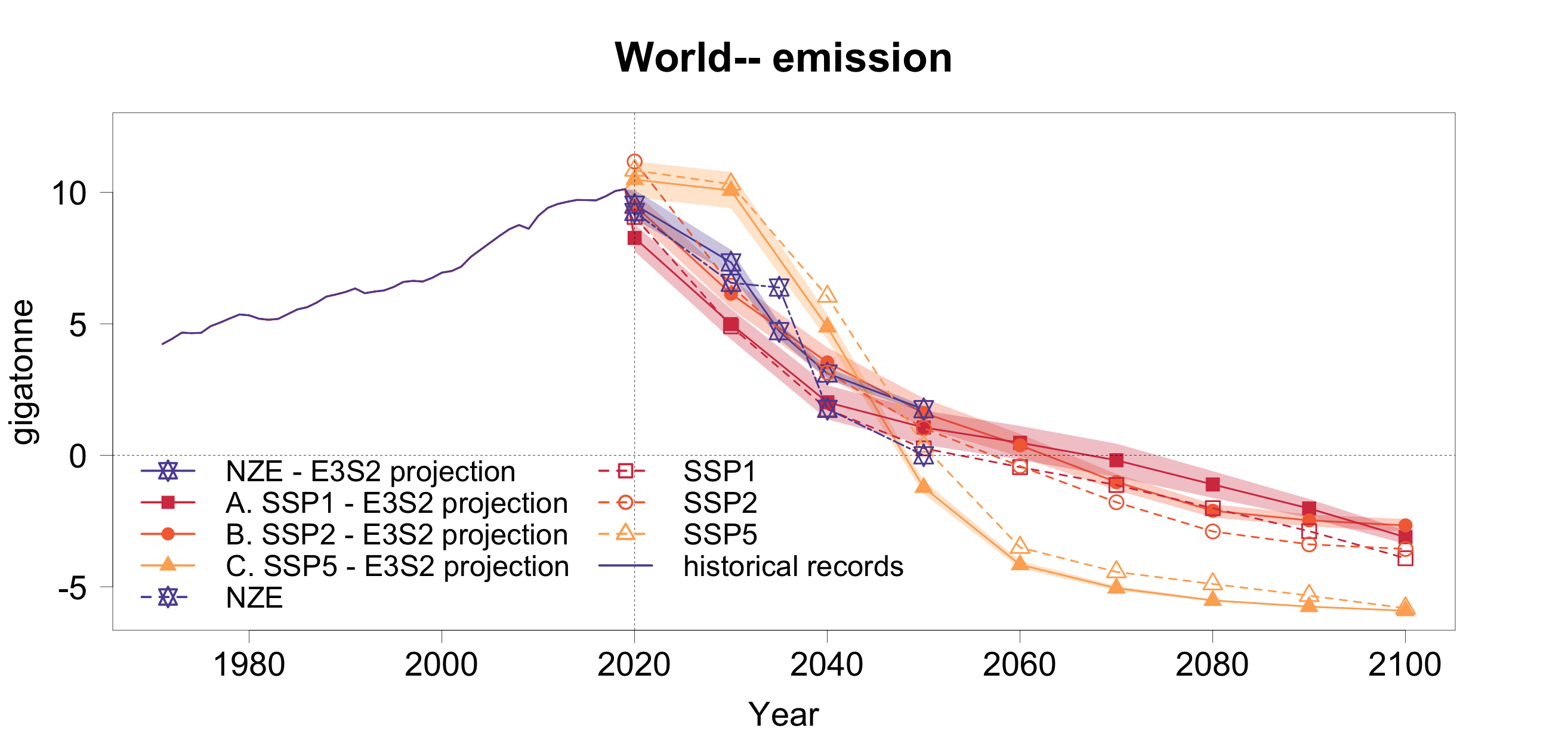}
	\end{subfigure}\hfill
	\begin{subfigure}{0.5\textwidth}
		\centering
		\subcaption{GDP}
		\includegraphics[width=\linewidth]{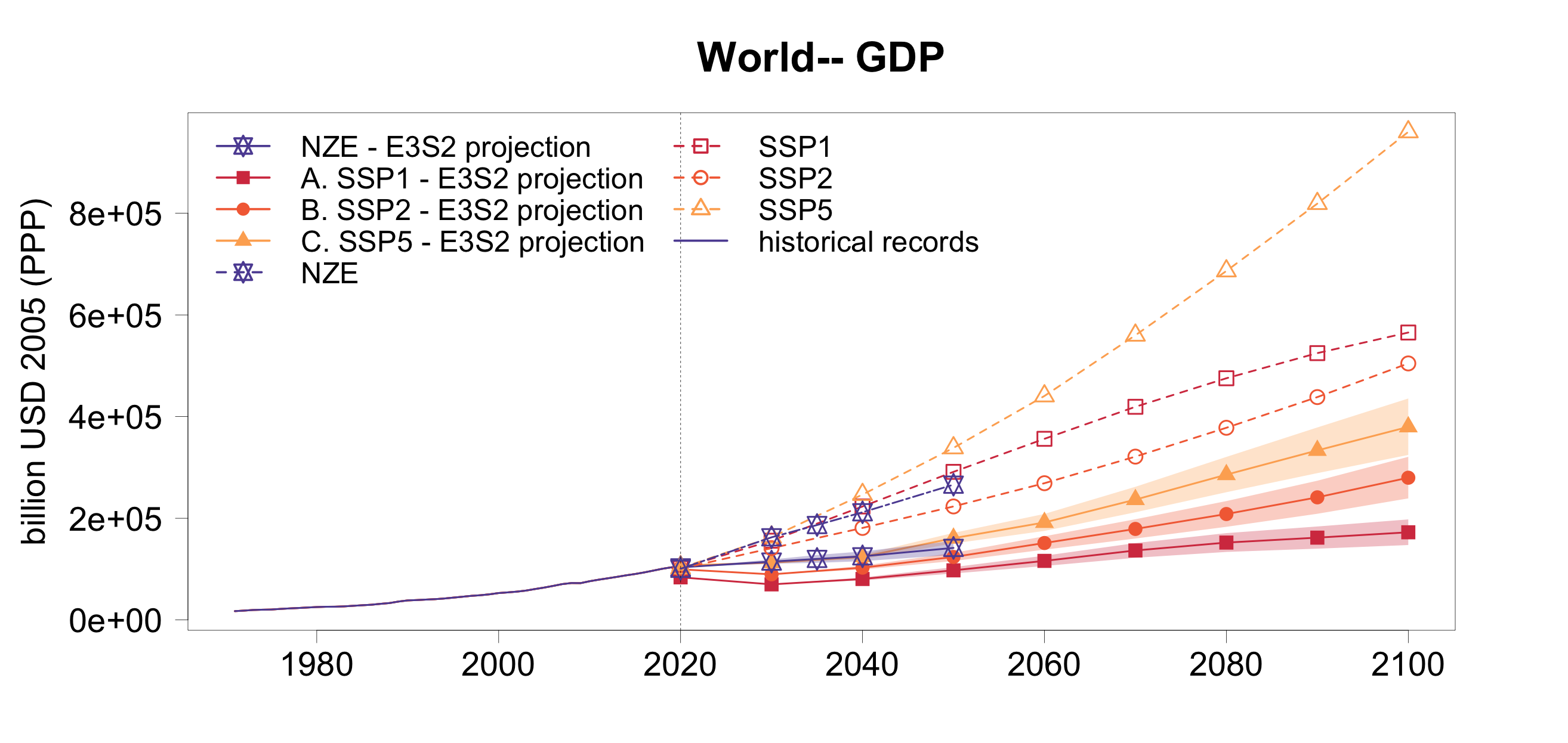}
	\end{subfigure}\\
	\begin{subfigure}{0.5\textwidth}
		\centering
		\subcaption{$\beta_{ Y}$}
		\includegraphics[width=\linewidth]{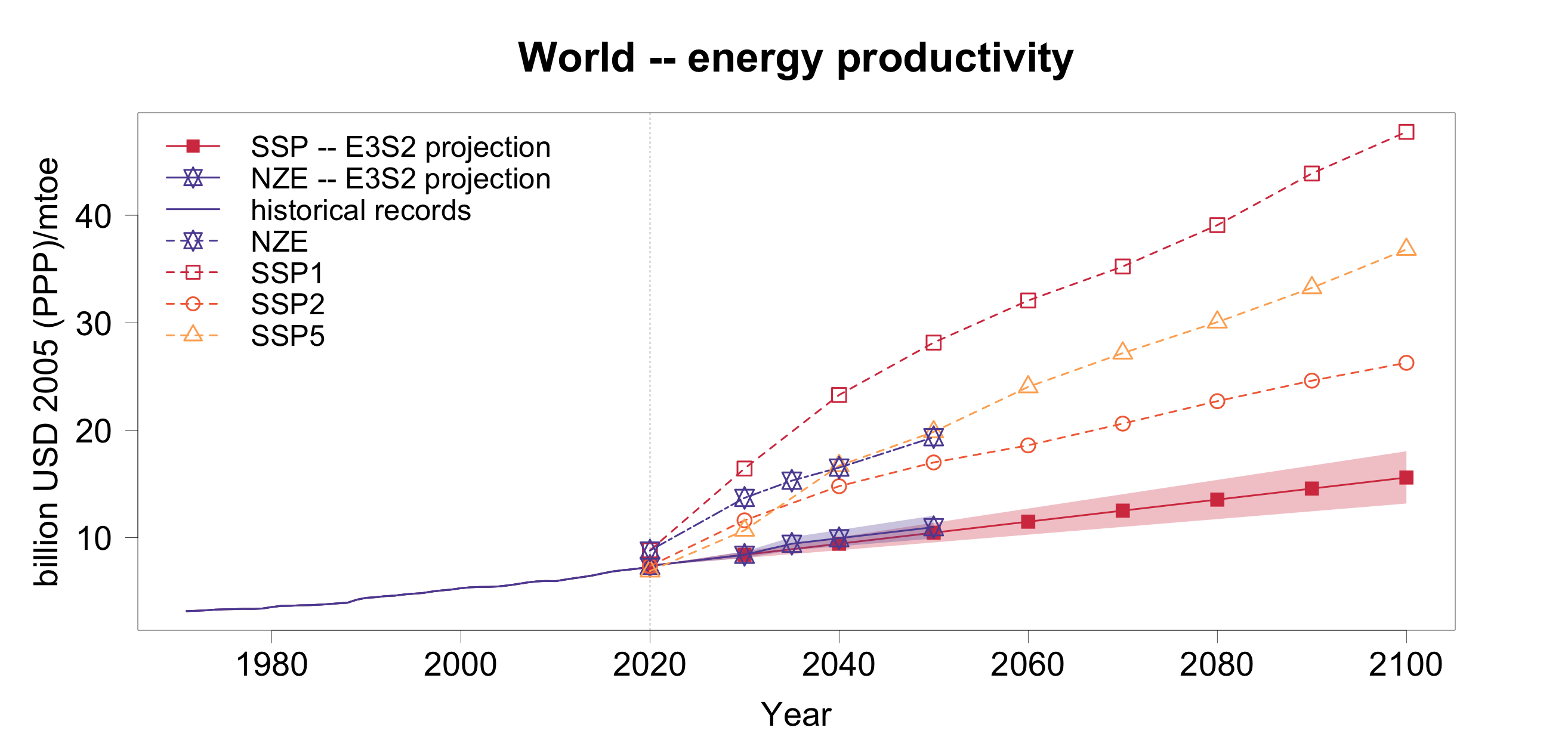}
	\end{subfigure}
	\label{IEAandSSP}

\end{figure}

We use the parameter estimates from fitting the E3S2 model to global data (Section \ref{empirical}) and adopt the same methodology for constructing projections introduced in Section \ref{approach}. Figure \ref{IEAandSSP} compares the long-term projections for CO$_2$ emissions, GDP, and $\beta_{Y}$ from our model, conditional on the energy pathways from SSP 1.9 W$\mathrm{m^{-2}}$ and IEA NZE. GDP projections from IEA NZE are in billion 2022 USD (PPP). We harmonize the series so that the value in 2020 matches the historical value, which is in USD 2005 billion (PPP), so that the NZE values can be compared to the historical dataset in this paper and the SSP values. 

Similar to the regional analysis, our model-generated CO$_2$ emissions projections are comparable to those provided by the scenarios. Up to 2030, the projection series from IEA NZE closely aligns with SSP2-1.9. However, beyond 2030, its decreasing trend accelerates, reaching net zero emissions in 2050. This is earlier than SSP1-1.9 and SSP2-1.9, but slightly later than SSP5-1.9.

The projected GDP and energy productivity ($\beta_{Y}$) from the E3S2 model once again exhibit significantly lower values compared to the corresponding projections from IEA NZE and SSP. The IEA NZE projection is most consistent with SSP1-1.9 in GDP projections and SSP5-1.9 in $\beta_{Y}$ projections.

\begin{figure}[H]
	\centering
	\caption{\footnotesize Comparison of global projections of CO$_2$ emission, GDP, and $\beta_Y$ generated from the E3S2-g model and the trajectories from SSP 1.9 W$\mathrm{m^{-2}}$ and IEA NZE scenarios. Confidence bands are pointwise at the 90\% level.}
	\begin{subfigure}{0.5\textwidth}
		\subcaption{Emissions}
		\centering
		\includegraphics[width=\linewidth]{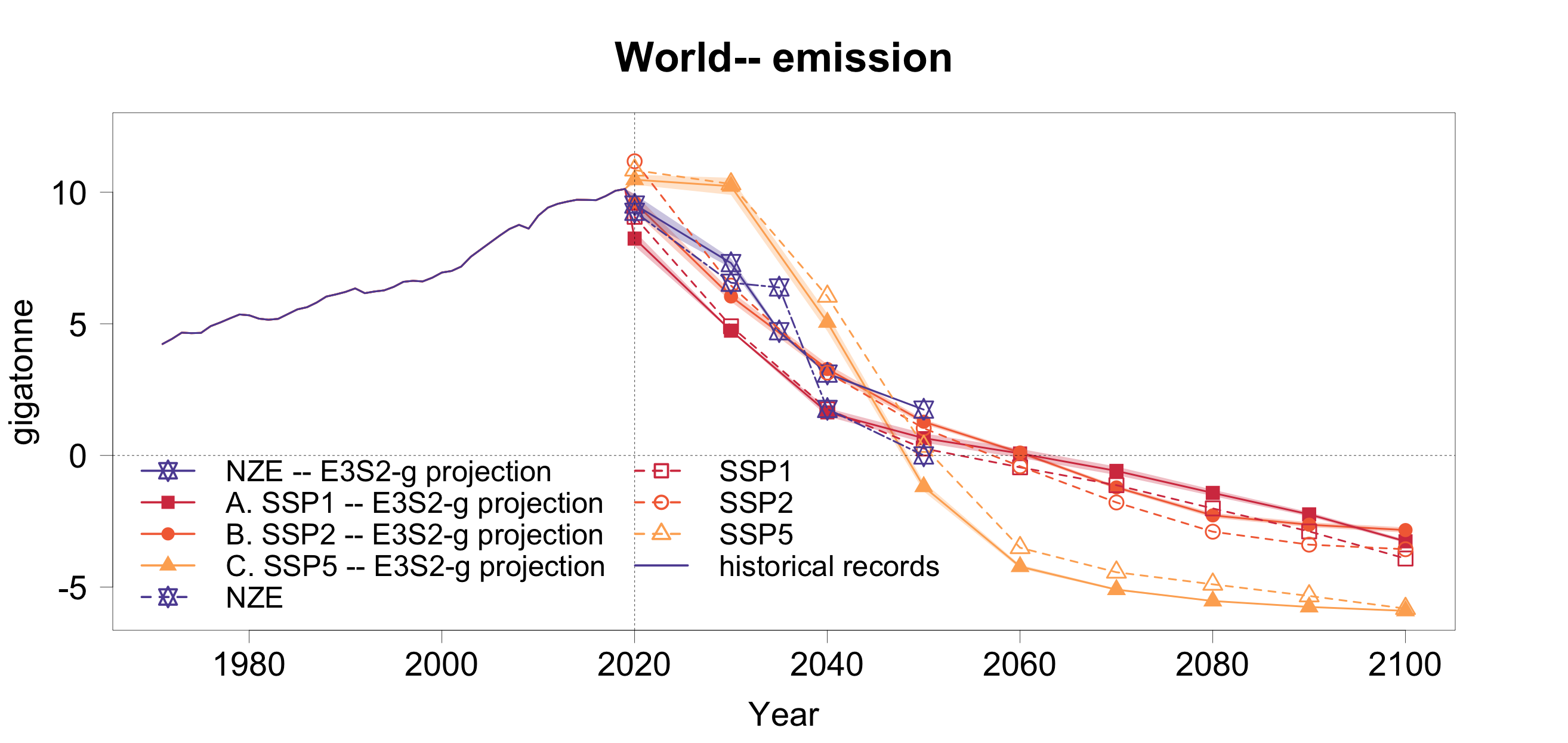}
	\end{subfigure}\hfill
	\begin{subfigure}{0.5\textwidth}
		\centering
		\subcaption{GDP}
		\includegraphics[width=\linewidth]{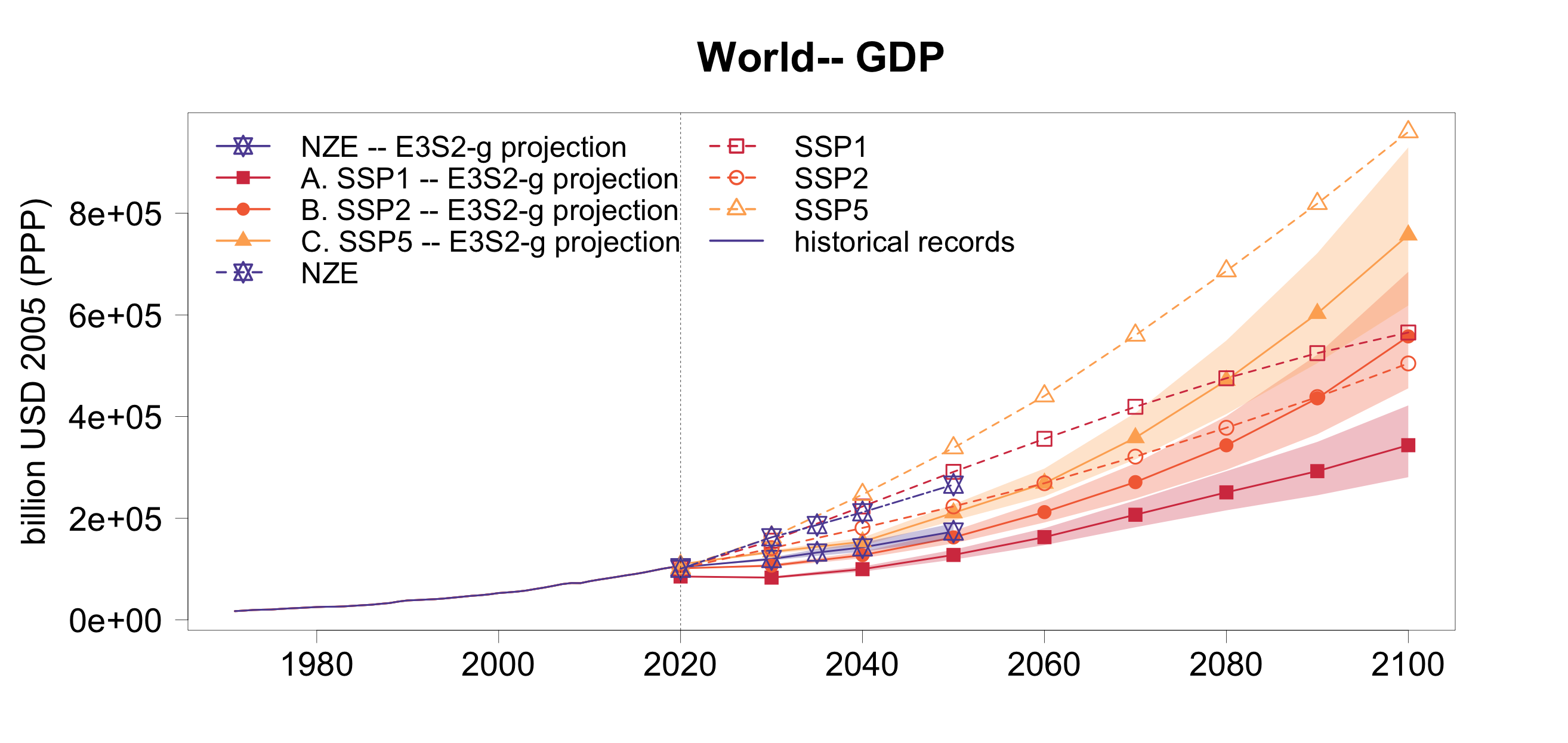}
	\end{subfigure}\\
	\begin{subfigure}{0.5\textwidth}
		\centering
		\subcaption{$\beta_{ Y}$}
		\includegraphics[width=\linewidth]{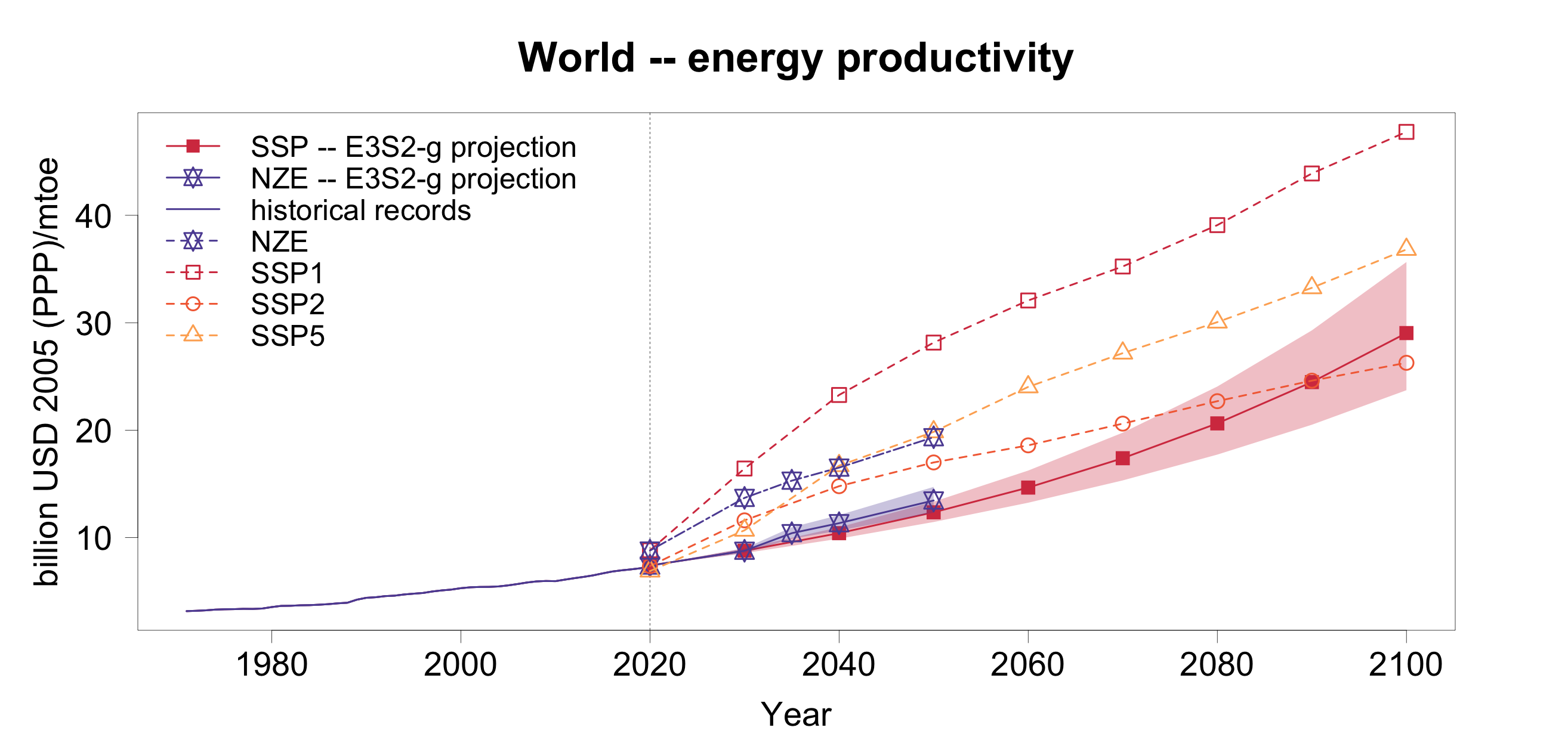}
	\end{subfigure}
	\label{IEAandSSP-g}
	
\end{figure}

Furthermore, we conduct an estimation and projection using the E3S2-g model based on global-level data to compare the geometric growth path with the projected trajectories from SSP1-1.9 and IEA NZE. Detailed parameter estimates and diagnostic statistics from this analysis are provided in Appendix \ref{estimates_EEE-g} and \ref{Diagnostis EEE-g}, respectively, alongside regional results. The comparative analysis of projections for emissions, GDP, and $\beta_{ Y}$ is presented in Figure \ref{IEAandSSP-g}.

A comparison of Panel (a) in Figure \ref{IEAandSSP} and Figure \ref{IEAandSSP-g} highlights a notable resemblance between the projection medians. This similarity is expected, given that the specification of geometric growth primarily influence GDP and $\beta_{ Y}$ projections. Nevertheless, the linear trend projection presented in Panel (a) of \ref{IEAandSSP} exhibits slightly wider confidence bands, due to larger uncertainty estimates.

Under geometric growth, the projections for GDP and energy productivity show considerable increases, more than doubling the values projected under a linear growth assumption. The trajectory of GDP generated by the E3S2-g model, when conditional on SSP2-1.9,  aligns with the path from the SSP assessment. However, in the two other SSP 1.9 scenarios,  the paths generated by the E3S2-g model still fall short of the SSP counterparts.

\clearpage

\section{Conclusion}
\label{Conclusion}

In this paper, we present a non-linear Gaussian state space system for CO$_2$ emissions, GDP, and the energy mix (E3S2 model). The E3S2 model consists of three modules: energy, emissions, and economy. The energy module takes the consumption quantities of coal, oil, gas, and nuclear energy as exogenously given and models a deterministic or stochastic trend in renewable energy. The emission module represents CO$2$ emissions as a weighted linear combination of fossil fuels, where we allow the weights, i.e., the emission conversion factors, to vary over time.  The economy module specifies GDP as the product of the energy productivity $\beta_{Y}$ and the aggregate of the consumption quantities of the five primary energy sources. 

The E3S2 model is a general form nesting many specifications. It is a flexible statistical model capable of capturing time-varying dynamics, addressing non-stationarity, and assessing estimation and projection uncertainty. It can be adapted to data from various geographic resolutions, including regions or countries.  

We follow a data-driven model selection procedure to choose an appropriate form for global analysis and the five regions considered in the SSP scenarios. The E3S2 model is estimated using maximum likelihood via the extended Kalman filter. We find that most of our emission conversion factor estimates are in line with benchmarks in the literature, especially when we take estimation uncertainty into consideration. Worldwide, and for the two regions MAF and LAM, the emission conversion factor of coal has been declining over time, while for OECD, it is gas that shows a downward trend. 

We perform long-term projections conditional on the fuel mix pathways from SSP 1.9 $W\mathrm{m^{-2}}$ scenarios and the IEA Net Zero by 2050 roadmap. The emissions projections align with those from these two scenarios. This shows that the emission conversion factors estimated by the E3S2 model are consistent with those used in the SSP and IEA-NZE projections. 

However, our GDP projections, constructed on the assumption of a linear trend in energy productivity, present a much less optimistic picture. In an alternative specification (E3S2-g model), where energy productivity has geometric growth, the GDP projections align better. For the three regions OECD, ASIA, and REF, the GDP projections from the E3S2-g model are compatible with the trajectories outlined in SSP 1.9, although they still fall short. 

Our results indicate that following the economic growth paths of SSP and IEA-NZE trajectories necessitates rapid growth trajectories for the energy productivity factor $\beta_Y$, deviating considerably from historical trends. If the green transition is to align in terms of GDP growth with the scenarios set forth in SSPs and by the IEA, drastic and transformative change in energy intensity of GDP is necessary.


\clearpage

\bibliographystyle{apacite}
\bibliography{Ref_Eric}

\clearpage

\appendix
\section*{Appendix}
\section{List of countries included in SSP regions but excluded in our analysis }
\label{listofcountries}
\begin{itemize}
	\item The five regions defined by the SSPs contain 190 countries, whereas our dataset contains 146 countries (primarily limited by the coverage of the IEA energy data). The 44 countries included in SSP regions but absent in our dataset are:
	\subitem \textbf{OECD}: Guam, Puerto Rico; 
	\subitem 
	\textbf{ASIA}: Afghanistan, Bhutan, Fiji, French Polynesia, Maldives, Federated States of Micronesia, New Caledonia, Papua New Guinea, Samoa, Solomon Islands, Timor-Leste, Vanuatu; 
	\subitem 
	\textbf{MAF}: Burkina Faso, Burundi, Cape Verde, Central African Republic, Chad, Comoros, Djibouti, Gambia, Guinea, Guinea-Bissau, Lesotho, Liberia, Malawi, Mali, Mauritania, Mayotte, Occupied Palestinian Territory, Réunion, Sierra Leone, Somalia, Western Sahara; 
	\subitem 
	\textbf{LAM}:  Aruba, Bahamas, Barbados, Belize, French Guiana, Grenada, Guadeloupe, Martinique, Virgin Islands (US). 
\end{itemize}

\section{The matrix form of the E3S2 model}
\label{EEEmatrixform}

The matrix form of a non-linear Gaussian state space model can be written as \shortcite{durbin2012time}, 
\begin{equation}
	\begin{aligned}
		\mathbf{X}_{t}=&\mathbf{T}_{t-1}\left(\mathbf{X}_{t-1}\right)+\mathbf{R}_{t-1}\left(	\mathbf{X}_{t-1}\right) \boldsymbol{\upeta}_{t-1} \\
		\mathbf{Y}_{t}=&\mathbf{Z}_t\left(\mathbf{X}_t\right)+\boldsymbol{\upvarepsilon}_t, 
	\end{aligned}
	\label{non-linear SSM}
\end{equation}
where $\mathbf{X}_t$, $\mathbf{Y}_t$, $\boldsymbol{\upeta}_{t-1}$, and $\boldsymbol{\upvarepsilon}_t$ denote vectors of state variables, observations, state disturbances, and measurement errors, respectively. The transition matrices $\mathbf{T}_{t-1}\left(\mathbf{X}_{t-1}\right)$, $\mathbf{R}_{t-1}\left(\mathbf{X}_{t-1}\right)$, and $\mathbf{Z}_t\left(\mathbf{X}_t\right)$ are functions of state variables. Expanding these matrix functions in Taylor series and neglecting the higher-order terms, Equation \eqref{non-linear SSM} can be linearized as: 
\begin{equation}
	\begin{aligned}
		\mathbf{X}_{t}=&\dot{\mathbf{T}}_{t-1}	\mathbf{X}_{t-1}+\mathbf{c}_{t-1}+\mathbf{R}_{t-1}\left(	\mathbf{x}_{t-1\mid t-1}\right) \boldsymbol{\upeta}_{t-1} \\
		\mathbf{Y}_{t}=&\dot{\mathbf{Z}}_t\mathbf{X}_t+\mathbf{d_t}+\boldsymbol{\upvarepsilon}_t, 
	\end{aligned}
	\label{linearized non-linear SSM}
\end{equation}
where
\begin{equation}
	\begin{aligned}
		\dot{\mathbf{Z}}_t= &\left.\frac{\partial \mathbf{Z}_t\left(\mathbf{X}_t\right)}{\partial \mathbf{X}_t^{\prime}}\right|_{\mathbf{X}_t=\mathbf{x}_t}, \quad \dot{\mathbf{T}}_{t-1}=\left.\frac{\partial \mathbf{T}_{t-1}\left(\mathbf{X}_{t-1}\right)}{\partial \mathbf{X}_{t-1}^{\prime}}\right|_{\mathbf{X}_{t-1}=\mathbf{x}_{t-1 \mid t-1}},\\
		\mathbf{d_t}=&\mathbf{Z}_t\left(\mathbf{x}_t\right)-\dot{\mathbf{Z}}_t \mathbf{x}_t, \quad \mathbf{c}_{t-1}=\mathbf{T}_{t-1}\left(	\mathbf{x}_{t-1\mid t-1}\right)-\dot{\mathbf{T}}_{t-1} \mathbf{x}_{t-1\mid t-1}, \\
		\boldsymbol{\upeta}_{t-1} \sim & \left(\boldsymbol{0}, \mathbf{Q}_{t-1}\left(\mathbf{x}_{t-1\mid t-1}\right)\right), \qquad \boldsymbol{\upvarepsilon}_t\sim\left(\boldsymbol{0}, \mathbf{H}_t\left(\mathbf{x}_t\right)\right),
		\label{jacobian}
	\end{aligned}
\end{equation}
and 
\begin{equation}
	\mathbf{x}_{t}=\mathbbm{E}\left(\mathbf{X}_{t} \mid \mathbf{X}_{t-1}\right), \qquad \mathbf{x}_{t\mid t}=\mathbbm{E}\left(\mathbf{X}_{t} \mid \mathbf{Y}_t\right).
\end{equation}
In the E3S2 model, the state equations are
\begin{equation}
	\scriptsize
	\begin{aligned}
		\underbrace{\left(\begin{array}{c}
				E_{t}^{*} \\
				Y_{t}^{*} \\
				\beta_{C,t} \\
				\beta_{O,t} \\
				\beta_{G,t} \\
				\beta_{Y,t}\\
				R_{t}^* \\
				d_{R,t} \\
				d_{\beta_Y,t}
			\end{array}\right)}_{\mathbf{X}_{t}}=	&
		\underbrace{\left(\begin{array}{c}
				\beta_{C,t-1} C_t + \beta_{O,t-1} O_t + \beta_{G,t-1} G_t \\
				\beta_{Y,t-1}\left(C_t + O_t + G_t + N_t + d_{R,t-1} + R^*_{t-1}\right) \\
				\beta_{C,t-1}\\
				\beta_{O,t-1} \\
				\beta_{G,t-1} \\
				d_{\beta_Y,t-1}+\beta_{Y,t-1}\\
				d_{R,t-1}+R_{t-1}\\
				d_{R, t-1}\\
				d_{\beta_Y, t-1}
			\end{array}\right) }_{\mathbf{T}_{t-1}\left(\mathbf{X_{t-1}}\right)}
		+\\
		&\underbrace{	{\begin{pmatrix}
					1 & 0 & C_t & O_t & G_t& 0& 0 & 0& 0\\
					0 & 1 & 0 & 0 & 0& 0& \beta_{Y,t-1\mid t-1} & 0& 0\\
					0 & 0 &	1  & 0 & 0& 0& 0 & 0& 0\\
					0 & 0 & 0 &	1 & 0& 0& 0 & 0& 0\\
					0 & 0 & 0 & 0 & 1 & 0& 0 & 0& 0\\
					0 & 0 & 0 & 0 & 0& 1& 0 & 0& 0\\
					0 & 0 & 0 & 0 & 0& 0& 1& 0& 0\\
					0 & 0 & 0 & 0 & 0& 0&0  & 1 & 0 \\
					0 & 0 & 0 & 0 & 0& 0& 0 & 0&  1 
		\end{pmatrix}}}_{\mathbf{R}_{t-1\mid t-1}\left(\mathbf{X_{t-1\mid t-1}}\right)}	\underbrace{\begin{pmatrix}
				\eta_{E,t-1} \\
				\eta_{Y,t-1} \\
				\eta_{\beta_{C},t-1} \\
				\eta_{\beta_{O},t-1} \\
				\eta_{\beta_{G},t-1} \\
				\eta_{\beta_Y,t-1}\\
				\eta_{R,t-1}\\
				\eta_{d_R,t-1}\\
				\eta_{d_{\beta_Y},t-1}
		\end{pmatrix}}_{\boldsymbol{\upeta}_{t-1}},
	\end{aligned}
\end{equation}
and the measurement equations are
\begin{equation}
	\footnotesize
	\underbrace{	\left(\begin{array}{c}
			E_{t} \\
			Y_{ t} \\
			R_{ t}
		\end{array}\right)}_{\mathbf{Y}_t}=\underbrace{\left(\begin{array}{c}
			E_t^* \\
			Y_t^*  \\
			R_t^*
		\end{array}\right)}_{\mathbf{Z}_t(\mathbf{X}_t)} +\underbrace{ \begin{pmatrix}
			\varepsilon_t^E \\
			\varepsilon_t^Y \\
			\varepsilon_t^R \\
	\end{pmatrix}}_{\boldsymbol{\upvarepsilon}_t}.
\end{equation}
The Jacobian matrix in Equation \eqref{jacobian} takes the form: 
\begin{equation}
	\scriptsize
	\dot{\mathbf{T}}_{t-1}=	\left(\begin{array}{ccccccccccccc}
		0 & 0 & 	C_t & 		O_t & 		G_t & 0 & 0& 0 & 0\\
		0 & 0 & 0 &0& 0& h_{t-1|t-1} & \beta_{Y,t-1\mid t-1} & \beta_{Y,t-1\mid t-1}\\
		0 & 0 & 	1  & 0 & 0 & 0 & 0 & 0 & 0\\
		0 & 0 & 0 & 	1  & 0 & 0 & 0& 0 & 0\\
		0 & 0 & 0 & 0 & 	1  & 0 & 0 & 0 & 0\\
		0 & 0 & 0 & 0 & 0 & 1 & 0& 0 & 1\\
		0 & 0 & 0 & 0 & 0 & 0 & 1& 1& 0\\
		0 & 0 & 0 & 0 & 0 & 0 & 0& 1	 & 0\\
		0 & 0 & 0 & 0 & 0 & 0 & 0& 0  & 1 
	\end{array}\right),
\end{equation}
where
\[
h_{t-1|t-1} = C_t + O_t + G_t + N_t + d_{R,t-1|t-1} + R^*_{t-1|t-1},
\]
and 
\begin{equation}
	\footnotesize
	\dot{\mathbf{Z}}_t =\mathbf{Z}\left(\mathbf{X_t}\right)=	\begin{pmatrix}
		\begin{array}{llllllllll}
			1 & 0 & 0 & 0 & 0 & 0 & 0 & 0& 0 \\
			0& 1 & 0 & 0 & 0 & 0 & 0& 0 & 0 \\
			0 & 0 & 0 & 0 & 0 & 0 & 1& 0 & 0
		\end{array}
	\end{pmatrix}.
	\label{EEEmatrix}
\end{equation}
The variance-covariance matrices of the state disturbances and measurement errors are independent of the state variables: 
\begin{equation}
	\footnotesize
	\begin{aligned}
		\mathbf{Q}_{t-1}\left(\mathbf{x}_{t-1\mid t-1}\right)=\mathbf{Q} =	{\begin{pmatrix}
				\sigma_{\eta_E}^2 & 0 & 0 & 0 & 0& 0& 0 & 0& 0\\
				0 & 	\sigma_{\eta_Y}^2& 0 & 0 & 0& 0& 0 & 0& 0\\
				0 & 0 &	\sigma_{\eta_{\beta_{ C}}}^2  & 0 & 0& 0& 0 & 0& 0\\
				0 & 0 & 0 &	\sigma_{\eta_{\beta_{ O}}}^2 & 0& 0& 0 & 0& 0\\
				0 & 0 & 0 & 0 & 	\sigma_{\eta_{\beta_{ G}}}^2 & 0& 0 & 0& 0\\
				0 & 0 & 0 & 0 & 0& 	\sigma_{\eta_{\beta_{ Y}}}^2& 0 & 0& 0\\
				0 & 0 & 0 & 0 & 0& 0& \sigma_{\eta_R}^2& 0& 0\\
				0 & 0 & 0 & 0 & 0& 0&0  & \sigma_{\eta_{d_{{{R}}}}}^2& 0 \\
				0 & 0 & 0 & 0 & 0& 0& 0 & 0&  \sigma_{\eta_{d_{{\beta_{ Y}}}}}^2
		\end{pmatrix}}
	\end{aligned}
\end{equation}
and 
\begin{equation}
	\footnotesize
	\mathbf{H}_t\left(\mathbf{x}_t\right)=\mathbf{H}=\begin{pmatrix}
		\sigma_{\varepsilon_{E}}^2 & 0 & 0 \\
		0 &	\sigma_{\varepsilon_{Y}}^2 & 0 \\
		0 & 0&		\sigma_{\varepsilon_{R}}^2  \\
	\end{pmatrix}.
\end{equation}

\section{Further results for the E3S2 model}
\label{KFoutput}
\begin{figure}[H]
	\caption{Smoothed states of $E^*$, $Y^*$, and $R^*$ (solid lines) compared to data (circles); smoothed energy productivity $\beta_Y$ and smoothed stochastic trend of $\beta_Y$ (if any) and of $R^*$ (if any); residuals (one-step ahead standardized prediction errors of $E^*$, $Y^*$, and $R^*$). }
	\begin{subfigure}{\textwidth}
				\centering
		\subcaption{OECD}
		\includegraphics[width=0.95\linewidth]{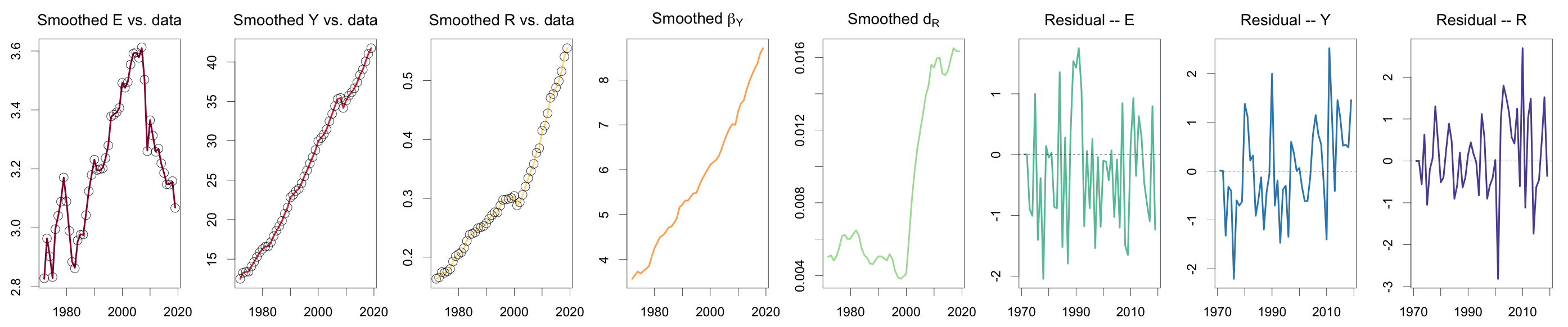}
	\end{subfigure}
	\begingroup\scriptsize
	\begin{subfigure}{\textwidth}
		\centering
		\subcaption{REF}
		\includegraphics[width=0.95\linewidth]{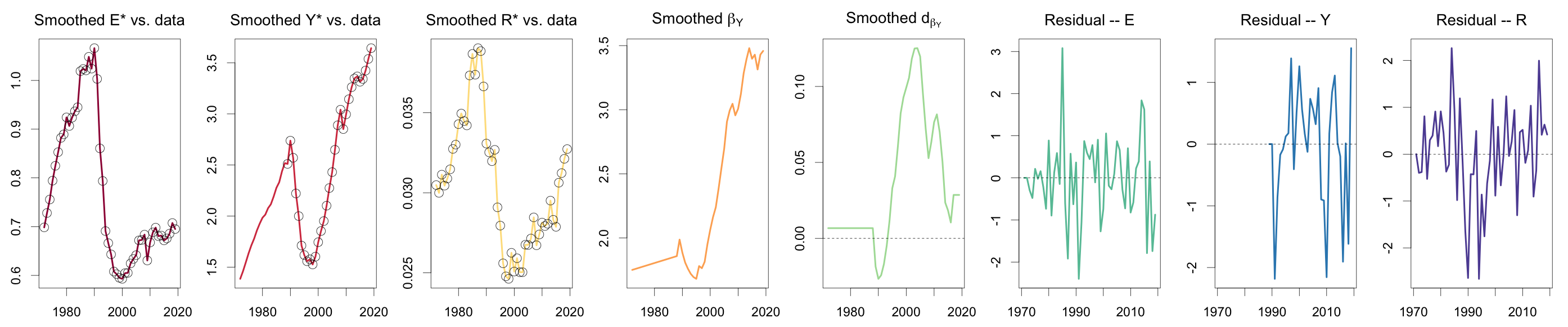}
	\end{subfigure}
	\begin{subfigure}{\textwidth}
				\centering
		\subcaption*{ASIA}
		\includegraphics[width=0.95\linewidth]{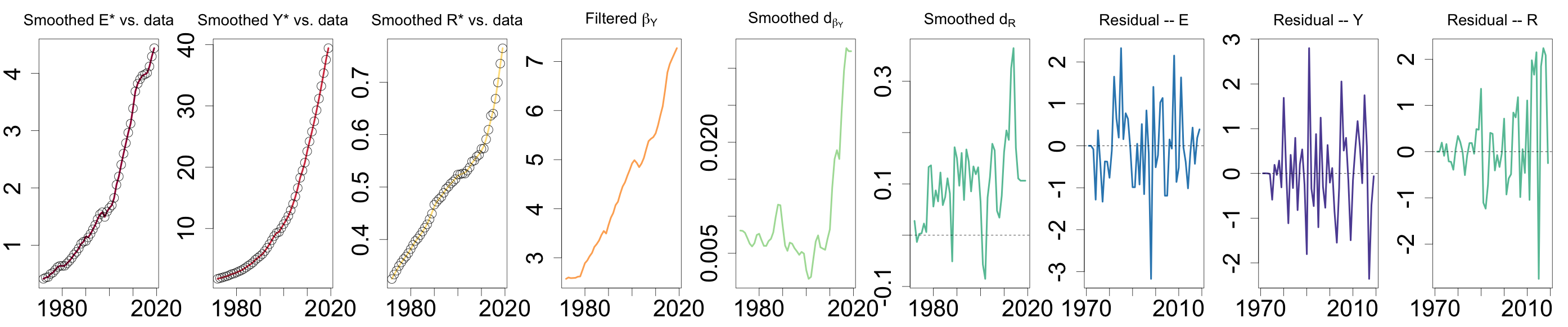}
	\end{subfigure}
	\endgroup
\label{KF}
\end{figure}
\begin{figure}\ContinuedFloat
	\begin{subfigure}{\textwidth}
		\centering
		\subcaption{MAF}
		\includegraphics[width=0.95\linewidth]{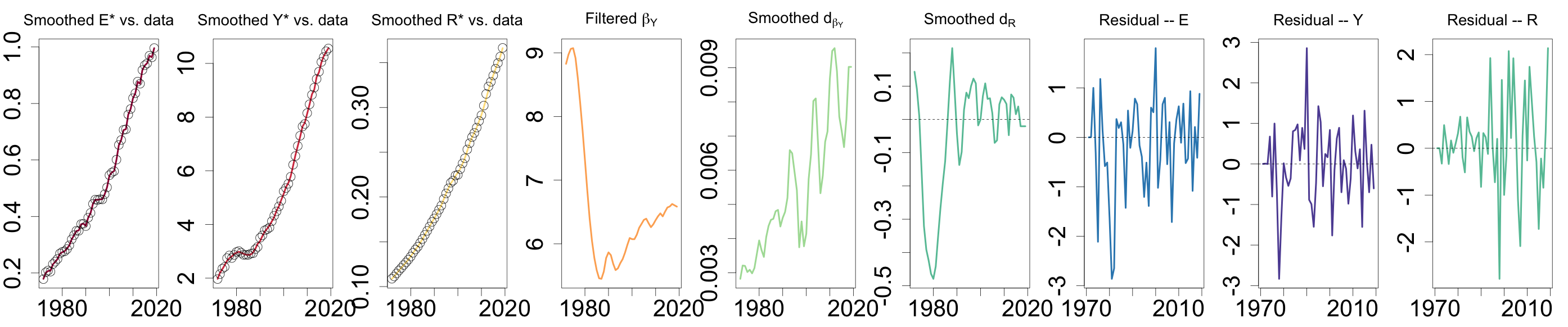}
	\end{subfigure}
	\begin{subfigure}{\textwidth}
		\centering
		\subcaption{LAM}
		\includegraphics[width=0.95\linewidth]{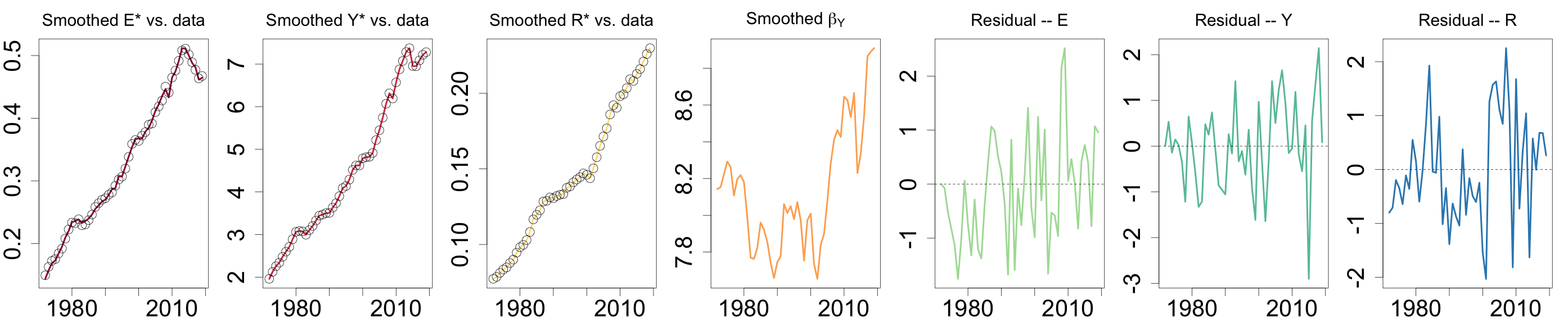}
	\end{subfigure}
	\begin{subfigure}{\textwidth}
				\centering
	\subcaption{WORLD}
	\includegraphics[width=0.95\linewidth]{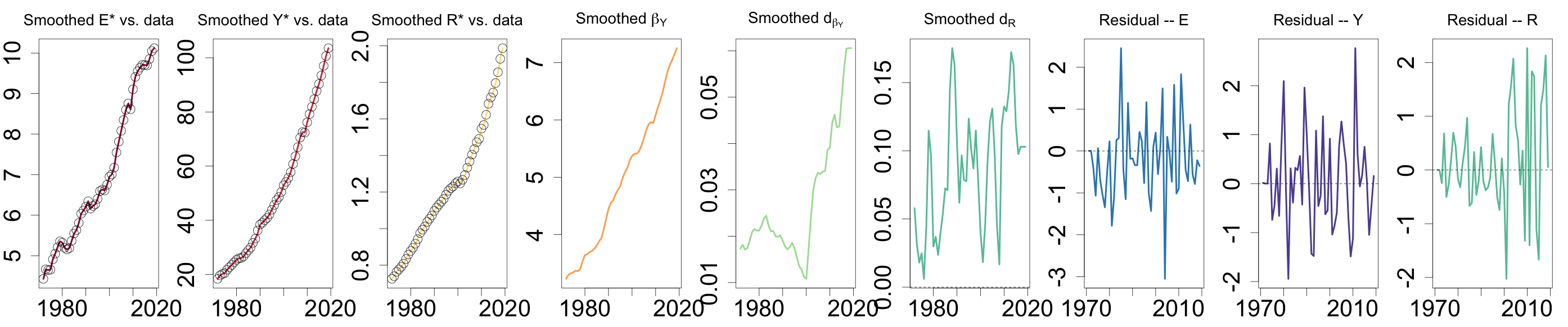}
\end{subfigure}
\end{figure}

\clearpage

\section{E3S2-g model}
\subsection{The matrix form of the E3S2-g model}
\label{EEEg}
\textit{State equation}:
\begin{equation}
	\scriptsize
	\begin{aligned}
		& \left(\begin{array}{c}
			E_{t}^* \\
			Y_{t}^* \\
			\beta_{C,t} \\
			\beta_{O,t} \\
			\beta_{G,t}\\
			\log	\beta_{Y,t} \\
			\alpha_{t}^* 
		\end{array}\right)=\left(\begin{array}{c}
			\beta_{C,t-1} C_t + 	\beta_{O,t-1} O_t + \beta_{G,t-1} G_t  \\
			\exp\left(\log \beta_{Y,t-1}\right)\left(C_t+O_t+G_t+N_t+R_t\right) \\
			\beta_{C,t-1}\\
			\beta_{O,t-1} \\
			\beta_{G,t-1} \\
			g+\log \beta_{Y,t-1} \\
			\phi	\alpha_{t-1}^* 
		\end{array}\right)
		& +\left(\begin{array}{ccccccccc}
			1& 0 & 	C_t & O_t & G_t &0 & 0  \\
			0 & 1 & 0 & 0 & 0 & 0 &0  \\
			0 & 0 & 1 & 0 & 0 & 0 &0  \\
			0 & 0 & 0 & 1 & 0 & 0 &0  \\
			0 & 0 & 0 & 0 & 1 & 0 &0  \\
			0 & 0 & 0 & 0 & 0 & 1 &0  \\
			0 & 0 & 0 & 0 & 0 & 0  & 1
		\end{array}\right)\left(\begin{array}{c}
			\eta_{E,t-1} \\
			\eta_{Y,t-1} \\
			\eta_{\beta_{C},t-1} \\
			\eta_{\beta_{O},t-1} \\
			\eta_{\beta_{G},t-1} \\
			\eta_{\beta_Y,t-1}\\
			\eta_{\alpha,t-1}
		\end{array}\right) \\
		&
	\end{aligned}.
\end{equation}
\textit{Measurement equation}:
\begin{equation}
	\scriptsize
	\left(\begin{array}{c}
		E_{t} \\
		Y_{ t} 
	\end{array}\right)=\left(\begin{array}{c}
		E_t^* + \alpha_{t}^*\\
		Y_t^*  
	\end{array}\right) + \begin{pmatrix}
		\varepsilon_t^E \\
		\varepsilon_t^Y 
	\end{pmatrix}.
\end{equation}
The Jacobian matrix $\dot{\mathbf{T}}_{t-1}$ takes the form:
\begin{equation}
	\scriptsize
\dot{\mathbf{T}}_{t-1} =
	\left(\begin{array}{ccccccccc}
		0 & 0 & C_t & O_t & G_t & 0 &0 \\
		0 & 0 & 0 & 0 & 0 & 	\exp\left(\log \beta_{Y,t-1|t-1}\right)\left(C_t+O_t+G_t+N_t+R_t\right)& 0\\
		0 & 0 &  1 & 0 & 0 & 0 &0 \\
		0 & 0 & 0 & 1 & 0 & 0 & 0 \\
		0 & 0 & 0 & 0 &  1& 0 &0 \\
		0 & 0 & 0 & 0 & 0 & 1 & 0 \\
		0 & 0 & 0 & 0 & 0 & 0 & \phi 
	\end{array}\right).
\end{equation}

\subsection{Parameter estimates}
	\label{estimates_EEE-g}
\begin{table}[H]
	\centering
	\scriptsize
	\setlength{\tabcolsep}{8pt} 
	\renewcommand{\arraystretch}{1} 
	\caption{\footnotesize Point estimates and standard errors (in parentheses) of CO$_2$ emission conversion factors $\beta_{ C}$, $\beta_{O}$, and $\beta_{G}$, of the geometric growth rate $g$, of the AR(1) coefficient $\phi$ for the AR term in the measurement equation (for the five regions) or the state equation (for WORLD), and of coefficients of dummies $D$ in the E3S2-g model. The superscript of $D$ refers to whether the dummy is included in the state equation ($S$) or the measurement equation ($M$), and the subscript refers to the variable name and time index of the dummy. Emission conversion factors are in the unit of tonne CO$_2$ equivalent per tonne of oil equivalent (tonne CO$_2$e/ toe).}
	\begin{threeparttable}
		\begin{tabular}{l|cccccccccccc}
			\hline \hline
			&  $\widehat{\boldsymbol{\beta}}_{C}$ & $\widehat{\boldsymbol{\beta}}_{O}$ & $\widehat{\boldsymbol{\beta}}_{G}$ & $\widehat{g}$ & $\widehat{\phi}$\\
			\hline
			\multirow{2}{*}{\textbf{LAM}}& 2.237 & 3.318 & 2.415 & 0.002 & 0.586 \\ 
			& (2.695) & (0.179) & (0.246) & (0.003) & (0.294) \\ 
			\hline
			& $\widehat{\boldsymbol{\beta}}_{C}$ & $\widehat{\boldsymbol{\beta}}_{O}$ & $\widehat{\boldsymbol{\beta}}_{G}$ &  $\widehat{g}$  & $\widehat{\phi}$& $\widehat{\boldsymbol{D}}^M_{E,2015}$ \\ 
				\hline 
					\multirow{2}{*}{\textbf{REF}}	&1.865 & 3.926 & 2.566 & 0.023 & 0.569 & 0.025 \\ 
			& (1.282) & (1.576) & (0.167) & (0.006) & (0.177) & (0.007) \\
			\hline
			&    $\widehat{\boldsymbol{\beta}}_{C}$ & $\widehat{\boldsymbol{\beta}}_{O}$ & $\widehat{\boldsymbol{\beta}}_{G}$ & $\widehat{g}$  & $\widehat{\phi}$& $\widehat{\boldsymbol{D}}^S_{E,1990}$  \\
			\hline
			\multirow{2}{*}{\textbf{OECD}}	&4.212 & 2.768 & 2.591 & 0.019 & 0.815 & 0.033 \\ 
			& (0.459) & (0.219) & (0.103) & (0.002) & (0.108) & (0.010) \\ 
			\hline
			&  $\widehat{\boldsymbol{\beta}}_{C}$ & $\widehat{\boldsymbol{\beta}}_{O}$ & $\widehat{\boldsymbol{\beta}}_{G}$& $\widehat{g}$ & $\widehat{\phi}$& $\widehat{\boldsymbol{D}}^S_{E,2004}$ & $\widehat{\boldsymbol{D}}^M_{E,1990}$\\
			\hline
			\multirow{2}{*}{\textbf{ASIA}}  &3.815 & 4.148 & 0.615 & 0.022 & 0.503 & $-0.075$ & $-0.072$\\ 
			&  (0.452) & (0.785) & (0.705) & (0.003) & (0.132) & (0.022) & (0.022)   \\ 
			\hline 
			&  $\widehat{\boldsymbol{\beta}}_{C}$ & $\widehat{\boldsymbol{\beta}}_{O}$ & $\widehat{\boldsymbol{\beta}}_{G}$& $\widehat{g}_1$ & $\widehat{g}_2$ &$\widehat{\phi}$& $\widehat{\boldsymbol{D}}^S_{Y,1982}$&$\widehat{\boldsymbol{D}}^M_{Y,1990}$& \\
			\hline
			\multirow{2}{*}{\textbf{MAF}}&6.132 & 1.32 & 3.569 & $-0.019$ & 0.004&0.979 & $-0.104$ & 0.212\\ 
			&  (18.713) & (1.018) & (0.532) & (0.007) &(0.006)& (0.06)& (0.080) & (0.070) \\ 
			\hline
				&  $\widehat{\boldsymbol{\beta}}_{C}$ & $\widehat{\boldsymbol{\beta}}_{O}$ & $\widehat{\boldsymbol{\beta}}_{G}$& $\widehat{g}$ & $\widehat{\phi}$& $\widehat{\boldsymbol{D}}^S_{E,1991}$&$\widehat{\boldsymbol{D}}^S_{Y,1990}$&  $\widehat{\boldsymbol{D}}^M_{Y,1989}$&\\
				\hline
				\multirow{2}{*}{\textbf{WORLD}}&3.788 & 3.188 & 2.385& 0.017 & 0.524&0.156& 1.086 & 1.040 \\ 
			&   (0.443) & (0.127) & (0.180) & (0.002) & (0.122) & (0.027) & (0.312) &(0.296)\\ 
			\hline
		\end{tabular}
	\end{threeparttable}
\end{table}

\subsection{Diagnostics}
	\label{Diagnostis EEE-g}
\begin{table}[H]
	\centering
	\caption{\footnotesize Diagnostic statistics of the standardized one-step ahead prediction errors of emissions and GDP for the E3S2-g model. It reports the first four moments (mean, standard deviation (Std), skewness (Skew), and kurtosis (Kurt) as well the test statistics of the Jarque-Bera test for non-normality \protect \cite{jarque1980efficient} (JB) and of the Ljung-Box test for autocorrelation \protect \cite{box1970distribution,ljung1978measure} (Q(1) corresponds to a lag order of 1 and Q(5)  lag order of 5). The null hypothesis of the Jarque-Bera is the skewness being 0 and the kurtosis being 3, and the null hypothesis of the Ljung Box test is that the residuals are serial uncorrelated when a fixed number of lags are included. }
	\setlength{\tabcolsep}{5pt} 
	\renewcommand{\arraystretch}{1.2} 
	\footnotesize
	\begin{tabular}{l|lcccccccccccccc}
		\hline \hline
		\multicolumn{9}{c}{\textbf{Geometric growth}} \\
		\hline 
		&	& Mean & Std & Skew & Kurt & JB & Q(1) & Q(5)  \\ 
		\hline
		\multirow{2}{*}{\textbf{LAM}}	&	Emissions & 0.036 & 1.010 & -0.145 & -0.807 & 1.194 & 0.022 & 1.711  \\ 
		&	GDP & 0.001 & 1.000 & -0.415 & 0.283 & 1.826 & 0.265 & 1.627\\ 
		\hline
		\multirow{2}{*}{\textbf{REF}}	&	Emissions & 0.091 & 1.014 & -0.044 & 2.364 & 0.480 & 0.736 & 9.032\\ 
		&	GDP &0.014 & 1.000 & -0.225 & 1.879 & 1.701 & 6.184$^*$ & 9.108 \\ 
		\hline
		\multirow{2}{*}{\textbf{OECD}}	&	Emissions	 & $-0.034$ & 1.009 & $-0.055$ & 2.136 & 1.517 & 0.169 & 3.757\\
		&	GDP & 0.000 & 1.000 & 0.565 & 4.075 & 4.862 & 1.114 & 8.146  \\ 
		\hline
		\multirow{2}{*}{\textbf{ASIA}}	&	Emissions & 0.040 & 1.009 & 0.174 & 3.277 & 0.397 & 0.192 & 2.569 \\ 
		&	GDP& 0.005 & 1.001 & -0.287 & 2.973 & 0.659 & 4.722$^*$ & 6.078 \\ 
		\hline 
		\multirow{2}{*}{\textbf{MAF}}&	Emissions  &  0.085 & 1.007 & -0.291 & 2.980 & 0.680 & 0.040 & 5.063  \\ 
		&	GDP  & 0.011 & 1.001 & 0.337 & 3.448 & 1.309 & 5.531 & 6.663 \\ 
			\hline 
		\multirow{2}{*}{\textbf{WORLD}}&	Emissions  &  0.008 & 1.011 & -0.108 & 3.128 & 0.127 & 0.001 & 4.862 \\ 
		&	GDP  & 0.005 & 1.000 & 0.478 & 3.080 & 1.838 & 10.071$^{**}$ & 14.932$^*$ \\ 
		\hline\hline
	\end{tabular}
\end{table}

\clearpage

\subsection{Projections}
\label{projection EEE-g}
\begin{figure}[h!]
	\centering
	\caption{\footnotesize Comparison of the CO$_2$ emissions projected from the E3S2-g model plus carbon capture and storage (CCS) and from SSP 1.9 W $\mathrm{m^{-2}}$. Confidence bands are pointwise at the 90\% level.}
	\begin{subfigure}{0.5\textwidth}
		\centering
		\includegraphics[width=\linewidth]{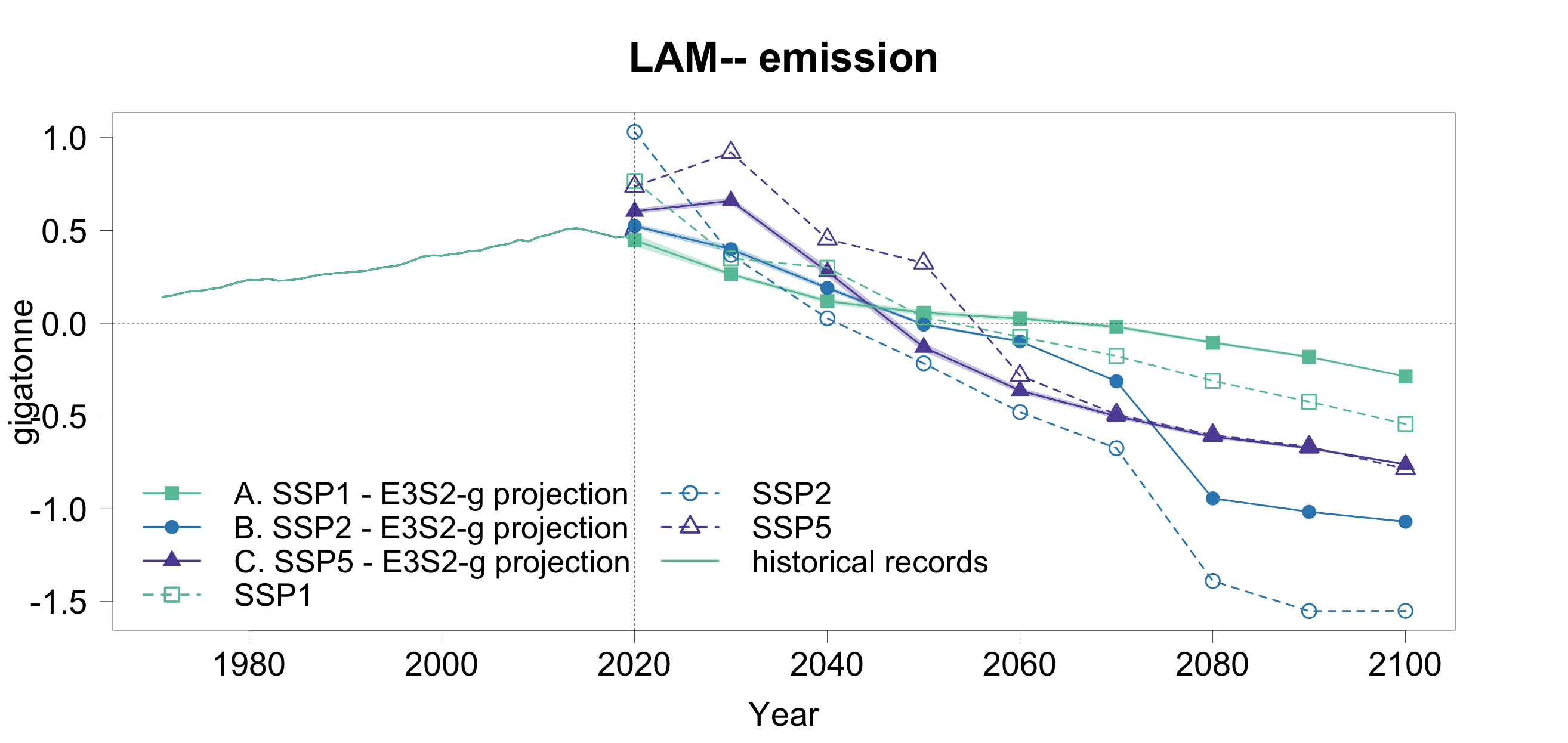}
	\end{subfigure}\hfill
		\begin{subfigure}{0.5\textwidth}
		\centering
		\includegraphics[width=\linewidth]{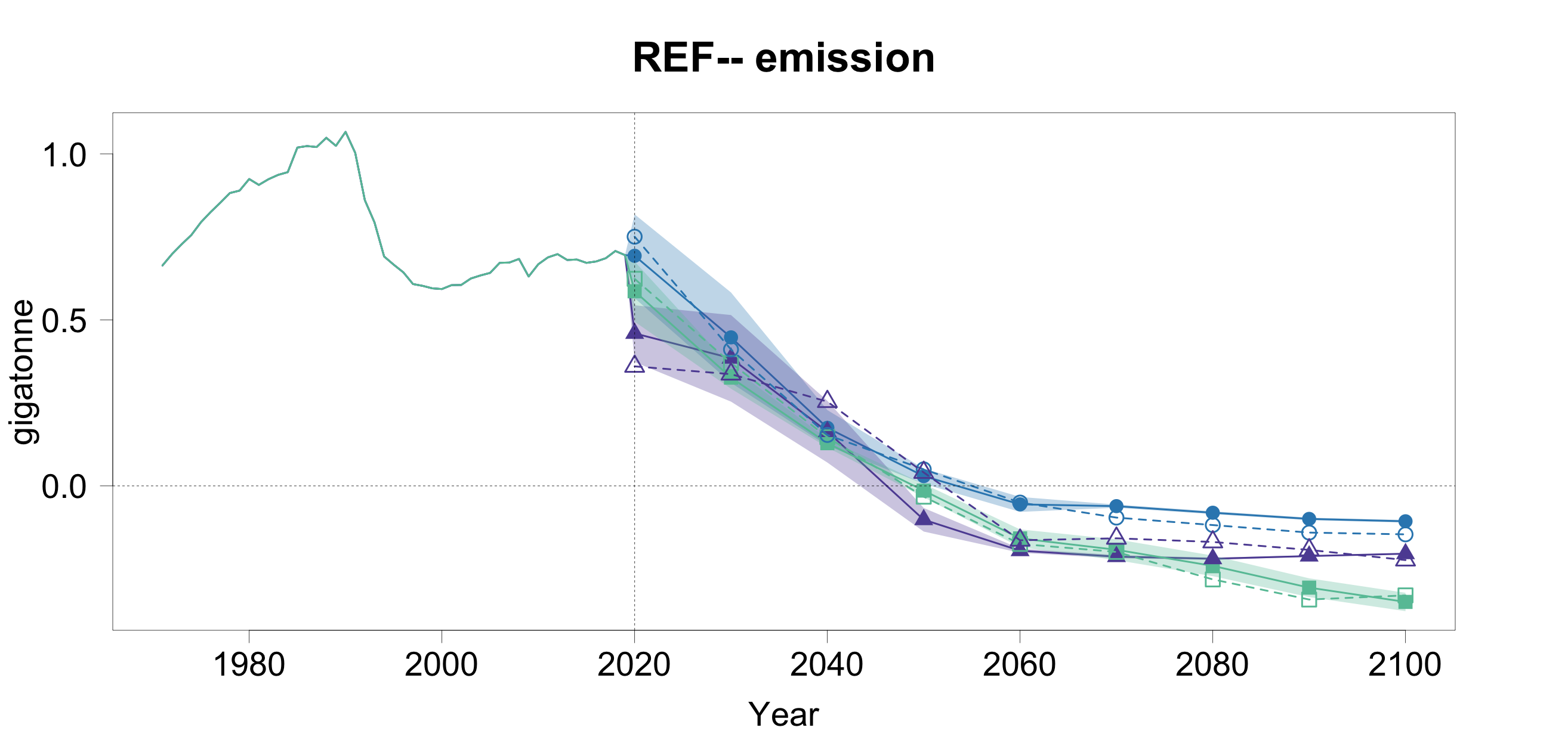}
		\end{subfigure}\\
	\begin{subfigure}{0.5\textwidth}
		\centering
		\includegraphics[width=\linewidth]{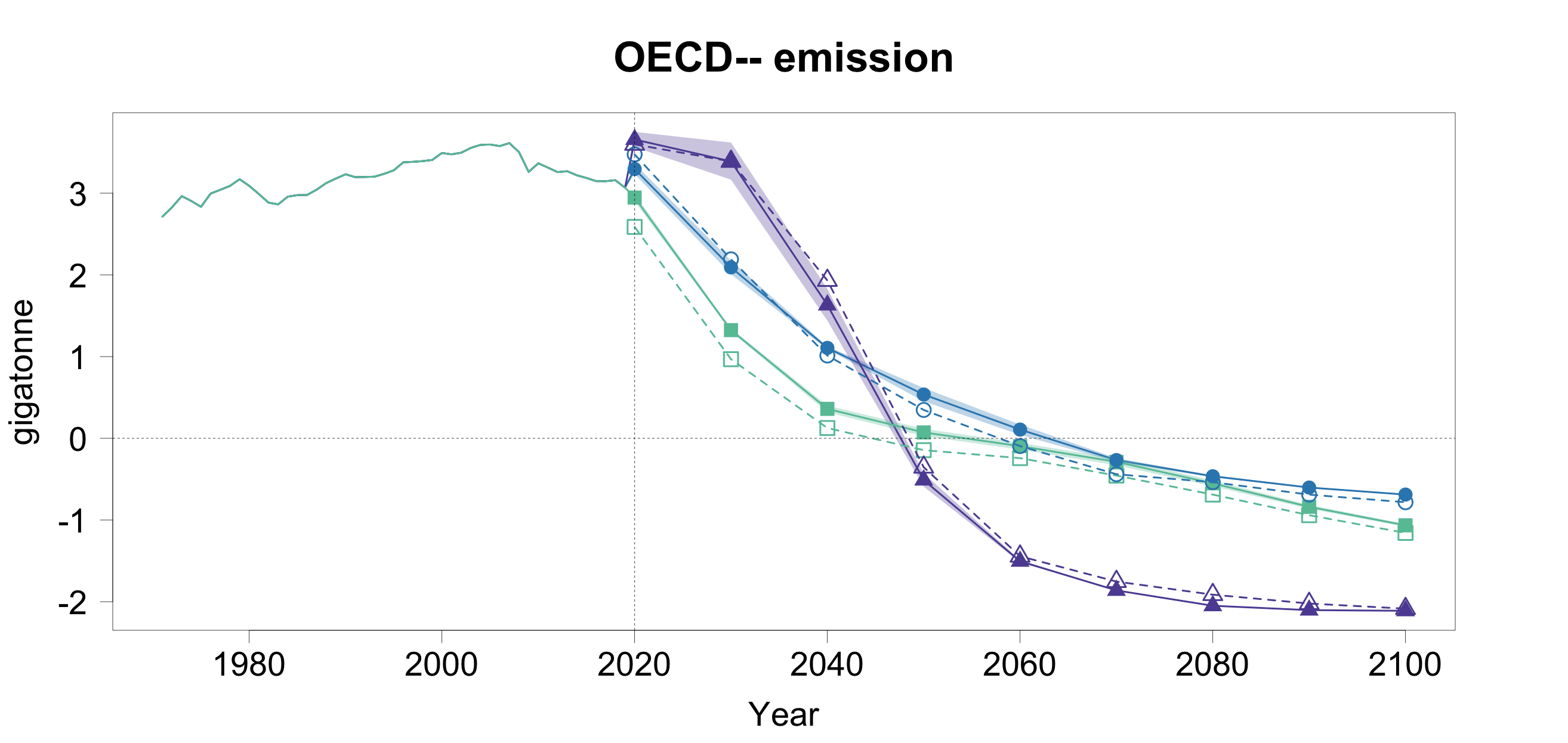}
	\end{subfigure}\\
	\begin{subfigure}{0.5\textwidth}
		\centering
		\includegraphics[width=\linewidth]{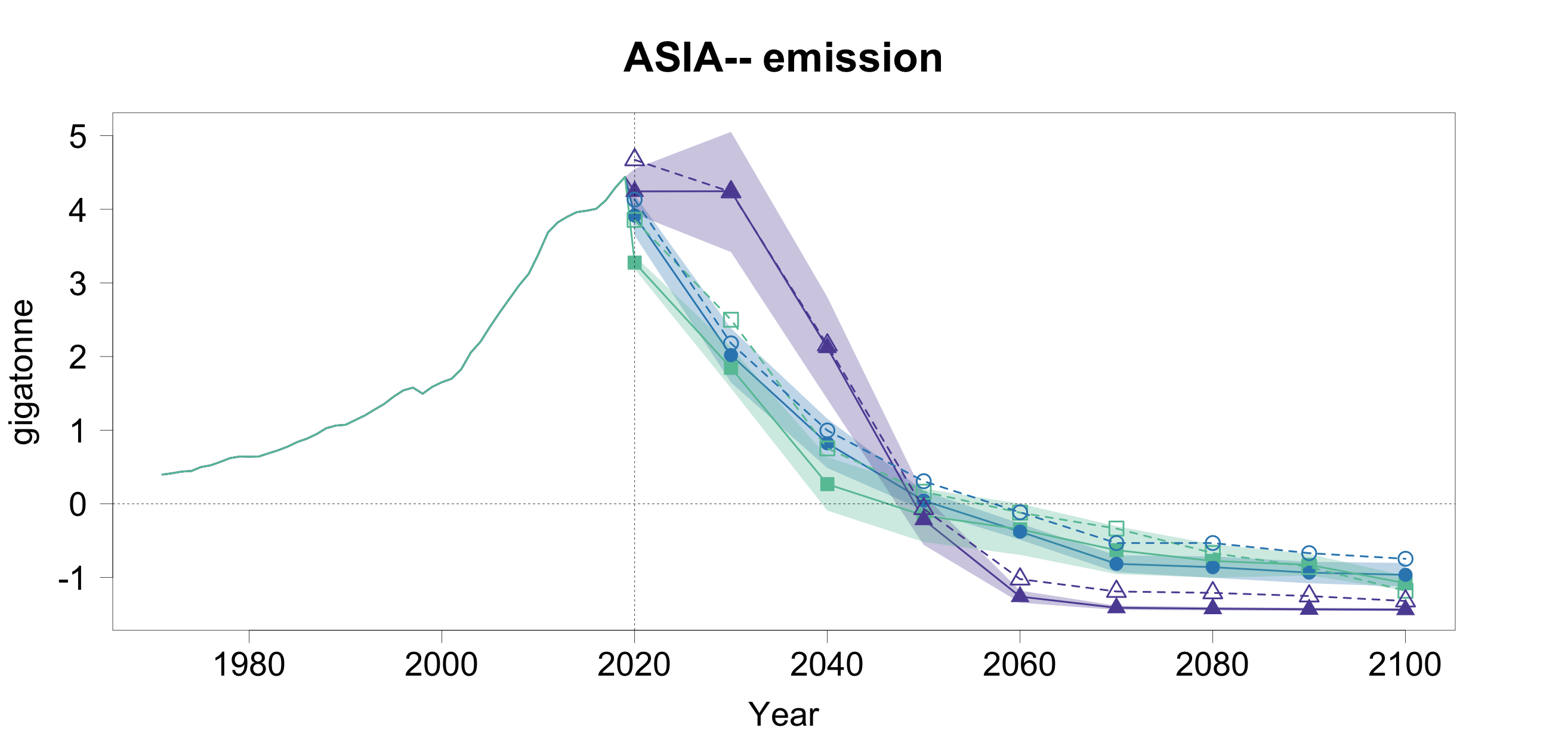}
	\end{subfigure}\hfill
	\begin{subfigure}{0.5\textwidth}
		\centering
		\includegraphics[width=\linewidth]{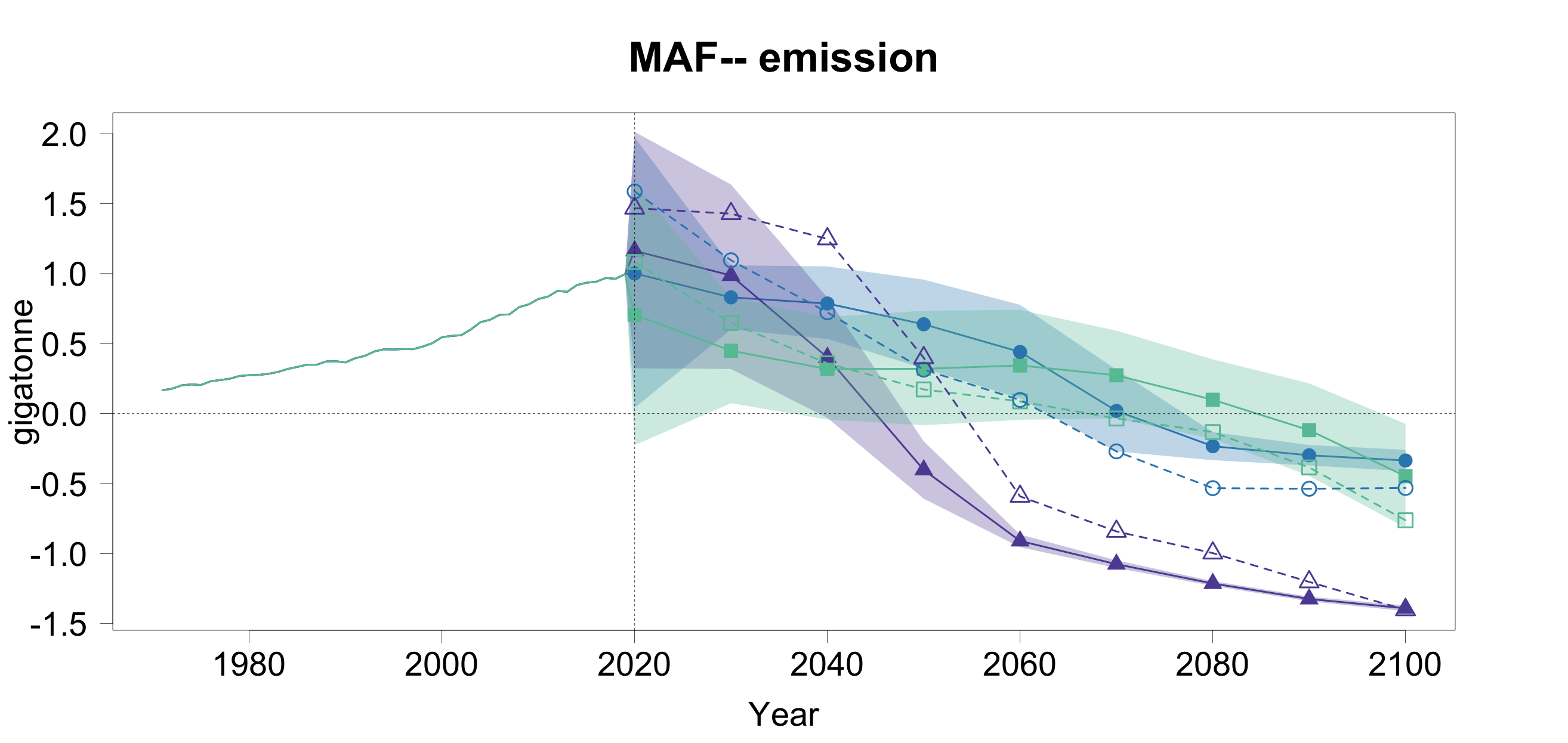}
	\end{subfigure}
	\label{emissionGeo}
\end{figure} 
\begin{figure}[h!]
	\centering
	\caption{Comparison of GDP projected by the E3S2-g model and SSP 1.9 W $\mathrm{m^{-2}}$. Confidence bands are pointwise at the 90\% level.}
	\begin{subfigure}{0.5\textwidth}
		\centering
		\includegraphics[width=\linewidth]{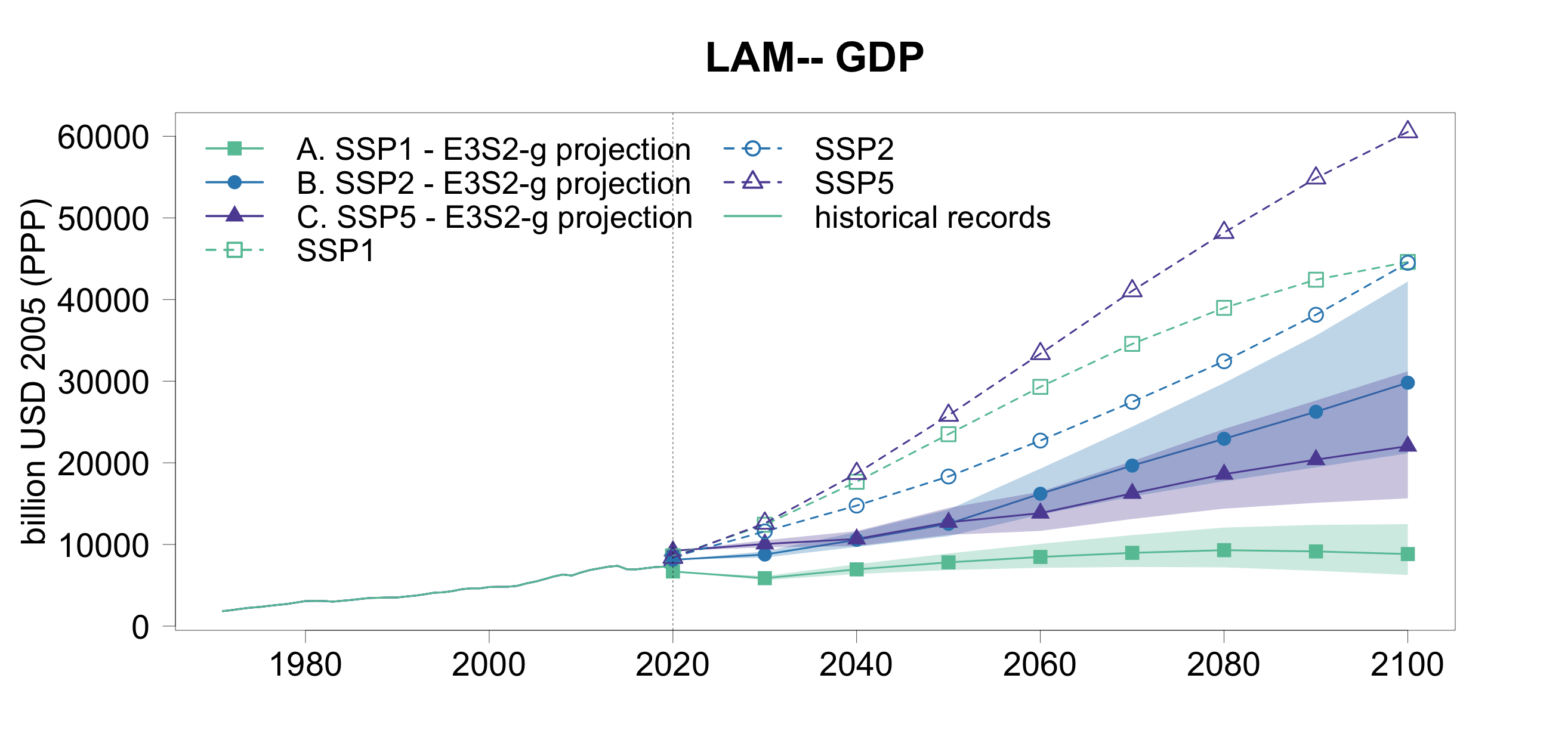}
	\end{subfigure}\hfill
		\begin{subfigure}{0.5\textwidth}
			\centering
			\includegraphics[width=\linewidth]{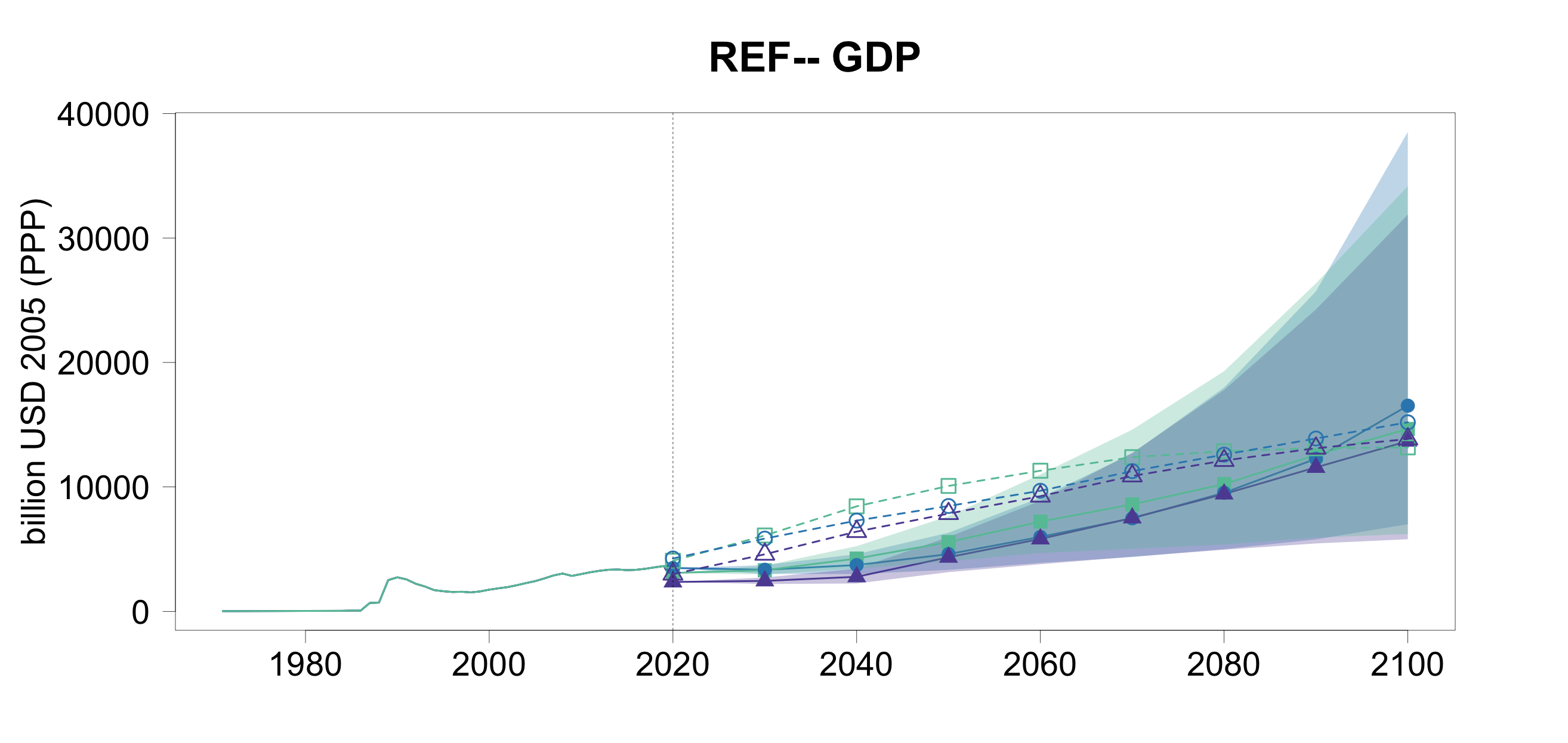}
		\end{subfigure}\\
	\begin{subfigure}{0.5\textwidth}
		\centering
		\includegraphics[width=\linewidth]{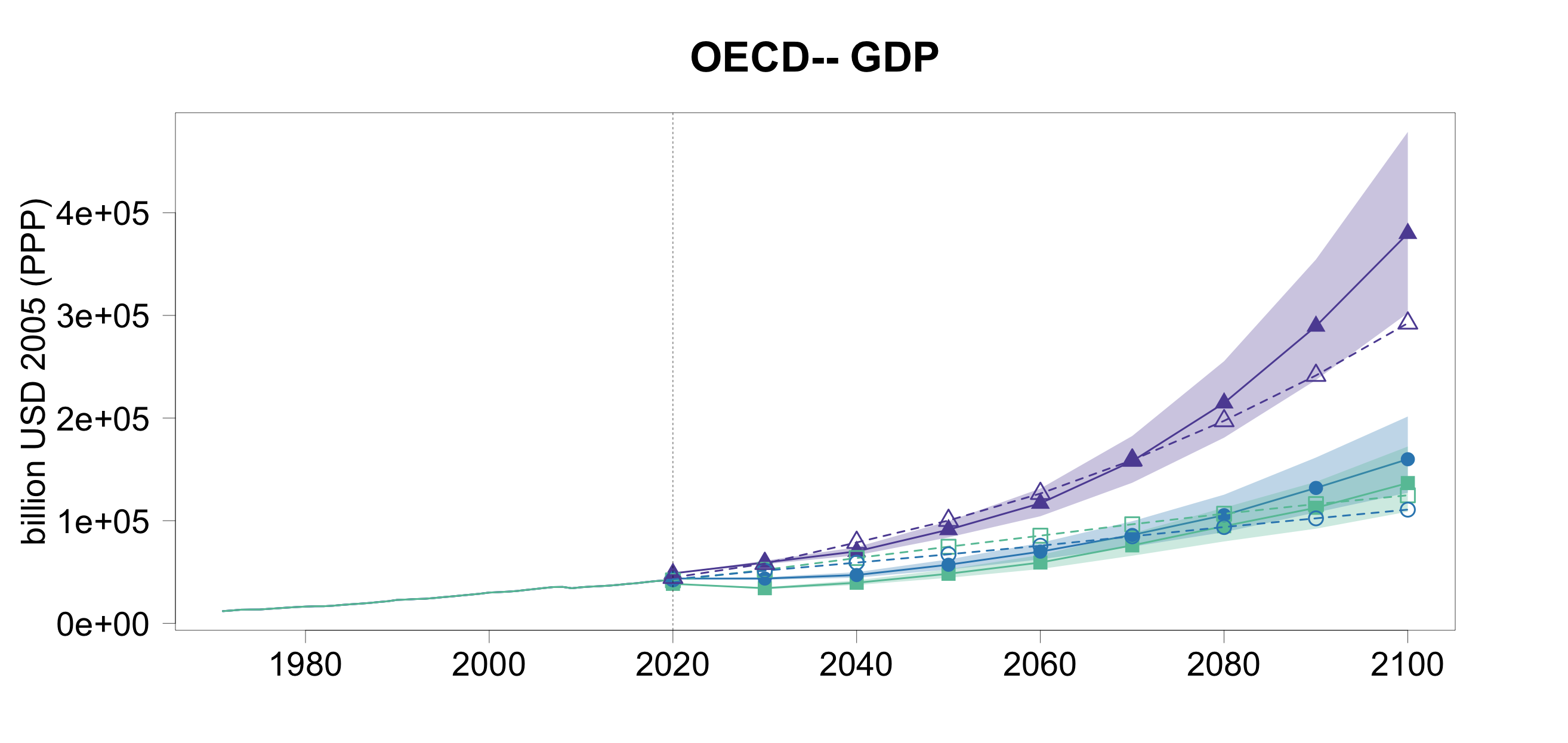}
	\end{subfigure}\\
	\begin{subfigure}{0.5\textwidth}
		\centering
		\includegraphics[width=\linewidth]{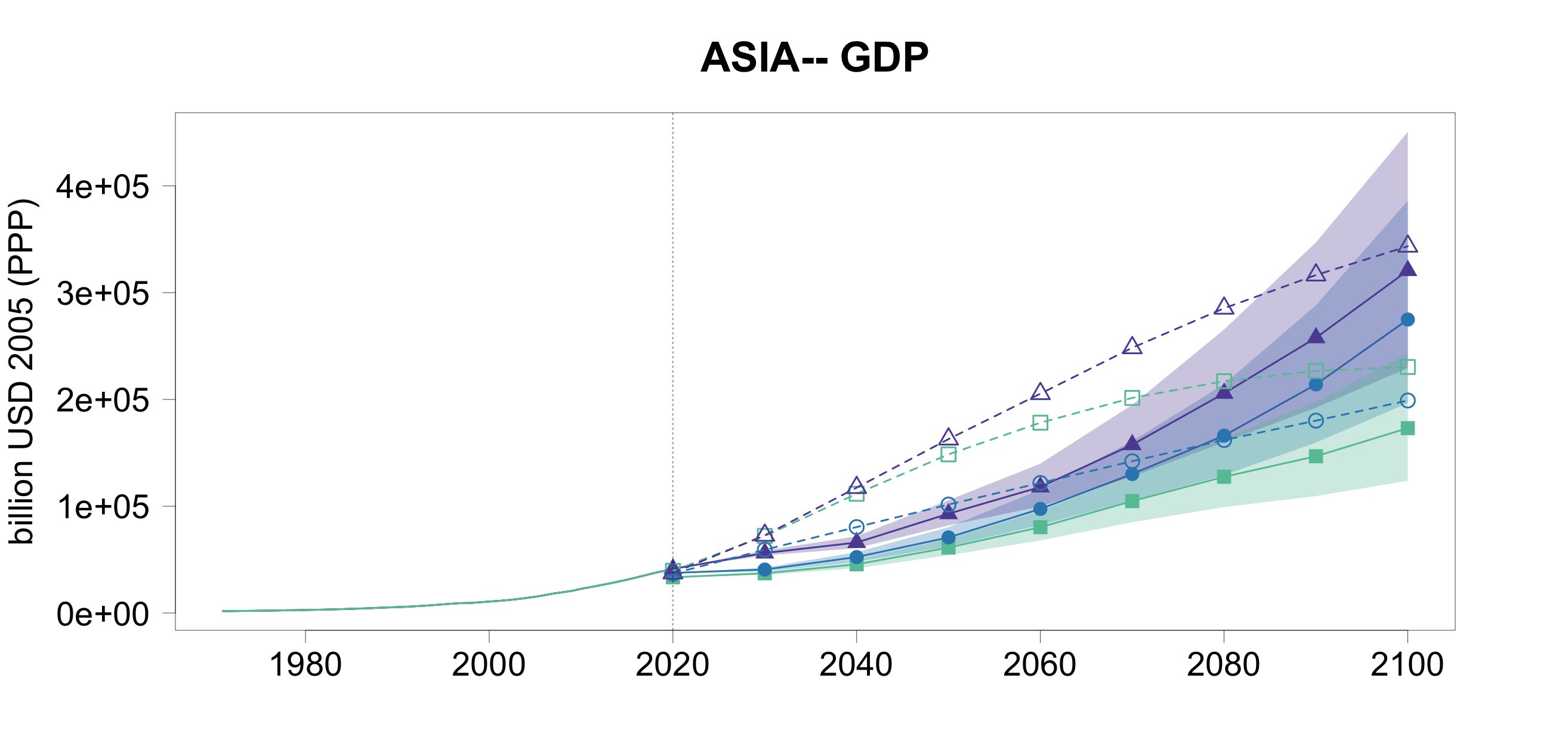}
	\end{subfigure}\hfill
	\begin{subfigure}{0.5\textwidth}
		\centering
		\includegraphics[width=\linewidth]{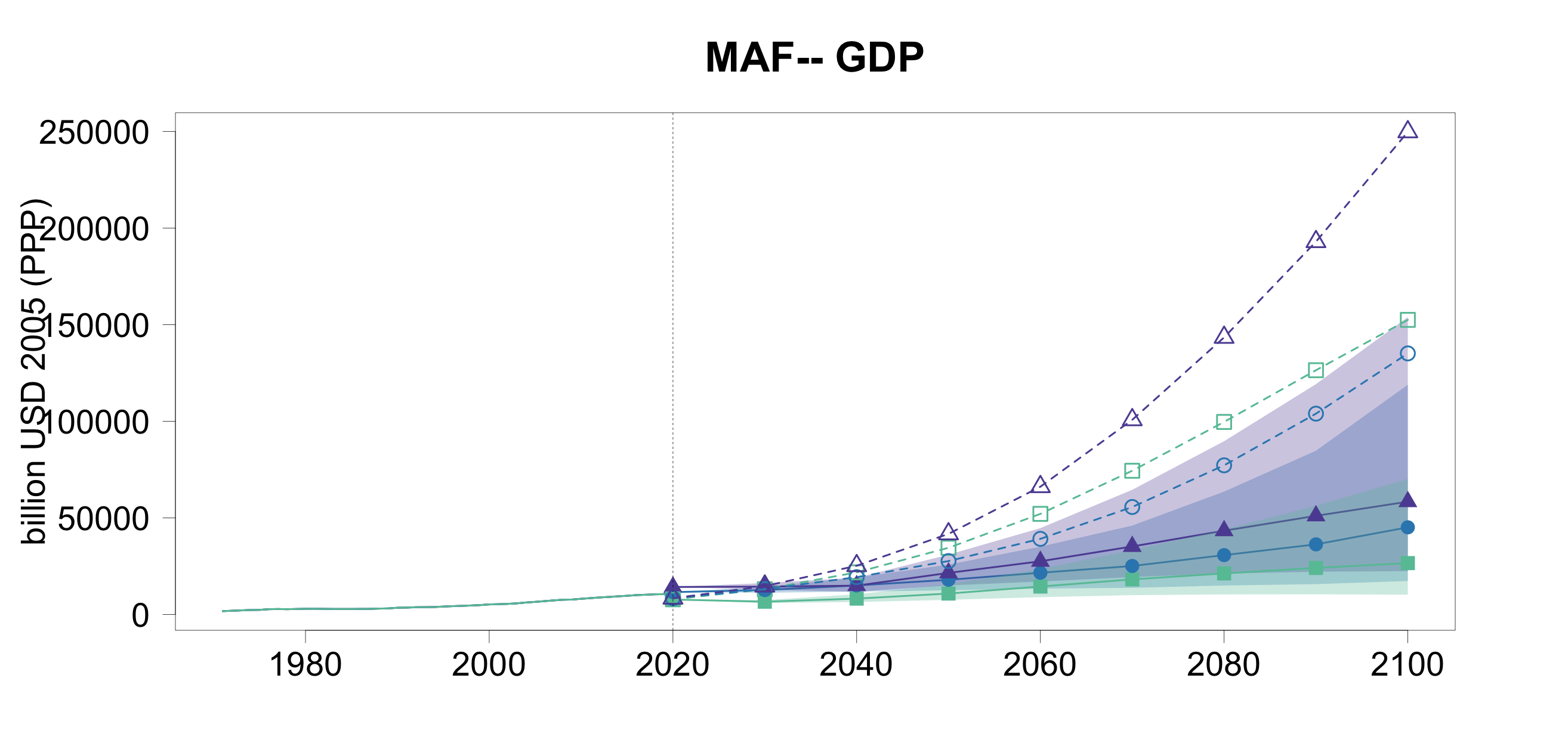}
	\end{subfigure}
	\label{GDPgeo}
\end{figure} 
\end{document}